\documentclass[
twocolumn,
amsfonts,
showpacs,
superscriptaddress,
amsmath,
amssymb,
aps,
longbibliography,
prx,
noeprint,
nofootinbib,
10pt
]{revtex4-2}

\usepackage{graphicx}
\usepackage{bm}
\usepackage[dvipsnames]{xcolor}
\usepackage[english]{babel}
\usepackage{dsfont}
\usepackage{comment}
\usepackage[caption=false]{subfig}
\usepackage[shortlabels]{enumitem}

\usepackage[colorlinks,%
            linkcolor=BrickRed,%
            citecolor=MidnightBlue,%
            urlcolor=MidnightBlue]{hyperref}

\newcounter{ls}

\newcommand{\bc}{\begin{center}}
\newcommand{\ec}{\end{center}}
\def\ba#1{\begin{array}{#1}\displaystyle}
\newcommand{\ea}{\end{array}}

\newcommand{\beq}{\begin{equation}}
\newcommand{\eeq}{\end{equation}}
\newcommand{\beqa}{\begin{eqnarray}}
\newcommand{\eeqa}{\end{eqnarray}}

\newcommand{\n}{\nonumber\\}
\newcommand{\bi}{\begin{itemize}}
\newcommand{\ei}{\end{itemize}}

\def\b#1{\bar{#1}}

\newcommand{\N}{{\mathbb{N}}}

\newcommand{\Tr}{{\rm Tr}}

\newcommand{\tth}{t_\mathrm{Th}}

\newcommand{\td}{t_{\mathrm{d},\alpha}}

\newcommand{\tramp}{t_{\mathrm{r},\alpha}}
\newcommand{\tH}{t_{\mathrm{H}}}

\newcommand{\ii}{{\rm i}}

\newcommand{\dd}{{\rm d}}

\renewcommand{\hat}[1]{\widehat{#1}}
\newcommand{\scW}{\mathcal{W}}
\newcommand{\scP}{\mathcal{P}}
\newcommand{\hatW}{\hat{\mathcal{W}}}
\newcommand{\hatF}{\hat{F}}

\newcommand{\rrangle}{\rangle\!\rangle}
\newcommand{\llangle}{\langle\!\langle}

\begin{document}

\title{Theory of Irreversibility in Quantum Many-Body Systems}

\author{Takato Yoshimura}
\email{takato.yoshimura@physics.ox.ac.uk}
\affiliation{All Souls College, Oxford OX1 4AL, U.K.}
\affiliation{Rudolf Peierls Centre for Theoretical Physics, University of Oxford, 1 Keble Road, Oxford OX1 3NP, U.K.}

\author{Lucas S\'a}
\email{ld710@cam.ac.uk}
\affiliation{TCM Group, Cavendish Laboratory, University of Cambridge, JJ Thomson Avenue, Cambridge CB3 0HE, U.K.\looseness=-1}

\begin{abstract}
We address the longstanding challenge in quantum many-body theory of reconciling unitary dynamics with irreversible relaxation. In classical chaos, the unitary evolution operator develops Ruelle-Pollicott (RP) resonances inside the unit circle in the continuum limit, leading to mixing. In the semiclassical limit, chaotic single-particle quantum systems relax with the same RP resonances. In contrast, the theory of quantum many-body RP resonances and their link to irreversibility remain underdeveloped. Here, we relate the spectral form factor to the sum of autocorrelation functions and, in generic many-body lattice systems without conservation laws, argue that all quantum many-body RP resonances converge inside the unit disk, highlighting the role of nonunitary and the thermodynamic limit. While we conjecture this picture to be general, we analytically prove the emergence of irreversibility in the random phase model (RPM), a paradigmatic Floquet quantum circuit model, in the limit of large local Hilbert space dimension. To this end, we couple it to local environments and compute the exact time evolution of autocorrelation functions, the dissipative form factor, and out-of-time-order correlation functions (OTOCs). Although valid for any dissipation strength, we then focus on weak dissipation to clarify the origin of irreversibility in unitary systems. When the dissipationless limit is taken after the thermodynamic limit, the unitary quantum map develops an infinite tower of decaying RP resonances---chaotic systems display so-called anomalous relaxation.  
We identify the exact RP resonances in the RPM and prove that the same RP resonances are obtained from operator truncation. We also show that the OTOC in the RPM can undergo a two-stage relaxation and that during the second stage, the approach to the stationary value is again controlled by the leading RP resonance. Finally, we demonstrate how conservation laws, many-body localization, and nonlocal interactions merge the leading RP resonance into the unit circle, thereby suppressing anomalous relaxation. Our findings clarify the microscopic origin of macroscopic relaxation in quantum systems and emphasize the role of RP resonances as a largely unexplored signature of many-body quantum chaos.
\end{abstract}

\maketitle

\section{Introduction}

The emergence of irreversible macroscopic physics from reversible microscopic dynamics is one of the oldest and most studied questions of statistical mechanics~\cite{brush2003}. 
The resolution of this apparent paradox lies in spontaneous symmetry breaking---the solutions of the equations of motion can have less symmetry than the equations themselves---in this case, of time-reversal symmetry. Statistical mechanics gives a more thorough grounding to these ideas by providing a mechanism for selecting the time-reversal-broken solutions~\cite{Gaspard_1998,gaspardReview}. 

In classical systems, the probability density of an ensemble of trajectories evolves in time under the action of the Frobenius-Perron (FP) operator. Because of the conservation of phase-space volume (Liouville's theorem), the evolution is unitary and the eigenvalues of the FP operator lie on the unit circle in the complex plane. However, if the system is chaotic, the spectrum of the FP operator becomes dense, and isolated singularities can emerge inside the unit circle upon analytic continuation, the so-called Ruelle-Pollicott (RP) resonances~\cite{ruelle1985,ruelle1986,pollicott1985,pollicott1986}. These resonances control the decay of different observables, with the associated timescales inversely proportional to their distance to the unit circle, and the leading RP resonance rules the late-time approach to a steady state. Thus, the singularities arising from chaos lead to irreversibility in purely unitary dynamics~\cite{Gaspard_1998,gaspardReview}.

The same mechanism is at play in quantum systems~\cite{manderfeld2001JPA,nonnenmacher2003,garcia-mata2003PRL,garcia-mata2004PRE,garcia-mata2018,prosen2002,prosen2004,prosen2007,mori2024,Yoshimura2024,znidaric2024,znidaric2024b,jacoby2024,zhang2024}. The density matrix of an ensemble of quantum systems evolves under a quantum channel, which is unitary for closed systems. Yet, correlation functions decay, and, correspondingly, some quantum analog of RP resonances must be hidden in the spectrum of the quantum channel. A first hurdle to this approach is that the spectrum of a quantum channel is not dense even if the system is chaotic---a finite-size quantum chaotic system has a discrete spectrum with eigenvalues correlated according to random matrix theory (RMT)~\cite{Bohigas_Characterization_1984,haake2013}.
Accordingly, to obtain a dense spectrum, one needs to take either the semiclassical~\cite{manderfeld2001JPA,nonnenmacher2003,garcia-mata2003PRL,garcia-mata2004PRE,garcia-mata2018} or the thermodynamic limit~\cite{prosen2002,prosen2004,prosen2007,mori2024,Yoshimura2024,znidaric2024,znidaric2024b,jacoby2024,zhang2024}. In single-particle system, the former recovers the classical RP resonances; in many-body systems, the latter is required. Given the practical impossibility of taking the strict thermodynamic limit, alternative mechanisms have been devised that provide a ``seed'' for quantum RP resonances, making them visible in large but finite systems. Prosen~\cite{prosen2002,prosen2004,prosen2007} first proposed coarse-graining in operator space, which renders the quantum channel nonunitary. With an appropriately designed algorithm~\cite{prosen2002,znidaric2024}, making the coarse-graining finer amounts to simultaneously taking the thermodynamic and unitary limits in the correct order and one can observe a quick convergence of eigenvalues to points \emph{inside} the unit disk---these are the quantum RP resonances. Remarkably, this procedure provided not only fundamental insight into the structure of many-body quantum systems but also a powerful numerical tool, as the decay of correlation functions converged faster to the thermodynamic limit with the coarse-graining procedure. The same idea is also behind subsequent efficient numerical methods~\cite{Rakovszky_Dissipation_2022,Keyserlingk_Operator_2022}.

More recently, it was realized that explicitly coupling the system of interest to an external bath provides an alternative seed for quantum RP resonances, with the advantage of also elucidating the physics of the weak-dissipation limit of chaotic open many-body quantum systems~\cite{sa2022PRR,garcia2023PRD,scheurer2023ARXIV,mori2024,Yoshimura2024}. These systems retain a finite (order one) relaxation rate even when the dissipation strength vanishes, provided that the thermodynamic limit is taken before the weak-dissipation limit---a phenomenon dubbed anomalous relaxation in Ref.~\cite{garcia2023PRD2}. Anomalous relaxation was first observed in the dissipative Sachdev-Ye-Kitaev model~\cite{sa2022PRR,garcia2023PRD2} and, later, in a dissipative quantum spin-liquid model~\cite{scheurer2023ARXIV}, the driven-dissipative transverse-field Ising model~\cite{mori2024}, and an open quantum circuit~\cite{Yoshimura2024}. It was conjectured to be a generic feature of chaotic systems and absent in integrable ones~\cite{garcia2023PRD2}, with Mori~\cite{mori2024} formalizing these ideas and connecting them to quantum RP resonances.
We note that Ref.~\cite{garcia2023PRD2} also put forward a possible holographic interpretation of anomalous relaxation---and, therefore, quantum RP resonances---in gravity, see also Ref.~\cite{dodelson2024}.

Despite all this recent progress, the theory of quantum many-body RP resonances is still rudimentary.
The existence of quantum RP resonances can be rigorously established in single-particle  systems~\cite{jaksic2002,nonnenmacher2003,garcia-mata2003PRL,garcia-mata2004PRE,garcia-mata2018} and is well-understood. Notwithstanding, the presence of local interactions in many-body quantum systems renders many-body quantum chaos fundamentally different from its single-particle counterpart. It thus remains unclear if the different manifestations of quantum many-body RP resonances that have recently appeared in the literature are equivalent and how they relate to the different timescales of many-body quantum chaos.
While most of the scarce evidence in interacting many-body systems is heuristic or numerical, the dissipative random phase model (DRPM), a quantum-chaotic dissipative quantum circuit model introduced by the same authors of the present paper in Ref.~\cite{Yoshimura2024}, is a notable exception.
In the limit of large local Hilbert space dimension, this model is exactly solvable for arbitrary system size and dissipation strength and constitutes, therefore, a perfect testbed to further develop the theory of irreversibility of many-body quantum systems. In this paper, we proceed in this direction with four distinct goals in mind.

First, we present a unified picture of dynamical and spectral many-body quantum chaos, and how RP resonances---and irreversible behavior---emerge from the spectrum of the unitary time evolution operator in the thermodynamic limit. We do this by expanding the spectral form factor (SFF) in terms of autocorrelation functions and identifying those operators whose autocorrelation functions decay in time (as dictated by RP resonances) and those whose autocorrelation functions grow at late times (giving rise to the ramp of the SFF) in the thermodynamic limit. We then show how dissipation and operator truncation equivalently remove the latter from the SFF, allowing us to use the SFF as a probe of the RP resonance spectrum. In the remainder of the paper, we then illustrate this general mechanism analytically in the DRPM and numerically for qubit brickwork random Floquet circuits (RFCs).

Second, we exploit the exact solvability of the DRPM in the limit of large local Hilbert space dimension to perform the first analytical computation of a full quantum many-body RP resonance spectrum. We obtain not only the leading RP resonance~\cite{Yoshimura2024} but an infinite tower of RP resonances that rule the decay of all possible autocorrelation functions in the closed-system limit of the DRPM, the random phase model (RPM)~\cite{Chan_Spectral_2018}. Furthermore, the analytical control we have over the model also provides a clear physical mechanism for the emergence of the quantum RP resonances, namely, the same domain walls responsible for the Thouless physics in the RPM~\cite{Chan_Spectral_2018}---the physics before the onset of universal random-matrix behavior. We conjecture this mechanism to be universal in quantum many-body systems.

Third, we study a so-far unexplored dynamical regime of many-body quantum chaos, the \emph{Ruelle regime}. Arguably, the two most striking features of chaos are sensitivity to initial conditions and mixing. If a quantum system has a well-defined semiclassical regime, the former manifests as an early-time exponential growth in out-of-time-order correlators (OTOCs), which defines a quantum Lyapunov exponent. The Lyapunov regime has been extensively investigated for both single-particle~\cite{larkin1969,berman1978,jalabert2018,garcia-mata2018} and many-body systems~\cite{kitaev2015,maldacena2016JHEP}, but many systems without an effective semiclassical parameter completely lack this behavior. On the contrary, the decay of correlation functions is a universal feature of all chaotic quantum systems---the quantum version of mixing. 
The characterization of chaos at intermediate timescales proportional to the system size---much later than the initial growth of correlations but much earlier than the ultralong timescales characterized by random matrix universaility---has remained essentially unexplored and, crucially, this is the regime where the effect of locality (i.e., many-body interactions) becomes the starkest.
To illustrate this point, we analytically demonstrate that, while the OTOC of the RPM has no Lyapunov regime, the same quantum RP resonances as before rule the saturation of the OTOC~\cite{polchinski2015} (and its decay in the presence of dissipation), which is identified as the Ruelle regime of the model.

Fourth, we clarify the conditions for the occurrence of anomalous relaxation in many-body quantum systems. We show that to obtain complex RP resonances inside the unit disk we need few-body interactions and extensive dissipation. Even when these conditions are satisfied, some RP resonances can be pushed to the unit circle in the presence of conserved quantities. We show this is the case analytically for the DRPM with a conserved $U(1)$ charge and numerically for a brickwork RFC in a many-body localized phase. Together with the general picture we developed before, these results provide a comprehensive understanding of the emergence of irreversibility in quantum many-body systems.

We start by giving an overview of our general theory and main results in Sec.~\ref{sec:general}. In Sec.~\ref{sec:DRPM}, we give a detailed illustration of the main results in the context of the RPM. Finally, we test the limits of anomalous relaxation both analytically and numerically in Sec.~\ref{sec:suppression}. Concluding remarks are presented in Sec.~\ref{sec:conclusion}.

{
\hypersetup{linkcolor=black}
\tableofcontents 
}

\section{General theory and main results}
\label{sec:general}

\subsection{Classical Ruelle-Pollicott resonances}\label{sec:classicalRP}

Ruelle-Pollicot (RP) resonances are known to govern late-time asymptotics in classical systems with Hamiltonian $H$. Let us briefly review how one can extract the RP resonances based on the resolvent formalism~\cite{Hasegawa_1992_unitarity,Gaspard_1998,gaspardReview}. In what follows, we assume that the Hamiltonian depends on time in a periodic way $H(t)=H(t+1)$ where the Floquet period is taken to be 1. We also measure time in units of the period, i.e., $t$ is a nonnegative integer.

Consider a classical chaotic system of $N$ particles in the $2d$-dimensional phase space $\mathcal{M}\subseteq\mathbb{R}^{2d}$ and denote its coordinates by $X=(q_1,\dots,q_N;p_1,\dots,p_N)$. The density distribution $f(X)=N^{-1}\sum_{i=1}^N\delta(X-X_i)$ is then evolved in time by the Liouville operator $\mathcal{L}_\mathrm{cl}(t)\bullet=\{H(t),\bullet\}$ as $f_t(X)=U^tf_0(X)$, where $U=e^{\int_0^1\dd\tau\,\mathcal{L}_\mathrm{cl}(\tau)}$ is the FP operator (i.e., the time-evolution operator). When convenient, we shall use the vectorized notation $(X|f)=(f|X)^*=f(X)$ and denote the expectation value of an observable $A$ by the inner product $(A|f)$. For example, the time-evolved density distribution can be expressed as
\begin{equation}
    f_t(X)=\sum_{\alpha}(X|\alpha)(\alpha|U^tf)=\sum_{\alpha,\alpha'}(X|\alpha)(\alpha|U^t|\alpha')(\alpha'|f),
\end{equation}
where $|\alpha)$ forms an orthonormal basis of $L^2(\mathcal{M})$ with $(\alpha|\alpha')=\delta_{\alpha\alpha'}$.

The FP operator is unitary when acting on $L^2(\mathcal{M})$, and thus its continuous spectrum lies on the unit circle $|z|=1$. However, such a spectrum tells us little about the late-time decay of correlation functions. To describe the decay, one needs to look at the behavior of the matrix elements of the FP operator $(\alpha|U^t|\alpha')$, which can be rewritten in terms of the resolvent $1/(z-U)$:
\begin{equation}
(\alpha|U^t|\alpha')=\frac{1}{2\pi\ii}\oint_C\dd z\,e^{-\ii zt}(\alpha|\frac{1}{z-U}|\alpha'),
\end{equation}
where the counterclockwise contour $C$ is placed just outside of the unit circle $z=1$. Now, consider shrinking the contour so that it passes the unit circle, at which moment the matrix elements $(\alpha|U^t|\alpha')$ undergo discontinuous changes. Importantly, however, these matrix elements remain finite as $|z|$ approaches 1, as there are no eigenoperators $|\alpha)\in L^2(\mathcal{M})$ that correspond to the eigenvalues constituting the continuous spectrum, which we will demonstrate shortly. This implies that the unit circle serves as a jump between two Riemann sheets, and poles in the second Riemann sheet $|z|<1$ would appear when analytically continuing from the first one. These poles are the RP resonances, which can be captured by shrinking the contour towards the origin. Assuming there are no singularities other than the isolated poles, the matrix elements can be expressed as
\begin{equation}
     (\alpha|U^t|\alpha')=\frac{1}{2\pi\ii}\sum_{k=0}^\infty\oint_{z=e^{-{\nu_k}}}\dd z\,e^{-\ii zt}(\alpha|\frac{1}{z-U}|\alpha'),
\end{equation}
where $e^{-\nu_k}$ with $\mathrm{Re}\,\nu_k>0$ is the position of a pole labeled by the nonnegative integer $k$ and the integration is done on a contour around $e^{-\nu_k}$ without hitting or encircling any other poles.

To proceed, it is convenient to introduce projection operators
\begin{equation}
    \Pi^{(k)}:=\frac{1}{2\pi\ii}\oint_{z=e^{-{\nu_k}}}\dd z\,\frac{1}{z-U},
\end{equation}
which commute with the FP operator $U \Pi^{(k)}= \Pi^{(k)}U$ and satisfy $ \Pi^{(k)} \Pi^{(k')}= \Pi^{(k)}\delta_{kk'}$~\cite{Hasegawa_1992_unitarity}. Note that since the integration contour encircles the simple pole $z=e^{-{\nu_k}}$ in $\Pi^{(k)}$, it follows that $U\Pi^{(k)}=e^{-{\nu_k}}\Pi^{(k)}$. The projector $\Pi^{(k)}$ can be further decomposed as $\Pi^{(k)}=|R_k)(L_k|$ where $|R_k)$ and $|L_k)$ are right and left generalized eigenstates satisfying
\begin{equation}\label{eq:classical_genstates}
    U|R_k)=e^{-{\nu_k}}|R_k),\quad (L_k|U=e^{-{\nu_k}}(L_k|.
\end{equation}
Let us explain why $|R_k)$ and $|L_k)$ are not actual eigenstates but generalized ones. To see it, observe first that $(R_k|U^\dagger U|R_k)=(R_k|R_k)$ since $U$ is a unitary operator. On the other hand, we also note that $(R_k|U^\dagger U|R_k)=e^{-2\mathrm{Re}\,\nu_k}(R_k|R_k)$ according to Eq.~\eqref{eq:classical_genstates}. These two relations contradict each other unless the norm $(R_k|R_k)$ is not normalizable, i.e., diverges, rendering the states not elements of $L^2(\mathcal{M})$. Together with a similar observation made for $(L_k|$, we thus conclude that they are not the eigenfunctions of $L^2(\mathcal{M})$ but rather Schwartz distributions~\cite{Hasegawa_1992_unitarity}.

Since 
\begin{equation}
    (\alpha|U_t|\alpha')=\sum_ke^{-\nu_kt}(\alpha|\Pi^{(k)}|\alpha')
\end{equation} 
holds for any matrix element, it is immediately observed that the late-time decay of the average of a traceless observable $A$ (i.e., a function $A$ whose integral over the whole phase space vanishes) with respect to an initial density distribution $f$, which we denote by $\langle A\rangle_t$, is ruled by the leading RP resonance:
\begin{equation}
     \langle A\rangle_t=\sum_{k=0}^\infty e^{-\nu^*_kt}(A|\Pi^{(k)}|f)\simeq e^{-\nu^*_1t}(A|\Pi^{(1)}|f).
\end{equation}
The resonance $\nu_0=0$ is associated with the uniform steady-state density distribution and, therefore, does not contribute to the dynamics of traceless observables.
Note that sometimes boundary conditions prohibit the presence of the leading RP resonance $\nu_1$, in which case the late time dynamics is controlled by the next leading one $\nu_2$ (we order the RP resonances according to their magnitudes of the real part $0\leq|\mathrm{Re}\,\nu_0|\leq|\mathrm{Re}\,\nu_1|\leq\cdots$), an example of which will be discussed in Sec.~\ref{sec:DRPM_autocorrelation}. 

While the exact RP resonances have been computed for simple models such as the Baker map~\cite{Hasegawa_1992_unitarity}, it is in general difficult to evaluate them by obtaining $|R_k)$ and $|L_k)$, which are well-defined on an appropriate space that is not $L^2(\mathcal{M})$. In practice, to avoid such an obstacle in computing the RP resonances directly, one adds a small noise to the dynamics, which in turn transforms the Liouville operator into the Fokker-Planck operator $\mathcal{L}_\mathrm{FP}=\mathcal{L}_\mathrm{cl}+D\nabla^2$~\cite{Malinin_nonlinear_2008}, which can be diagonalized more easily. The eigenvalues of the Fokker-Planck operator $\mathcal{L}_\mathrm{FP}$ are known to coincide with those of the original Liouville operator in the noiseless limit $D\to0$~\cite{Gaspard_bifurcation_1995}.

\subsection{Quantum Ruelle-Pollicott resonances}
\label{sec:quantumRP}

The previous construction holds, essentially unaltered, for single-particle quantum systems, e.g., in low-dimensional quantum maps~\cite{nonnenmacher2003,garcia-mata2003PRL,garcia-mata2004PRE,garcia-mata2018} or quantum billiards~\cite{pance2000PRL,sridhar2002JSP}. In the semiclassical limit, correlations decay with a rate given by the RP resonances of the limiting classical dynamical system. In contrast, truly quantum RP resonances occur only for quantum systems in the thermodynamic limit, even without any classical limit; how these emerge from the unitary dynamics of isolated systems is the focus of the rest of the paper.

The first attempt to generalize the concept of RP resonances to many-body quantum settings was made by Prosen in Ref.~\cite{prosen2002} where a course-graining procedure using operator truncation was introduced to define a nonunitary time-evolution operator. This procedure was inspired by analytic continuation in the classical case, and the eigenvalues of the nonunitary operator were then identified as the quantum RP resonances, which dictate the late-time dynamics of the system. Recently, another way of defining the quantum RP resonances, this time paralleling the stochastization of the Liouville operator, was proposed by Mori~\cite{mori2024}. Namely, it was suggested that the eigenvalues of the time-evolution operator in the presence of weak dissipation are related to quantum RP resonances and, in particular, they coincide precisely with the RP resonances in the limit of vanishing dissipation after taking the thermodynamic limit. In this paper, we shall primarily use the second approach to define the quantum RP resonances. We shall demonstrate that the above two approaches follow from the same physical mechanism and, hence, give rise to the same RP resonances.

Let us illustrate how RP resonances defined in this way generically govern the late-time asymptotics of correlation functions in a many-body Floquet system without conservation laws. Let us consider a Floquet lattice model with $L$ sites and local Hilbert space dimension $q$ such that the total Hilbert space dimension is $D=q^L$.
If the system is closed, then the discrete time evolution is implemented by the unitary operator $W$ so that the density matrix $\varrho$ evolves as
\begin{equation}
\label{eq:closed_channel}
    \varrho(t)=W^t \varrho(0)(W^\dagger)^t,
\end{equation}
This is no longer true in an open system, where, if the dynamics is discrete in time, the quantum channel $\mathcal{W}$ evolves the state $\varrho$ as 
\begin{equation}
\label{eq:open_channel}
\varrho(t)=\mathcal{W}^t[\varrho(0)], 
\qquad 
\mathcal{W}[\varrho(0)]=\sum_{j}K_{j}\varrho(0) K_{j}^\dagger,
\end{equation}
where $K_{j}$ forms a set of Kraus operators satisfying $\sum_j K_j^\dagger K_j=I_D$ (with $I_D$ the $D\times D$ identity matrix), which encode the dissipators used and are parametrized by a dissipation strength $\gamma\geq0$. In the limit of a closed system ($\gamma\to0)$, we have a single Kraus operator $K_{j}=W$ and we recover Eq.~(\ref{eq:closed_channel}). 

Since, in general, $\mathcal{W}$ is a nonunitary operator, its spectrum lies within the unit circle except for the eigenvalue corresponding to the infinite-temperature stationary state, which is equal to $1$. The eigenvalues of $\mathcal{W}$ are thus different from those of its conjugate $\mathcal{W}^\dagger$, giving rise to the biorthogonal basis $\{|R_k\rrangle,|L_k\rrangle\}_{k=0}^{D^2-1}$ that satisfies:
\begin{equation}
    \mathcal{W}|R_k\rrangle=e^{-\nu_k(\gamma)}|R_k\rrangle,\quad \llangle L_k|\mathcal{W}=\llangle L_k|e^{-\nu_k(\gamma)}.
\end{equation}
We order the eigenvalues as $0\leq |\mathrm{Re}\,\nu_0(\gamma)|\leq\cdots\leq|\mathrm{Re}\,\nu_{D^2-1}(\gamma)|$.
In terms of the basis operators, any observable $\mathcal{O}$ can be then decomposed as $|\mathcal{O}\rrangle=\sum_kc_k|L_k\rrangle$ where $c_k=\llangle R_k|\mathcal{O}\rrangle=\Tr(R^\dagger_k \mathcal{O})$, which implies that 
\begin{equation}
\mathcal{O}(t)=(\mathcal{W}^\dagger)^t[\mathcal{O}(0)]=\sum_kc_ke^{-\nu_k^*(\gamma)t}|L_k\rrangle.
\end{equation}
Importantly, we show below that if the dissipationless limit is taken after the thermodynamic limit, some of $e^{-\nu_k(\gamma)}$ remain within the unit circle~\cite{mori2024}. We shall thus call 
\begin{equation}
\label{eq:def_quantum_RP}
\nu_k=\lim_{\gamma\to0}\lim_{L\to\infty}\nu_k(\gamma)
\end{equation}
(quantum) RP resonances, which are quantum many-body counterparts to classical RP resonances. 
These resonances dictate the late time dynamics; in particular, the autocorrelation function $\langle \mathcal{O}(t)\mathcal{O}(0)\rangle$ of a traceless operator $\mathcal{O}$ at late times, where $\langle\cdots\rangle=D^{-1}\Tr(\cdots)$ is the infinite-temperature average appropriate for Floquet systems, is controlled by the leading RP resonance $\nu_1$:
\begin{equation}
    \langle \mathcal{O}(t)\mathcal{O}(0)\rangle\sim e^{-\nu_1^*t}.
\end{equation}

The central goal of this paper is to provide analytical and numerical evidence that, for generic chaotic lattice man-body systems, the RP resonances defined in Eq.~(\ref{eq:def_quantum_RP}) converge to inside the unit disk and, thus, elucidate the origin of irreversibility in isolated quantum many-body systems. The key to our study of irreversibility and RP resonances is the relation between the SFF and autocorrelation functions, which we develop in the next section.

\subsection{Irreversibility in Floquet many-body quantum systems}

\subsubsection{Autocorrelation functions}

To specify operators, we choose a Hermitian basis $P_\alpha=P_{\alpha^1}\otimes\cdots\otimes P_{\alpha^L}$, where $\alpha^x=0,\dots,q^2-1$ and the onsite basis operator $P_{\alpha^x}$ acting on site $x$ is normalized as $q^{-1}\Tr(P_{\alpha^x}P_{\beta^x})=\delta_{\alpha^x\beta^x}$ so that $D^{-1}\Tr(P_{\alpha}P_{\beta})=\delta_{\alpha\beta}$. For convenience, we also assume that the nonidentity basis operators $P_{\alpha^x}$ are normalized such that $P_{\alpha^x}^2=I_q$ (which is always possible at the expense of considering non-Hermitian basis operators~\cite{gottesman1998}; to use Hermitian basis operators, throughout this paper, we will focus our attention on the case $q=2^b$ for some integer $b$, in which case $P_\alpha$ can be taken as a tensor product of Pauli matrices).
Note that the operator $P_\alpha$ is generally nonlocal and made of $n$ clusters of nonidentity operators. We denote the size of the operator, i.e., the number of sites on which it acts nontrivially, by $a$.

In this paper, we shall be mainly concerned with the temporal behavior of the autocorrelation functions of basis operators $P_\alpha$,
\begin{equation}
\label{eq:def_Caa}
    C_{\alpha\alpha}(t)=\langle P_\alpha(t) P_\alpha\rangle,
\end{equation} 
where $P_\alpha(t)=(\mathcal{W}^\dagger)^t[P_\alpha]$ is the Heisenberg evolution of $P_\alpha$ and we recall that $\langle\cdots\rangle=D^{-1}\Tr(\cdots)$ is the infinite-temperature average. Note that Eq.~(\ref{eq:def_Caa}) is valid for both open and closed systems; in the latter case, it reduces to $P_\alpha(t)=(W^\dagger)^t P_\alpha W^t$. We denote the average of the autocorrelation function $C_{\alpha\alpha}(t)$ over an ensemble of similar systems by $ \overline{C_{\alpha\alpha}(t)}$.

\subsubsection{The spectral form factor and its dissipative generalization}

In closed quantum systems, the SFF is the quintessential object connecting the dynamics with spectral correlations. Indeed, the SFF can be defined dynamically as the square of the trace of the unitary evolution operator $W^t$,
\begin{equation}
\label{eq:def_SFF}
K(t)=|\Tr W^t|^2;    
\end{equation}
or given a spectral definition as the Fourier transform of the spectral two-point correlation function. 
The equivalence of these two definitions has played a pivotal role in establishing the theoretical footing~\cite{berry1985,sieber2001,sieber2002,heusler2004,muller2004,muller2005,Kos_ManyBody_2018,Garratt_Local_2021} behind the Bohigas-Giannoni-Schmit conjecture~\cite{Bohigas_Characterization_1984}, the central result of quantum chaos. Eigenvalue repulsion and spectral rigidity in ergodic systems are evidenced by the appearance of a ramp in the SFF at late times, a key signature of quantum chaos.

In open quantum systems, a precise connection between spectral correlations (as given by non-Hermitian RMT) and dissipative quantum dynamics is still unclear, as illustrated by the fact that the spectral and dynamical definitions of the SFF no longer coincide in the presence of dissipation and evidenced by the myriad of different proposed ``form factors'' in non-Hermitian physics~\cite{can2019JPhysA,kawabata2023PRB,braun2001,fyodorov1997,li2021PRL,garcia2023PRD,shivam2023PRL,gosh2022PRB,chan2022NatComm,xu2019PRL,xu2021PRB,cornelius2022PRL,matsoukas-roubeas2023ARXIV,kos2021PRL,kos2021PRB,Yoshimura2024}. Since we are interested in the dynamics, we will focus on the dissipative form factor (DFF)~\cite{can2019JPhysA}, the trace of the quantum channel $\scW^t$. The late-time behavior of the DFF is controlled by the spectral gap (which coincides with the leading RP resonance in the dissipationless limit, as we shall see below).
Vectorizing $\mathcal{W}=\sum_{j} K_{j} \otimes K^*_{j}$, the DFF for the Floquet quantum circuit can be expressed as~\cite{can2019JPhysA,Yoshimura2024,vikram2022ARXIV,kos2021PRB,Garratt_Local_2021}:
\begin{equation}\label{eq:DFF_circuit}
    F(t)=\Tr\scW^t
    =\sum_{j_1,\dots,j_t}|\Tr K_{j_1}\cdots K_{j_t}|^2.
\end{equation}
In the limit $\gamma\to0$, when there is a single $K_j=W$, we recover Eq.~(\ref{eq:def_SFF}).

\subsubsection{Autocorrelation-function expansion of the SFF}
\label{sec:general_sff_auto}

The key observation in relating the spectral and dynamical properties of the system is that the SFF can be decomposed in terms of autocorrelation functions~\cite{roberts2017JHEP,cotler2017JHEP,Gharibyan_Onset_2018,yoshimura2023operator}:
\begin{equation}
\label{eq:SFF_autocorrelation}
K(t)
=\sum_{P_\alpha\in\mathcal{P}}C_{\alpha\alpha}(t),
\end{equation}
where the sum is over $\mathcal{P}$, the entire set of basis operators. 
This can be easily obtained by noting that $D^{-1}\sum_{P_\alpha\in\mathcal{P}}[P_\alpha]_{ab}[P_\alpha]_{cd}=\delta_{ad}\delta_{bc}$, i.e., $D^{-1}\sum_{P_\alpha\in\mathcal{P}}P_\alpha\otimes P_\alpha$ is the \textsc{Swap} operator. The same identity also holds between the DFF and autocorrelation functions in open systems, i.e., the sum of autocorrelators equals the DFF even in the presence of dissipation modeled by quantum channels.

\begin{figure}[t]
    \centering
    \includegraphics[width=\columnwidth]{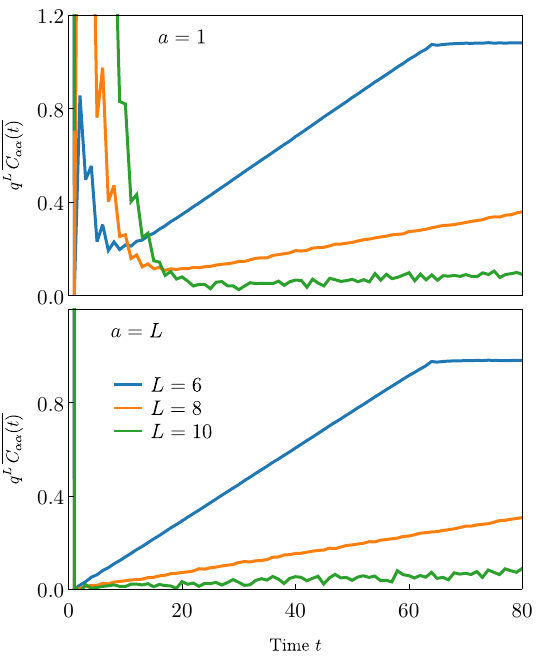}
    \caption{\textbf{Dynamics of ensemble-averaged autocorrelation functions in a qubit brickwork RFC.} We numerically computed the exact autocorrelation functions for $P_\alpha=\sigma_1^z$ (top panel) and $P_\alpha=\prod_{x=1}^L\sigma_x^z$ (bottom panel), where $\sigma_x^z$ is the third Pauli matrix at site $x$, using the circuit defined in Sec.~\ref{sec:models_brickwork} with $q=2$. The average was done over $6\times10^6$, $2.5\times10^5$, and $7.15\times10^4$ realizations for $L=6$, $8$, and $10$, respectively. To make the results for different sizes comparable, we have rescaled the vertical axis by $q^L$. For both local and nonlocal operators, there is a clear ramp-plateau structure (since the plateau time $\tH=q^L$ grows exponentially with $L$, the plateau is only visible for $L=6$, but we have checked that for larger sizes $\overline{C_{\alpha\alpha}(t)}$ also plateaus). For local operators only, there is also a Thouless peak before the ramp, and the timescale for the onset of the ramp grows with $L$ (see also Fig.~\ref{fig:ramp_scaling}).}
    \label{fig:brickwork_autocorrelation}
\end{figure}

In Ref.~\cite{yoshimura2023operator}, it was shown that ensemble-averaged autocorrelation functions of generic many-body Floquet systems have a three-stage evolution, resembling the famous dip-ramp-plateau structure of the SFF, see Fig.~\ref{fig:brickwork_autocorrelation} for the numerical calculation of the autocorrelation function of a local operator in a qubit brickwork RFC (defined below in Sec.~\ref{sec:models}).

First, they decay from their initial value $\overline{C_{\alpha\alpha}(0)}=1$ to a local minimum after a characteristic decay timescale $\td$ of the order of the Floquet period, which follows from the cancellation of system-dependent fluctuations upon ensemble-averaging. 
Second, the decay does not extend indefinitely and $\overline{C_{\alpha\alpha}(t)}$ grows sharply after $\td$, before decaying again. This peak, which we call the Thouless peak~\cite{Yoshimura2024}, is due to interactions of neighboring sites and is an exclusive signature of many-body physics. 
Third, after the Thouless peak decays to a value $\overline{C_{\alpha\alpha}}\ll q^{-L}$, the autocorrelation functions start \emph{regrowing} linearly after a ramp timescale $\tramp$, before eventually plateauing at $\overline{C_{\alpha\alpha}(t>\tH)}=q^{-L}$ at the Heisenberg time $\tH=q^L$. 

Above, we have made explicit that the decay and ramp times are operator-dependent by adding a subscript $\alpha$, whereas the Heisenberg time is the same for all operators and depends only on the system size and the local Hilbert space dimension. In the thermodynamic limit $L\to\infty$, the plateaus of individual autocorrelation functions are exponentially suppressed, and consequently any $\overline{C_{\alpha\alpha}(t)}$ decays to zero. However, there are also exponentially many individual autocorrelation functions, which compensate for the decay, and their sum (the SFF) displays a ramp-plateau structure with height $q^L$.

\begin{figure}[t]
    \centering
    \includegraphics[width=\columnwidth]{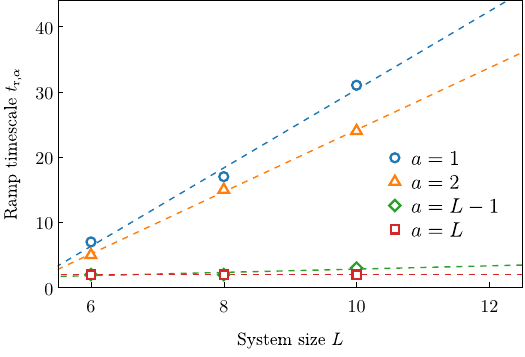}
    \caption{\textbf{Scaling of the ramp timescale as a function of system size for different types of operators.} We computed the time $\tramp$ which minimizes $\overline{C_{\alpha\alpha}(t)}$ \emph{after the Thouless peak}, using the numerically exact ensemble-averaged autocorrelation functions in qubit brickwork RFCs, as described in Fig.~\ref{fig:brickwork_autocorrelation}. To reduce the fluctuations for $L=10$, we performed an additional running-time average over three timesteps. We considered two local operators---$P_\alpha=\sigma_1^z$ ($a=1$) and $P_\alpha=\sigma_1^z\sigma_2^z$ ($a=2$)---and two extensive operators---$P_\alpha=\prod_{x=1}^{L-1}\sigma_x^z$ ($a=L-1$) and $P_\alpha=\prod_{x=1}^{L}\sigma_x^z$ ($a=L$). The dashed lines are fits to the numerical data to guide the eye and clearly show that $\tramp$ grows for local operators and stays constant for operators differing from the system size by a constant.}
    \label{fig:ramp_scaling}
\end{figure}

From the previous discussion, it becomes clear that to obtain RP resonances (i.e., the decay rates of individual correlation functions) from eigenvalues of the evolution operator (i.e., the late-time behavior of the SFF) we must suppress the ramp of the SFF. It is crucial, therefore, to identify which operators contribute to the ramp of the SFF in the thermodynamic limit and devise a scheme to suppress their growth. 

The key ingredient is the scaling of the ramp time $\tramp$: if $\tramp$ diverges in the thermodynamic limit, then the autocorrelation functions decay indefinitely and do not contribute to the ramp; if, instead, $\tramp$ is an $L$-independent constant, the corresponding operators have an extensive time available to grow and contribute to the ramp. Building on 
Ref.~\cite{yoshimura2023operator}, we conjecture that $\tramp$ grows for all operators except if their size differs from the system size $L$ by at most a constant. 
Consequently, only these operators contribute to the ramp and should be suppressed from the SFF to compute RP resonances.
In Sec.~\ref{sec:DRPM_autocorrelation}, we will prove that in the large-$q$ RPM, only fully nonlocal operators ($a=L$) contribute to the ramp. Moreover, Fig.~\ref{fig:ramp_scaling} provides numerical support for this claim in qubit brickwork RFCs: the ramp time of local operators (sizes $a=1$ and $a=2$) grows linearly, while for ultra-nonlocal operators (sizes $a=L-1$ and $a=L$) it is constant. Moreover, from Fig.~\ref{fig:brickwork_autocorrelation}, we can also see that the short operators are contributing most to the Thouless peak, whereas the ultralong operators have no second peak before the ramp. 

\subsubsection{RP resonances from weak dissipation}

Having identified long operators as those contributing to the ramp, we now turn to suppressing their growth. 
This can be achieved by introducing weak dissipation to the system, e.g., by adding a local quantum channel with strength $\gamma$ at each site after every Floquet period. In a many-body system, dissipation suppresses operator growth proportionally to their size~\cite{schuster2023PRL,mori2024}, i.e., a basis operator of size $a$ decays as $e^{-c a \gamma t}$, for some constant $c$. In the thermodynamic limit, we thus completely suppress all operators of extensive size (i.e., such that $\lim_{L\to\infty}a=\infty$), while finite operators have a finite dissipative correction. Since we have explicitly introduced dissipation into the system, we should, henceforth, consider the DFF, defined in Eq.~(\ref{eq:DFF_circuit}), which decays and has no ramp.

Let us take the dissipationless limit $\gamma\to0$ after the thermodynamic limit. The contribution of nonextensive operators (responsible for the Thouless peak and not the ramp) is unaffected compared to the case with strict $\gamma=0$. At the same time, the extensive operators that give rise to the ramp are completely suppressed and no longer contribute to the expansion. 
Schematically, we have
\begin{equation}
    ``\lim_{\gamma\to0}\lim_{L\to\infty} \mathrm{DFF}= \mathrm{SFF}- \mathrm{ramp}\,\text{''}
\end{equation}
and we can compute the RP resonances from the decay of $\lim_{\gamma\to0}\lim_{L\to\infty}\overline{F(t)}$.\footnote{Above, we argued that only operators whose size differs from $L$ by a constant contribute to the ramp, but dissipation suppresses all extensive operators, including those whose length is proportional to $L$ yet still differ from $L$ by an extensive amount (the ramp timescale $\tramp$ of these operators, therefore, diverges in the thermodynamic limit). However, we conjecture that, generally, these operators have a very small contribution to the Thouless peak. The location of the RP resonances in the complex plane is unaffected and only their weight (which we do not consider in detail in this paper) has a small correction.}
Just as in the classical case, the resonance spectrum of the quantum channel can be revealed through the noncommutativity of the thermodynamic and noiseless limits.

\subsubsection{RP resonances from operator truncation}

Another possibility for suppressing the ramp is to directly discard any operators that grow to a size larger than a threshold $r$, a method known as operator trunctation~\cite{prosen2002,prosen2004,prosen2007,znidaric2024,znidaric2024b}. 
To this end, we introduce a coarse-graining in the operator space by truncating an operator $A=\sum_{P_\alpha} c_\alpha P_\alpha$ to a finite support size, namely,
\begin{equation}
\label{eq:truncation_proj}
    A\to\scP_r(A)
    =\sum_{P_\alpha:a(\alpha)\leq r}c_\alpha P_\alpha,
\end{equation}
i.e., we discard any basis operators $P_\alpha$ of size $a$ larger than a coarse-graining length $r$. Let us consider the adjoint unitary time-evolution operator $(\scW^\dagger)^t$, under which operators evolve, $A(t)=(\scW^\dagger)^t[A]$, for some initial operator $A$.
We introduce the finite-size approximate time-evolution operator $\hatW_t^{(r)}=\scP_r(\scW^\dagger)^t$, by truncating operators down to size $r$ after the time evolution, i.e., starting with an operator of size less or equal to $r$, $\hatW_t^{(r)}$ discards any contributions to the operator growth of size greater than $r$. This coarse-graining introduces nonunitary into the dynamics (because truncations of unitary matrices are not unitary). Since $\lim_{r\to L}\hatW_t^{(r)}=(\scW^\dagger)^t$, we can study the spectrum of $(\scW^\dagger)^t$ by successively approximating it with finer approximations $\hatW_t^{(r)}$ of increasing $r$.

For certain systems, one can numerically compute the full spectrum of $\hatW_t^{(r)}$ numerically~\cite{prosen2004,prosen2007,znidaric2024}. 
Alternatively, we define the truncated form factor (TFF):
\begin{equation}
\label{eq:TFF}
    \hatF_r(t)=\Tr\hatW_t^{(r)}.
\end{equation}
Using Eq.~(\ref{eq:truncation_proj}) and the definition of the autocorrelation functions, we find
\begin{equation}
\label{eq:TFF_auto}
    \hatF_r(t)=\sum_{\alpha:a(\alpha)\leq r}
    C_{\alpha\alpha}(t),
\end{equation}
which is the autocorrelation-function expansion of the SFF truncated to operators of size less or equal to $r$. If the limits $L\to\infty$ and $r\to\infty$ are taken such that $L$ is always greater than $r$ plus any constant (i.e., we take the thermodynamic limit first), then we never have contributions from operators whose length differs from $L$ by a constant and there is thus no ramp in the form factor. Consequently, if $\scW$ has RP resonances inside the unit disk, then $\lim_{r\to L}\hatF_r(t)$ decays exponentially with a rate given by the leading RP resonance. 
The $r\to\infty$ limit thus has the same effect as the $\gamma\to0$ (when taken after the thermodynamic limit) and the RP resonance spectrum computed by the two methods is the same.

\subsection{Models}
\label{sec:models}

\begin{figure}[t]
    \centering
    \includegraphics[width=\columnwidth]{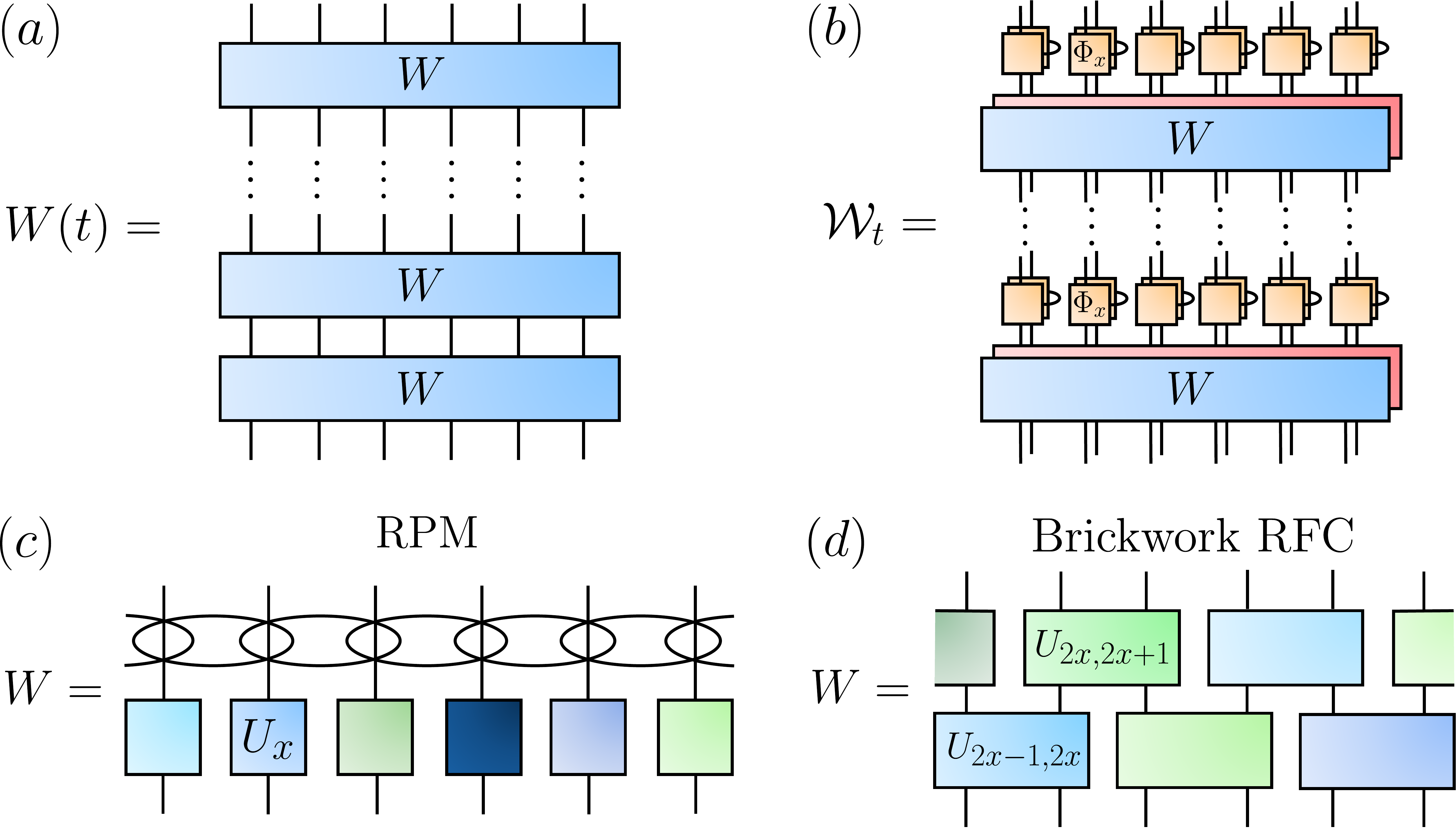}
    \caption{\textbf{Quantum circuit models.} (a) We consider closed Floquet quantum circuits where the time evolution is given by a repeated application of the unitary $W$ (blue rectangles). Each line represents a $q$-dimensional local Hilbert space. (b) When the system is open, it evolves under the quantum channel $\mathcal{W}^t$. Each Floquet step now has two copies of $W$ (blue and pink rectangles), followed by single-site quantum channels $\Phi_x$ that couple the two copies (yellow squares). (c)--(d) Circuit architecture of the unitary evolution operator $W$ with PBCs. (c) The RPM, where squares represent $q\times q$ Haar-random unitaries (the different colors denote that all gates are independently sampled), and ellipses the diagonal phase interactions between neighboring sites. (d) Brickwork RFC, composed of alternating layers of independent $q^2\times q^2$ Haar-random unitaries.}
    \label{fig:circuits}
\end{figure}

We now define the quantum circuit models that we will use in the remainder of the paper to illustrate the general picture constructed above, see Fig.~\ref{fig:circuits}. First, we introduce the DRPM, for which all the quantities defined above can be computed analytically at large $q$. The analytic computations will then be complemented by numerical results for qubit brickwork RFCs.

\subsubsection{(Dissipative) random phase model}
\label{sec:models_DRPM}

The paradigmatic example of generic open quantum many-body systems that we consider in this paper is the DRPM. We introduced the DRPM in Ref.~\cite{Yoshimura2024} as a generalization of the RPM~\cite{Chan_Spectral_2018}, a spatially extended isolated Floquet circuit under the influence of bulk dissipation. 
The discrete time evolution of the model is generated by the unit-time Floquet operator $W=W_2W_1$, where $W_1=U_1\otimes\dots\otimes U_L$ consists of $q\times q$ onsite scramblers $U_x$ that are Haar distributed, while $W_2$ induces interactions between adjacent sites and acts on the computational basis $|a_1,a_2,\cdots,a_L\rangle\in(\mathbb{C}^q)^L=\mathcal{H}$, where $a_x=0,\cdots,q-1$, diagonally with the phase $\exp\left(\ii\sum_{x=1}^L\varphi_{a_x,a_{x+1}}\right)$. Each $\varphi_{a_x,a_{x+1}}$ is independently Gaussian distributed with mean zero and variance $\varepsilon>0$.

Based on the structure of the RPM, in the DRPM each Floquet step is accompanied by a quantum channel that introduces dissipation. For simplicity, we assume that the quantum channel has a site-resolved form $\Phi=\bigotimes_x\Phi_x$. The action of the single-site channel $\Phi_x$ on the state $\varrho_x$ at site $x$ can be expressed in terms of the Kraus decomposition $\Phi_x(\varrho_x)=\sum_{i=0}^{k-1}M_i\varrho_xM^\dagger_i$, where $k$ is the number of channels, with Kraus operators $M_i$ that are normalized as $q^{-1}\Tr(M_iM^\dagger_j)=\eta_i\delta_{ij}$ with some channel-dependent $0<\eta_i<1$. They satisfy the sum rule $\sum_{i=0}^{k-1}M^\dagger_iM_i=I_q$ where $I_q$ is the $q\times q$ identity matrix.
For example, the depolarizing channel with depolarizing probability $p$, $\Phi_x(\varrho_x)=(1-p)\varrho_x+pI_q/q$, has $\eta_0=1-p(q^2-1)/q^2$ ($\eta_0\approx 1-p$ when $q\to\infty$) and $\eta_i= p/q^2$~\cite{Yoshimura2024}. It is then convenient to define an effective dissipation strength $\mathbb{R}^+\ni\gamma=-\varepsilon^{-1}\log \eta_M$, where $\eta_M$ is the largest of the $\eta_i$. $\gamma=0$ corresponds to the dissipationless limit and $\gamma\to\infty$ to infinitely-strong dissipation.
With a slight abuse of notation, let us also denote the quantum channel acting on vectorized states by $\Phi$. The whole time evolution of the circuit can then be represented by $\mathcal{W}^t$ where $\mathcal{W}=\Phi(W\otimes W^*)$.

\subsubsection{Brickwork random Floquet circuits}
\label{sec:models_brickwork}

The second type of circuit we consider is brickwork RFCs. In this case, the Floquet time step comprises two layers of unitary gates alternating in a brickwork fashion. More precisely, $W=W_\mathrm{e}W_\mathrm{o}$; for even $L$ and periodic boundary conditions (PBCs), $W_{\mathrm{o}}=U_{1,2}\otimes U_{3,4}\otimes\cdots\otimes U_{L-1,L}$ couples odd and even sites and $W_\mathrm{e}=\mathbb{T}\widetilde{U}_{2,3}\otimes\cdots\otimes \widetilde{U}_{L-2,L-1}\otimes \widetilde{U}_{L,1}\mathbb{T}^\dagger$, with $\mathbb{T}$ the one-site shift operator,\footnote{$\mathbb{T}$ is defined by its action on basis states $\mathbb{T}|a_1,a_2,\cdots ,a_L\rangle=|a_L, a_1, a_2, \cdots, a_{L-1}\rangle$.} couples even and odd sites; for even $L$ and open boundary conditions (OBCs), we remove $\widetilde{U}_{L,1}$ from $W_\mathrm{e}$; for odd $L$, the circuit is only defined with OBCs and we have $W_{\mathrm{o}}=U_{1,2}\otimes\cdots\otimes U_{L-2,L-1}\otimes I_q$ and $W_{\mathrm{e}}=I_q\otimes \widetilde{U}_{2,3}\otimes \cdots\otimes \widetilde{U}_{L-1,L}$. To model generic chaotic systems, the two-site gates $U_{x,x+1}$ and $\widetilde{U}_{x,x+1}$ are chosen as independent $q^2\times q^2$ Haar-random unitaries.
For the open circuit model, we consider the same local channels as in the DRPM. For the numerics in Secs.~\ref{sec:general_sff_auto} and \ref{sec:suppression_numerics}, we chose the depolarizing channel with depolarization strength $p$. 

\subsection{Analytic results}
For the RPM, we can compute the entire RP resonance spectrum analytically in the large-$q$ limit, as we show in Sec.~\ref{sec:DRPM} (due to its one-parameter structure, the large-$q$ limit of the RPM is nontrivial, contrary to the case of brickwork RFCs). Proceeding diagrammatically, we compute all ensemble-averaged autocorrelation functions in the DRPM for both PBCs and OBCs. In this section, we focus on the former in the limit $\gamma\to0$. Illustrating our earlier general reasoning, ensemble-averaged autocorrelation functions of fully nonlocal operators with support on all sites (i.e., $a=L$) were shown in Ref.~\cite{yoshimura2023operator} to not decay and are responsible for the ramp of the SFF:
\begin{equation}
      \overline{C_{\alpha\alpha}(t)}\simeq q^{-2L} (t-1).
\end{equation}
On the other hand, operators with size $a<L$ (and a number of clusters $n$) decay as
\begin{equation}
    \overline{C_{\alpha\alpha}(t)}
    =q^{-2a}(t-1)^n \sum_{m=0}^{a-n}\binom{n}{m}(t-2)^me^{-(2n+m)\varepsilon t},
\end{equation}
whence we infer that the RP resonances for PBCs are 
\begin{equation}
\label{eq:tower_RP}
    \nu_k=(k+1)\varepsilon,\quad k=1,2,\dots.
\end{equation}
We argue in Sec.~\ref{sec:DRPM_autocorrelation} that $k+1$ counts the number of domain walls between many-body configurations and $\varepsilon$ is the cost of a single domain wall. The leading RP resonance is, thus, $2\varepsilon$ (for PBCs, the number of domain walls is lower-bound by $2$, while there can be a single one for OBCs). 

The leading RP resonance can be directly obtained from the DFF using the prescription described before. By taking the thermodynamic limit at finite dissipation, the $a=L$ operators, and hence the ramp, are suppressed and the leading RP resonance emerges in the spectrum. The noncommutativity of the limits manifests itself in the analytic expression for the gap ruling the decay of the DFF~\cite{Yoshimura2024}:
\begin{equation}
    \Delta=\varepsilon\min\{2+\gamma,\gamma L\}.    
\end{equation}
This result was obtained in Ref.~\cite{Yoshimura2024} (we review it Sec.~\ref{sec:DFF}) as the leading term in a domain wall expansion. By keeping more terms, we can reconstruct the whole tower of RP resonances in Eq.~(\ref{eq:tower_RP}). The same RP resonance spectrum can be explicitly computed via operator truncation in the RPM, as we show in Sec.~\ref{sec:DRPM_Truncation}.

We further emphasize the role of RP resonances in relaxation by computing the OTOC of the RPM and DRPM at large $q$. In Sec.~\ref{sec:OTOC}, we define the OTOC as an averaged square commutator, which leads to two contributions:
\begin{equation}
\mathcal{C}(x,y;t)=\langle \mathcal{O}(y,t)\mathcal{O}(y,t)\rangle-\langle \mathcal{O}(y,t)P_{\alpha^x}\mathcal{O}(y,t)P_{\alpha^x}\rangle, 
\end{equation}
where $\mathcal{O}(y,t)$ is the time evolution of a local traceless operator $\mathcal{O}$, which, at $t=0$, is supported at position $y$. In closed systems, like the RPM, the norm $\langle \mathcal{O}(y,t)\mathcal{O}(y,t)\rangle$ is a constant (for simplicity, we normalize $\mathcal{O}$ such that $\langle \mathcal{O}^2\rangle=1$) to which the OTOC relaxes. The relaxation rate is thus determined by the decay of the four-point contribution $\langle \mathcal{O}(y,t)P_{\alpha^x}\mathcal{O}(y,t)P_{\alpha^x}\rangle$. We find that the OTOC shows two-stage relaxation \cite{Bensa_phantom_2021,Bensa_relaxation_2022,znidaric2022,Znidaric_relaxation_2023,znidaric2023_PRR,bensa2024,bensa2024_PRA,Jonay_thermalization_2024,jonay2024twostagerelaxationoperatorsdomain} for PBCs but not for OBCs. Namely, when PBCs are imposed, the relaxation rate during the first stage, which takes place over the timescale $t\lesssim L/v_B$ where $v_B$ is the butterfly velocity, is given by $2\varepsilon$, whereas in the second stage, for $t\gtrsim L/v_B$, it increases to $4\varepsilon$. The large-$q$ OTOC during the second stage thus reads asymptotically
\begin{equation}
     \overline{\mathcal{C}(x,t)}\simeq1-\frac{(1-e^{-2\varepsilon})^{L-2}}{(L-x-1)!(x-1)!}t^{L-2}e^{-4\varepsilon t}.
\end{equation}
Note that since this contribution involves two copies of $\mathcal{O}(y,t)$, it is indeed expected that the OTOC decays with the rate $4\varepsilon$, which is twice the leading RP resonance ($2\varepsilon$) for PBCs in the thermodynamic limit according to Eq.~(\ref{eq:tower_RP}).
This is in stark contrast to the case of OBCs where the OTOC in $q\to\infty$ always decays with the same rate $2\varepsilon$, i.e., twice the leading RP resonance for OBCs. The difference between these behaviors can be understood from a heuristic picture of operator spreading as explained in Sec.~\ref{sec:OTOC}.

In open systems such as the DRPM, the norm $\langle \mathcal{O}(y,t)\mathcal{O}(y,t)\rangle$ also decays. In the large-$q$ limit of the DRPM, we evaluate it exactly in terms of a transfer matrix and find that the decay rate in the dissipationless limit $\gamma\to0$ after taking the thermodynamic limit $L\to\infty$ is indeed generically given by $4\varepsilon$ for PBCs and either $2\varepsilon$ or $4\varepsilon$ for OBCs depending on the value of $y$.

\section{Case study: the (dissipative) random phase model}
\label{sec:DRPM}

In this section, we analytically compute, for arbitrary onsite quantum channels, different dynamical quantities that characterize the evolution of the DRPM, defined in Sec.~\ref{sec:models_DRPM}: autocorrelation functions of local operators, the dissipative form factor, the truncated form factor, and out-of-time-order correlators. We are particularly interested in the late-time decay of these quantities and extracting quantum RP resonances from them.

\subsection{Autocorrelation functions}
\label{sec:DRPM_autocorrelation}

We shall compute autocorrelation functions diagrammatically in the large-$q$ limit by applying the method developed in Ref.~\cite{yoshimura2023operator} for evaluating the same objects in the dissipation-free RPM. The main idea is that since both Haar unitaries and Kraus operators act on each site individually in the DRPM, we can take Haar averages at every site independently. Onsite Haar averaging generally induces pairings between the indices of unitaries $[U_x]_{aa'}$ and their conjugates $[U^*_x]_{bb'}$, possibly at different times. In particular, it is known in the RPM that only a subset of the $(t!)^2$ possible pairings, called {\it cyclic pairings}, contribute at large $q$, which we label with $s=0,\dots,t-1$~\cite{Chan_Spectral_2018}. The $t$ cyclic pairing degrees of freedom therefore give rise to the $t\times t$ transfer matrix 
\begin{equation}
\label{eq:def_T}
    T=(1-e^{-\varepsilon t})I+e^{-\varepsilon t} \mathtt{E}
\end{equation}
that acts on the $t$-dimensional vector space spanned by the cyclic pairing states $|s\rangle$. Here, $\mathtt{E}$ is the constant matrix of 1s.
We can thus replace the (1+1)-dimensional evolution of the physical degrees of freedom with $q$-dimensional local Hilbert space by a one-dimensional evolution of pairing degrees of freedom with $t$-dimensional local Hilbert space; these pairing degrees of freedom evolve in space under the action of a transfer matrix $T$.
Because of the Gaussian phase interaction between neighboring sites, the transfer matrix acts in a very simple way: if the two adjacent sites have the same pairing, then the transfer matrix does nothing; if the pairings are different---i.e., there is a pairing \emph{domain wall}---, they incur an entropic cost $e^{-\varepsilon t}$. It is this cost of domain walls that causes the decay of correlation functions in the RPM.

\begin{figure}
    \centering
    \includegraphics[width=0.8\columnwidth]{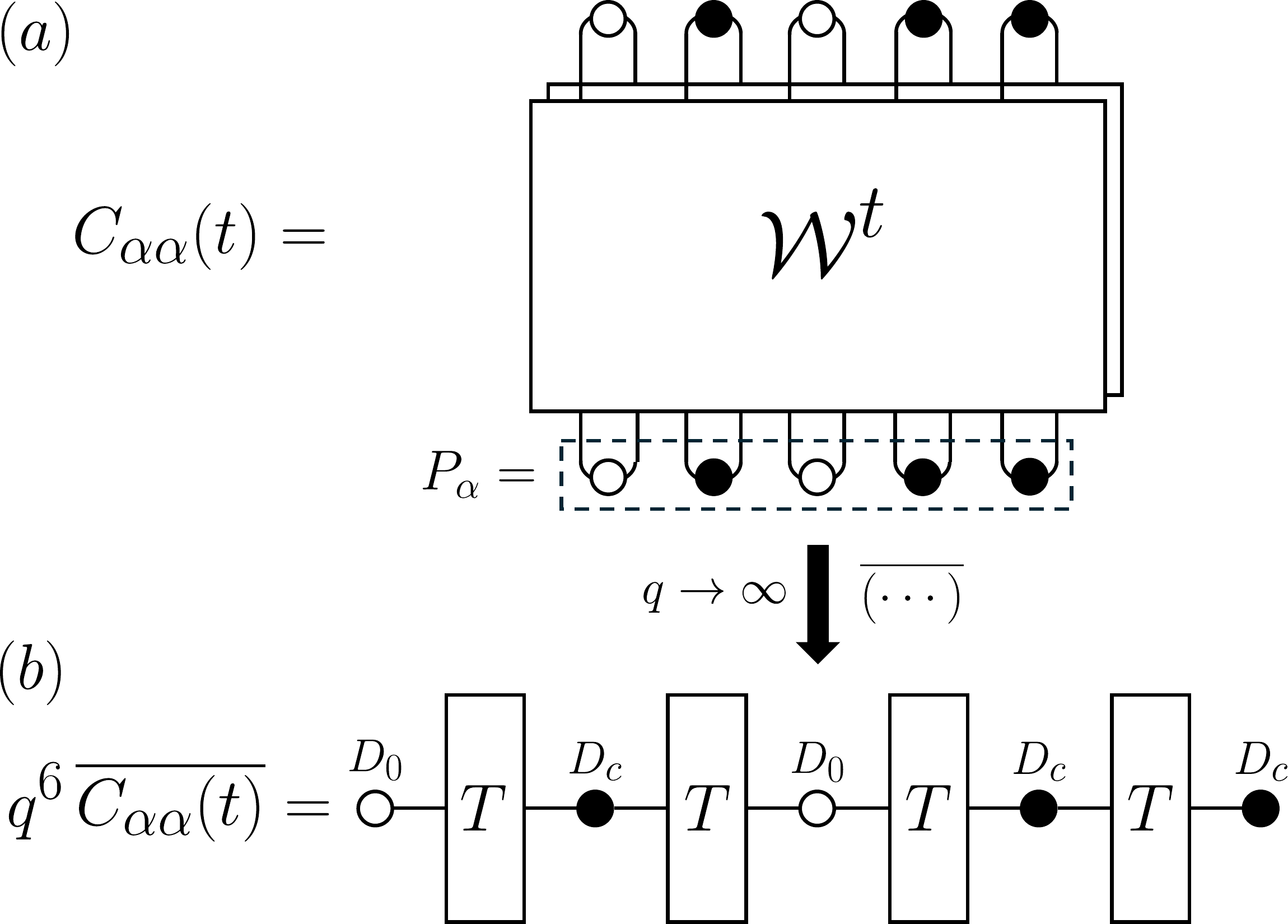}
    \caption{\textbf{Schematic representation of the large-$\bm{q}$ computation of an autocorrelation function in the DRPM.} For definiteness, we illustrate $L=5$ with OBCs. (a) Exact autocorrelation function $C_{\alpha\alpha}(t)$ for $P_{\alpha}=I_q\otimes P_{\alpha^2}\otimes I_q\otimes P_{\alpha^4}\otimes P_{\alpha^5}$ that is nontrivial (black dots, $\alpha^x\neq0$) on sites $2$, $4$, and $5$ and the identity (white circles, $\alpha^x=0$) on sites $1$ and $3$. It is computed by evolving $P_\alpha$ in time under the quantum channel $\mathcal{W}^t$ and then computing the overlap with the initial operator. (b) In the large-$q$ limit, the ensemble average autocorrelation function is computed by evolving the pairing degrees of freedom $s=0,\dots, t-1$ in the spatial direction with the transfer matrix $T$. If, in the original picture, the operator on site $x$ is nontrivial, the corresponding pairing state is projected into one of $s\neq0$ pairing states (black dots) and suppressed with dissipative weight $\kappa(t)$; if the original operator is trivial, then we project to the $s=0$ pairing (white dots) and there is no suppression.}
    \label{fig:autocorrelation}
\end{figure}

The same mechanism is at play in the DRPM. In this case, it can be easily shown~\cite{Yoshimura2024} that diagrams with pairings $s\neq0$ additionally carry the weight $\kappa(t)$, where $0<\kappa(t)=\sum_i\eta_i^t<1$ quantifies the ``amount of dissipation'' at time $t$ (with $\kappa=1$ corresponding to the dissipationless case and $\kappa=0$ to the fully-dissipated case). 
At late times, $\kappa(t)\sim \eta_M^t=e^{-\varepsilon \gamma t}$, where $\eta_M$ is the largest of the $\eta_i$ (for simplicity, we assume $\eta_M$ to be nondegenerate; otherwise, all correlation functions must be trivially multiplied by its degeneracy). Recall that the dissipation strength is $\gamma=-\varepsilon^{-1} \log\eta_M$.
Since no such change due to dissipation happens for diagrams associated with the $s=0$ pairing, this indicates that the presence of dissipation manifests itself simply as an additional prefactor $[\kappa(t)]^a\sim e^{-a\varepsilon\gamma t}$ to autocorrelation functions in the RPM.

Because the basis operators $P_{\alpha^x}$ are traceless, the $s=0$ pairing does not contribute if $\alpha^x\neq0$, and we apply a projector $D_\mathrm{c}=\kappa\sum_{s=1}^t|s\rangle\langle s|=\kappa(I_t-|0\rangle\langle0|)$ at site $x$. Analogously, one can show that only the $s=0$ pairing contributes at $x$ if $\alpha^x=0$~\cite{yoshimura2023operator}, in which case we insert a projector $D_0=|0\rangle\langle 0|$ at site $x$. The different sites are then connected by the transfer matrix $T$. Finally, with PBCs, we additionally connect the first and last site by a transfer matrix and take the trace to compute autocorrelation functions at large $q$:
\begin{equation}
    \overline{C_{\alpha\alpha}(t)}\simeq q^{-2a}\mathrm{Tr}\big[\prod_{x=1}^L TD_x\big]\qquad \text{(PBC)},
\end{equation}
where $a$ is the number of nonidentity operators in the operator string $P_\alpha$, and $D_x=D_0$ if $\alpha^x=0$ and $D_x=D_c$ if $\alpha^x\neq0$.
For OBCs we instead apply the equal weight superposition $|\phi\rangle= \sum_s|s\rangle$ at the boundaries:
\begin{equation}
    \overline{C_{\alpha\alpha}(t)}\simeq q^{-2a}\langle\phi|D_1\prod_{x=2}^L TD_x|\phi\rangle \qquad \text{(OBC)}.
\end{equation}
The calculation of autocorrelation functions using the transfer matrix is schematically illustrated in Fig.~\ref{fig:autocorrelation}.

As we shall see in the following, single-domain-wall terms with the entropic cost $e^{-\varepsilon t}$, which turn out to correspond to the leading RP resonance, are allowed to be present only with OBCs. To illustrate this, we first focus on the case of PBCs.

\subsubsection{Periodic boundary conditions}

First, we consider the autocorrelation function of operators that have an identity on at least one site, $a<L$. Without loss of generality, by using the PBCs, we can fix the first operator to be the identity, $\alpha^1=0$. Then, using the transfer-matrix method outlined above for an operator $P_\alpha$ with $n$ nontrivial clusters, each of size $a_m$ (such that $\sum_{m=1}^na_m=a$), we obtain
\begin{equation}\label{eq:correlation1}
\begin{split}
    \overline{C_{\alpha\alpha}(t)}
    &\simeq q^{-2a}\prod_{m=1}^n[(TD_c)^{a_m}T]_{00}
    \\
    &=q^{-2a}[\kappa(t)]^a e^{-2n\varepsilon t}(t-1)^n[\chi(t)]^{a-n},
\end{split}
\end{equation}
where $\chi(t)=1+(t-2)e^{-\varepsilon t}$ and we have assumed that the same quantum channel acts on every site---if dissipation is inhomogeneous, then one should replace $[\kappa(t)]^a$ with $\prod_x \kappa_x(t)$. Strikingly, the expression Eq.~\eqref{eq:correlation1} is essentially the same as the one in the RPM except for the prefactor $[\kappa(t)]^a$ that accounts for dissipation.

At late times, two contributing factors induce the decay of autocorrelation functions. One is the entropic cost $e^{-\varepsilon t}$ associated with a domain wall of $t$ pairing degrees of freedom, and the other one is the prefactor $[\kappa(t)]^a\sim e^{-a\varepsilon\gamma t}$.
Then, the autocorrelation functions asymptotically approach
\begin{equation}\label{eq:corr_local}
     \overline{C_{\alpha\alpha}(t)}\simeq q^{-2a}(t-1)^ne^{-(2n+\gamma a)\varepsilon t},
\end{equation}
from which we infer that operator strings that decay most slowly are those with $a=n=1$, i.e., local operators.

If instead we consider a maximally-nonlocal operator, $a=L$ (which also implies $n=1$), the autocorrelation functions are given by
\begin{equation}\label{eq:corr_nonlocal}
\begin{split}
    \overline{C_{\alpha\alpha}(t)}
    &\simeq q^{-2L}\sum_{s=1}^{t-1}[(TD_c)^{L-1}T]_{ss}
    \\
    &=q^{-2L}[\kappa(t)]^L([\chi(t)]^L+(t-2)[\lambda(t)]^L),
\end{split}
\end{equation}
where $\lambda(t)=1-e^{-\varepsilon t}$, and decay at late times as
\begin{equation}
      \overline{C_{\alpha\alpha}(t)}\simeq q^{-2L} (t-1)e^{-L\varepsilon\gamma t}\left[1+\frac{L(L-1)(t-2)}{2}e^{-2\varepsilon t}\right].
\end{equation}

As we discuss in Sec.~\ref{sec:DFF}, operators that are either strictly local ($a=n=1$) or maximally nonlocal $a=L$ turn out to fully control the late-time asymptotics of the dissipative form factor at large $q$.

\subsubsection{Open boundary conditions}\label{sec:corr_open}

As seen above, for OBCs, the behavior of autocorrelation functions differs depending on whether nonidentity operators act on the boundaries. Let us consider the different possibilities. If $P_\alpha$ acts on both boundaries with the identity operator, the averaged autocorrelation functions are still given by Eq.~\eqref{eq:correlation1}. 

When $a<L$ and a nontrivial operator string $P_\alpha$ acts on both boundaries of the system ($x=1,L$), the ensemble-averaged autocorrelation functions at large $q$ are given by
\begin{align}
    \overline{C_{\alpha\alpha}(t)}
    &\simeq q^{-2a}\sum_{s,s'=1}^{t-1}(D_cT)^{a_1}_{s0}\!\left[\prod_{m=2}^{n-1}[(TD_c)^{a_m}T]_{00}\right]\!(TD_c)^{a_n}_{0s'} \n
    &=q^{-2a}[\kappa(t)]^ae^{-2(n-1)\varepsilon t}(t-1)^n[\chi(t)]^{a-n}.
\end{align}

Instead, if $P_\alpha$ acts with the identity operator on only one boundary, say $x=L$ but not $x=1$, the autocorrelation functions read
\begin{align}
    \overline{C_{\alpha\alpha}(t)}
    &\simeq q^{-2a}\sum_{s=1}^{t-1}(D_cT)^{a_1}_{s0}\prod_{m=2}^{n}[(TD_c)^{a_m}T]_{00} \n
    &=q^{-2a}[\kappa(t)]^ae^{-(2n-1)\varepsilon t}(t-1)^n[\chi(t)]^{a-n}.
\end{align}
As a special case, when a strictly local operator ($a=n=1$) acts nontrivially on either of the boundaries, we have
\begin{align}
\label{eq:RP_eps}
   \overline{C_{\alpha\alpha}(t)}
   =q^{-2}(t-1)e^{-(1+\gamma)\varepsilon t}
\end{align}
which, unlike the autocorrelators of operators supported on the bulk, decays with exponent $\varepsilon$.

In the fully-nonlocal case, $a=L$ and $n=1$, the autocorrelation functions instead behave as
\begin{align}
     \overline{C_{\alpha\alpha}(t)}
     &=q^{-2L}\sum_{s,s'=1}^{t-1}[D_c(TD_c)^{L-1}]_{ss'}
    \n
    &=q^{-2L}[\kappa(t)]^L(t-1)[\chi(t)]^{L-1}.
\end{align}

\subsubsection{Entropic cost of domain walls as RP resonances}
\label{sec:DRPM_domain_walls}

Having evaluated autocorrelation functions for both PBCs and OBCs, it is now clear that domain walls have a one-to-one correspondence with the tower of RP resonances. According to the general result in Sec.~\ref{sec:quantumRP}, autocorrelation functions $\overline{C_{\alpha\alpha}(t)}$ in the RPM are expected to decay as $\overline{C_{\alpha\alpha}(t)}\sim\sum_{k=1}^{q^L-1}d_ke^{-\nu_kt}$, where we assume that the RP resonances are self-averaging and $d_k\in\mathbb{R}$ for $k=1,\dots,q^L-1$.
For PBCs, Eq.~\eqref{eq:correlation1} indicates that autocorrelation functions in the limit of vanishing dissipation behave at late times as 
\begin{equation}
    \overline{C_{\alpha\alpha}(t)}\sim e^{-2n\varepsilon t}\sum_{m=0}^{a-n}\binom{n}{m}(t-2)^me^{-m\varepsilon t}.
\end{equation}
We thus infer that the RP resonances for PBCs are 
\begin{equation}
    \nu_k=(k+1)\varepsilon,\quad k=1,2,\dots,
\end{equation}
implying that the number of domain walls $k+1$ determines the $k$-th RP resonance (for a basis operator with $n$ nonidentity clusters, the minimal number of domain walls is $k=2n$). Note that due to PBCs, pairing configurations with a single domain wall are prohibited. The leading RP resonance for PBCs is, therefore, $2\varepsilon$.

When instead OBCs are imposed, the situation is similar except that a single domain wall is allowed to appear, provided that the operator $P_\alpha$ acts on only one of the boundaries of the system nontrivially. In this case, the leading RP resonance becomes $\varepsilon$, see Eq.~(\ref{eq:RP_eps}).

\subsection{Dissipative form factor}
\label{sec:DFF}

The structure of the ensemble-averaged DFF of the DRPM was first extensively studied by the same authors of this paper in Ref.~\cite{Yoshimura2024}, which we briefly recapitulate now. Upon Haar averaging and imposing PBCs, the DFF at large $q$ is given in a neat way,
\begin{equation}\label{eq:DFF_exact}
    \overline{F(t)}=\Tr\,\hat{T}^L,\quad \hat{T}=TD,
\end{equation}
where $T$ is defined in Eq.~(\ref{eq:def_T}) and $D=D_0+D_c=\kappa I+(1-\kappa)|0\rangle\langle 0|$. Eq.~(\ref{eq:DFF_exact}) can also be obtained by summing over all possible autocorrelation functions in Eqs.~(\ref{eq:correlation1}) and (\ref{eq:corr_nonlocal}).

The profile of the DFF is characterized by the presence of three peaks~\cite{Yoshimura2024}. The first peak occurs initially $t=0$ with $\overline{F(0)}=q^L$, after which it quickly decays to an $O(1)$ value. The second peak then takes place due to the interplay between locality and the underlying RMT spectral statistics over a timescale comparable to the Thouless time in the RPM. Finally, the third peak can appear depending on the dissipation strength $\kappa$: if $1-\kappa$ is small enough the ramp is sustained over a timescale that is parametrically larger than the Thouless time, but if it is large enough, the peak is essentially suppressed and merges with the Thouless peak.

Although writing out a compact expression for the full DFF Eq.~\eqref{eq:DFF_exact} is not possible in general, its late-time behavior can be captured by a sum of a few terms, each of which has a clear physical origin. This can be obtained by noting that the expression Eq.~\eqref{eq:DFF_exact} can be organized in terms of domain walls in the space of pairing degrees of freedom, where each domain carries the statistical cost $e^{-\varepsilon t}$. It is therefore the configurations with fewer domain walls that control the asymptotic behavior of the DFF at late times. In particular, due to PBCs, the leading corrections to the asymptotic value $\overline{{F}(t\to\infty)}=1$ should come from the terms with zero or two domain walls. We thus obtain the following late-time expansion of the DFF:
\begin{equation}\label{eq:DFF_expansion}
    \overline{F(t)}=1+F_\mathrm{d}(t)+F_\mathrm{DW}(t)+\cdots,
\end{equation}
where 
\begin{equation}
    F_\mathrm{d}(t)=(t-1)e^{-\gamma L\varepsilon t}
\end{equation}
and
\begin{align}
\label{eq:DFF_FDW}
     F_\mathrm{DW}(t)&=(t-1)Le^{-\varepsilon t(2+\gamma)}\n
     &\quad+\frac{(t-1)(t-2)L(L-1)}{4}e^{-\varepsilon t(2+\gamma L)}.
\end{align}
The first nontrivial term $F_\mathrm{d}(t)$ comes purely from dissipation, and decays at a rate $\varepsilon\gamma L$. We call the peak associated with $F_\mathrm{d}(t)$ the dissipation peak. On the other hand, $F_\mathrm{DW}(t)$ primarily stems from domain-wall physics, hence the peak associated with it shall be called the Thouless peak. From Eqs.~(\ref{eq:DFF_expansion})--(\ref{eq:DFF_FDW}), it immediately follows that
\begin{equation}
    \Delta=\varepsilon\min\{2+\gamma,\gamma L\}.
\end{equation}

Before analyzing the behavior of the asymptotic expansion Eq.~\eqref{eq:DFF_expansion}, let us determine which autocorrelation functions are responsible for its different terms using Eq.~(\ref{eq:SFF_autocorrelation}).
First, it is readily inferred that the dissipative contribution $F_\mathrm{d}(t)$
is coming from autocorrelation functions of operators that are maximally nonlocal ($a=L$), Eq.~\eqref{eq:corr_nonlocal}. While this could be somewhat counterintuitive, it happens simply because when the dissipation strength is weak enough, the dissipative decay induced by the coupling to boundaries of the operator clusters, which is avoided in the maximally nonlocal operators, overwhelms that induced by the quantum channels. The first term in $F_\mathrm{DW}(t)$, on the other hand, competes with these terms and originates from autocorrelation functions for strictly local operators with $a=n=1$ as in Eq.~\eqref{eq:corr_local}. Since there are $(q^2-1)^a$ different configurations of onsite operators in these operators, the prefactor $q^{2a}$ in both Eq.~\eqref{eq:corr_local} and Eq.~\eqref{eq:corr_nonlocal} cancels out.

In Ref.~\cite{Yoshimura2024}, we argued that the behavior of the DFF can be best appreciated by assuming the scaling form of the effective dissipation strength $\gamma=\gamma_0L^\alpha$ where $\alpha\in\mathbb{R}$. While it was shown there that
the DFF has three distinctive regimes separated by the value of $\alpha$, here we only review the case $\alpha=0$, which is the most relevant to the present work.

At $\alpha=0$, the Thouless and dissipative peaks overlap and, as a result, the gap $\Delta=\varepsilon(2+\gamma_0)$ does not close even in the dissipationless limit, provided that the thermodynamic limit $L\to\infty$ is taken first. Such anomalous relaxation has been known to occur in different models~\cite{sa2022PRR,garcia2023PRD2,scheurer2023ARXIV,mori2024}, but the exact calculation for the DRPM in Ref.~\cite{Yoshimura2024} clarified its microscopic origin for the first time. Namely, anomalous relaxation in the DRPM is triggered by the fact that the two relaxation timescales of the DFF induced by quantum chaos and dissipation could coincide for some choice of $\alpha$. This indicates that if the timescale associated with quantum chaos, the Thouless time, scales differently with system size, anomalous relaxation would be absent. This scenario typically occurs when the system has a conservation law, implying that the presence of anomalous relaxation is intimately tied to that of conserved charges. We will clarify this point by looking into the DRPM with a $U(1)$ conservation law in Sec.~\ref{sec:suppression_u1}, but before doing so, let us study the relaxation dynamics of other quantities in the DRPM.

\subsection{Operator truncation and the truncated form factor}
\label{sec:DRPM_Truncation}

Previously we claimed that the RP resonances obtained in the weak-dissipation limit of the DRPM are an intrinsic property of the closed RPM. Then, the value of the RP resonance must be independent of the ``seed'' of nonunitarity. In this section, we show that this is indeed the case by using a different source of nonunitary, namely, operator truncation (or coarse-graining), as originally proposed by Prosen~\cite{prosen2004,prosen2007} for the kicked Ising model; see also Refs.~\cite{znidaric2024,znidaric2024b} for recent studies. For the RPM we perform the computation analytically and find the same value of the RP resonance, thus proving the agreement of the two methods for the first time.

Now, let us extract the leading RP resonance from the asymptotic behavior of the TFF, defined in Eq.~(\ref{eq:TFF}) and expanded in terms of autocorrelation functions in Eq.~(\ref{eq:TFF_auto}), in the DRPM at large $q$. 
While we can perform the computation for arbitrary $r$, we are particularly interested in the limit $\lim_{r\to\infty}\lim_{L\to\infty}\hatF_r(t)$.
Using the late-time expansion of the large-$q$ autocorrelation functions Eq.~\eqref{eq:correlation1}, we have
\begin{equation}
     \overline{\hatF_r(t)}=1+\sum_{a=1}^r(L-a+1)(t-1)e^{-2\varepsilon t}
\end{equation}
from which we infer that the leading RP resonance in PBCs is $2\varepsilon$, which precisely agrees with the leading RP resonance obtained from introducing weak dissipation to the RPM. Note that at late times the asymptotic behavior of the TFF becomes identical to that of the partial spectral form factor (PSFF) in the limit where $L_A$ goes as $L_A\to L$ but does not strictly coincide with $L$---see App.~\ref{sec:PSFF} for the details of the calculation.

\subsection{OTOCs and their relaxation}
\label{sec:OTOC}

So far we have discussed how the quantum RP resonances govern the decay of the ensemble-averaged autocorrelation functions. Now we turn to the OTOCs, which are another important probe of quantum chaos. We demonstrate that quantum RP resonances also control their saturation to the stationary value one using the exact OTOCs of the large-$q$ RPM obtained in Ref.~\cite{yoshimura2023operator}. We then show that dissipation induces an overall decay of the OTOCs in the DRPM, which is again dictated by the leading RP resonance.

The OTOC is defined as the average square commutator
\begin{equation}\label{eq:def:OTOC}
    \mathcal{C}_{\mathcal{O}_1\mathcal{O}_2}(x,y;t)
    =-\frac{1}{2}\langle[\mathcal{O}_1(y,t),\mathcal{O}_2(x,0)]^2\rangle,
\end{equation}
where $\mathcal{O}_i(x,t)$ is an arbitrary operator initially supported on site $x$ and evolved to time $t$.

Expanding the right-hand side of Eq.~(\ref{eq:def:OTOC}) for $\mathcal{O}_1(y,t)=\mathcal{O}(y,t)$ with $\langle\mathcal{O}(y,0)\mathcal{O}(y,0)\rangle=1$ and $\mathcal{O}_2(x,0)=P_{\alpha^x}$, the OTOC we shall evaluate reads
\begin{equation}\label{eq:def:OTO2}
\mathcal{C}(x,y;t)=\langle \mathcal{O}(y,t)\mathcal{O}(y,t)\rangle-\langle \mathcal{O}(y,t)P_{\alpha^x}\mathcal{O}(y,t)P_{\alpha^x}\rangle.
\end{equation}
Note that the norm $\langle \mathcal{O}(y,t)\mathcal{O}(y,t)\rangle$ is unity in closed systems, the case we consider first.

\subsubsection{Relaxation of the OTOC in the RPM}\label{sec:relax_otoc}

Here we show that PBCs induce two-stage relaxation of the OTOC while OBCs do not. To see it, let us first briefly recap how one can obtain the exact large-$q$ formula of the OTOC at finite system size for PBCs. For the complete derivation, see Ref.~\cite{yoshimura2023operator}.

In Ref.~\cite{yoshimura2023operator}, the building block $\overline{\langle \mathcal{O}(y,t)P_{\alpha^x}\mathcal{O}(y,t)P_{\alpha^x}\rangle}$ was calculated by Haar-averaging each onsite diagram as in autocorrelation functions. The key observation was that, at large $q$, upon Haar averaging there are $2t+1$ leading pairings that carry the same weight (as opposed to $t$ cyclic leading pairings in autocorrelation functions) due to the presence of two replicas of both the Haar unitary $U_y$ and its conjugate $U_y^*$. This then gives rise to the transfer matrix $S$ that acts on the ($2t+1$)-dimensional vector space spanned by $s=0,\dots,2t$ pairings (labels $s=0,\dots,t$ and $s=t+1,\dots,2t$ correspond to Gaussian and non-Gaussian pairings, respectively~\cite{yoshimura2023operator}),
\begin{equation}
    S=\begin{pmatrix}
        S_1 & S_2 \\
        S_2^\mathrm{T} & S_3
    \end{pmatrix},
\end{equation}
where $S_1$, $S_2$, and $S_3$ are $(t+1)\times(t+1)$, $(t+1)\times t$, and $t\times t$ matrices, respectively. Their matrix elements are given by 
\begin{equation}
\begin{split}
    &[S_1]_{ab}=\delta_{ab}+\rho^{|a-b|-1}(1-\delta_{ab}),\\
    &[S_2]_{ab}=\delta_{ab}+\rho^{|a-b|-1}\left[1-\delta_{ab}+\Theta(b-a)(\rho-1)\right],\\
    &[S_3]_{ab}=\delta_{ab}+\rho^{|a-b|}(1-\delta_{ab}),
\end{split}
\end{equation}
where $\rho=e^{-2\varepsilon}$, indices $a,b$ run from $0$ to $t$, and $\Theta(a)$ is the step function with $\Theta(0)=0$. 
In addition to this, each pairing also carries a weight of either $q$ or $-q$ depending on whether the pairing is Gaussian or non-Gaussian, respectively, giving rise to the onsite $(2t+1)\times(2t+1)$ matrix $qE$ where $E=\mathrm{diag}\,\omega_s$ with $\omega_s=1$ for $s=0,\dots,t$ and $\omega_s=-1$ for $s=t+1,\dots,2t$.

With the transfer matrix $S$ and the onsite matrix $E$ at our disposal, we can evaluate the four-point function $\overline{\langle \mathcal{O}(y,t)P_{\alpha^x}\mathcal{O}(y,t)P_{\alpha^x}\rangle}$ by applying $E$ to every site and $S$ to every edge that connects adjacent sites. In addition to this, we also have to insert the projector onto the $s=0$ and $s=t$ pairings at site $y$ and $x$, respectively. 

For PBCs, assuming $x>y$ for simplicity, we therefore have
\begin{align}
    \label{eq:OTOC_OPOP}
    &\overline{\langle \mathcal{O}(y,t)P_{\alpha^x}\mathcal{O}(y,t)P_{\alpha^x}\rangle}\n
    &=\mathrm{Tr}[(ES)^{L-(x-y)}|0\rangle\langle0|(ES)^{x-y}|t\rangle\langle t|] \n
    &=\hat{S}^{L-(x-y)}_{t0}\hat{S}^{x-y}_{0t},
\end{align}
where $\hat{S}=ES$, which yields the OTOC:
\begin{equation}
    \overline{\mathcal{C}(x,y;t)}=1-\hat{S}_{t0}^{L-(x-y)}\hat{S}_{0t}^{x-y}.
\end{equation}
The matrix element $\hat{S}_{0t}^\ell$ is given by~\cite{yoshimura2023operator}
\begin{equation}
\label{eq:hatS_sum}
    \hat{S}_{0t}^\ell=\hat{S}_{t0}^\ell=1-\sum_{i=0}^{t-\ell-1}
        \begin{pmatrix}
        t-1\\
        i
        \end{pmatrix}
        \rho^i(1-\rho)^{t-i-1}.
\end{equation}
It was shown in Ref.~\cite{yoshimura2023operator} that the large-$q$ OTOC in the thermodynamic limit is given by $\overline{\mathcal{C}(x,y;t)}=1-\hat{S}_{0t}^x$ because $\lim_{L\to\infty}\hat{S}_{t0}^{L-(x-y)}=1$. It has the following asymptotic form on the hydrodynamic scale $x-y,t\gg1$:
\begin{equation}
    \overline{\mathcal{C}(x,y;t)}\simeq\Phi\left(\frac{v_Bt-r}{\sqrt{2\mathcal{D}t}}\right),\quad r:=x-y,
\end{equation}
where $v_B=1-\rho$ is the butterfly velocity, $\mathcal{D}=\rho(1-\rho)/2$ is the diffusion constant, and $\Phi(x)=\int_{-\infty}^x\dd y\,e^{-y^2/2}/\sqrt{2\pi}$.

We now show that the OTOC with finite system size undergoes two-stage relaxation when PBCs are used. The two stages during relaxation are divided by the timescale $t_\mathrm{relax}= L/v_B$, over which both operator fronts start having some overlap with the operator $P_{\alpha^x}$ due to PBCs. When $t\lesssim t_\mathrm{relax}$, the OTOC can be approximated by the infinite-volume result, for which the sum in Eq.~(\ref{eq:hatS_sum}) can be carried out explicitly, yielding
\begin{equation}
\begin{split}
\overline{\mathcal{C}(x,y;t)}
\simeq1+&(1-\rho)^r\rho^{t-r-1}\binom{t-1}{t-r-1}
\\&\times\left(1-{}_2F_1(1,-r,t-r,\rho/(\rho-1)\right),
\end{split}
\end{equation}
with ${}_2F_1(a,b,c;z)$ the hypergeometric function and $r>0$. Under the latter condition, the hypergeometric function is a polynomial with a finite number of terms that approaches one in the large-$t$ limit. For fixed $r$ and at late times, we therefore have
\begin{equation}
   \overline{\mathcal{C}(x,y;t)}\simeq1-\frac{(1-\rho)^{r-1}}{(r-1)!}t^{r-1}\rho^t,
\end{equation}
which decays to one with the rate $2\varepsilon$. 

Next, we turn to the second stage of relaxation $t\gtrsim t_\mathrm{relax}$, in which case the operator $\mathcal{O}(y,t)$ has scrambled over the entire system and we can no longer set $\hat{S}_{t0}^{L-(x-y)}=1$. In this case, we have $r,L-r< t_\mathrm{relax}$, and following the same argument as above, we obtain
\begin{equation}
     \overline{\mathcal{C}(x,y;t)}\simeq 1-\frac{(1-\rho)^{L-2}}{(L-r-1)!(r-1)!}t^{L-2}\rho^{2t}.
\end{equation}
We thus infer that during the second stage of relaxation, the decay rate is replaced by $4\varepsilon$, which is twice as large as that during the first stage. Note that this rate can also be obtained by simply invoking the relation between RP resonances and late-time dynamics, as discussed in Sec.~\ref{sec:quantumRP}. Namely at late times we can effectively assume $|\mathcal{O}(y,t)\rrangle\approx e^{-\nu^*_1t}|L_1\rrangle$. Putting this into Eq.~\eqref{eq:def:OTO2}, we immediately observe that the ensemble-averaged OTOC decays to one with the rate $2\nu^*_1$, which reads $4\varepsilon$ for the RPM with PBCs. Thus, also the saturation of the OTOC in the large-$q$ RPM is ruled by (twice) the leading RP resonance, which is in agreement with the assertion that the RP resonances are intrinsic properties of the closed system. 

Having clarified the origin of two-stage relaxation when PBCs are imposed, let us move on to the OTOC with OBCs.
In this case, as in autocorrelation functions, there are two major changes compared to systems with PBCs; namely, the trace along the spatial direction has to be replaced by the boundary state $|\phi\rangle=\sum_{s=0}^{2t}|s\rangle$ and the behavior of the OTOC can differ qualitatively depending on where the initial operator $\mathcal{O}(y,0)$ is supported. Again assuming $x>y$ without loss of generality, the connected part of the OTOC reads
\begin{align}
    &\overline{\langle \mathcal{O}(y,t)P_{\alpha^x}\mathcal{O}(y,t)P_{\alpha^x}\rangle}\n
    &=\langle\phi|\hat{S}^{y-1}|0\rangle\langle0|\hat{S}^{x-y}|t\rangle\langle t|\hat{S}^{L-x}E|\phi\rangle\n
    &=\sum_{s,s'}\hat{S}^{y-1}_{s0}\hat{S}^{x-y}_{0t}\hat{S}^{L-x}_{ts'}\omega_{s'}.
\end{align}
The last line can be further simplified by noting the identities $\sum_s\hat{S}^{\ell}_{s0}=\sum_{s'}\hat{S}^{\ell}_{ts'}\omega_{s'}=1$ for any $\ell>0$, yielding 
\begin{equation}
\label{eq:OTOC_relax_OBC}
    \overline{\langle \mathcal{O}(y,t)P_{\alpha^x}\mathcal{O}(y,t)P_{\alpha^x}\rangle}=\hat{S}^{x-y}_{0t}.
\end{equation}
This indicates that the OTOC always decays with the same rate $2\varepsilon$ for any system size and the value of $y$, hence the two-stage relaxation is absent with OBCs. Note that, as in the case of PBCs, the rate associated with the long-time decay with finite $L$ can be deduced solely based on the fact that late-time dynamics is controlled by the leading RP resonance, which is $\varepsilon$ for the RPM with OBCs. Following the same argument above for PBCs, the OTOC then has to decay with the rate $2\varepsilon$ [since it has two time-evolved operators $\mathcal{O}(0,t)$], which is precisely the case in Eq.~(\ref{eq:OTOC_relax_OBC}).

On a physical ground, the presence or absence of doubling of the rate may be understood in the following way: Supposing that the system is subject to PBCs and $x>0$, the saturation of the OTOC $\overline{\mathcal{C}(x,y;t)}$ to its stationary value one during the initial stage is driven by the right end of the (sum of) operator string $\mathcal{O}(y,t)$ that starts to cover the position $x$. This process persists until the second stage, where its left end also starts to have some overlap with $\mathcal{O}(x,0)$ due to PBCs over the time scale $\sim L/v_B$ after wrapping around the system, thereby accelerating saturation with a twice higher rate. This mechanism is absent in systems with OBCs, as only one end of the time-evolved operator $\mathcal{O}(0,t)$ will overlap with the operator at $x$ even at sufficiently late times. 
We note that while the absence of two-stage relaxation in generic random brickwork circuits with OBCs was observed in Ref.~\cite{Bensa_relaxation_2022}, it was shown in Ref.~\cite{Bensa_relaxation_2022, jonay2024twostagerelaxationoperatorsdomain} that the dual-unitary circuits are somewhat exceptional and display the change of relaxation rates even with OBCs.

The existence of such two-stage relaxation in the OTOC~\cite{Bensa_relaxation_2022,Znidaric_relaxation_2023,jonay2024twostagerelaxationoperatorsdomain,bensa2024} and other related quantities such as the purity~\cite{Bensa_phantom_2021,znidaric2022,Znidaric_relaxation_2023,znidaric2023_PRR,bensa2024_PRA,Jonay_thermalization_2024} has been known and studied in recent years. While these studies have argued the origin of the two-stage relaxation from different perspectives such as the so-called ``phantom'' eigenvalues~\cite{Bensa_phantom_2021}, the pseudospectrum~\cite{znidaric2022,znidaric2023_PRR}, and the entanglement membrane theory~\cite{jonay2024twostagerelaxationoperatorsdomain}, a unifying microscopic understanding of it is still missing. Here, our analysis explains that it is intimately related to how operators spread differently depending on boundary conditions, and why the rate has to {\it double} instead of changing to a different rate when entering the second stage. We also showed that at late times the OTOC decays with the leading RP resonance allowed by the boundary condition used. However, the nature of the rate in the first stage (when PBCs are imposed), in particular how the pseudospectrum~\cite{znidaric2022,znidaric2023_PRR} and RP resonances can be related in a precise way (given the similarities of their definitions in terms of noncommutativity of the thermodynamic and noiseless limits) remains unclear. We leave this investigation for future study.

\subsubsection{Decay of the OTOC in the DRPM}\label{ref:decay_drpm}

So far we have studied the relaxation behavior of the OTOC in the RPM. Next we move on to investigate the effect of dissipation on the relaxation of the OTOC in the DRPM.

As in the DFF, it can be easily seen that the effect of dissipation manifests itself in the diagonal onsite matrix $E$ of size $2t+1$, which in the presence of dissipation reads $E_\mathrm{d}=\mathrm{diag}\,\omega_s$, where now $\omega_s=\omega^{t-s-1}$ for $s=0,\dots,t-1$, $\omega_t=1$, and $\omega_s=-\omega^{2t-s}$ for $s=t+1,\dots,2t$, with $\omega=e^{-2\varepsilon \gamma}$ (e.g., for the depolarizing channel, $\omega=(1-p)^2$). For example, the onsite matrix for $t=3$ is
\begin{equation}
    E_\mathrm{d}=\begin{pmatrix}
        \omega^2&0&0&0&0&0&0\\
        0&\omega&0&0&0&0&0\\
        0&0&1&0&0&0&0\\
        0&0&0&1&0&0&0\\
        0&0&0&0&-\omega^2&0&0\\
        0&0&0&0&0&-\omega&0\\
        0&0&0&0&0&0&-1\\
    \end{pmatrix}.
\end{equation}
With this, we can now repeat the same argument as in the closed case, and the only change is to replace $E$ with $E_\mathrm{d}$. 

Let us first focus on PBCs, for which, defining $\hat{S}_\mathrm{d}=E_\mathrm{d}S$, we have 
\begin{equation}
    \overline{\langle \mathcal{O}(y,t)P_{\alpha^x}\mathcal{O}(y,t)P_{\alpha^x}\rangle}=[\hat{S}^{L-(x-y)}_\mathrm{d}]_{t0}[\hat{S}^{x-y}_\mathrm{d}]_{0t}.
\end{equation}
Similarly, we can also express the norm $\overline{\langle \mathcal{O}(y,t)\mathcal{O}(y,t)\rangle}$, which is independent of $y$ for PBCs, as
\begin{equation}
\label{eq:OTOC_OO}
\overline{ \langle \mathcal{O}(y,t)\mathcal{O}(y,t)\rangle}=\mathrm{Tr}[(E_\mathrm{d}S)^L|0\rangle\langle0|]=[\hat{S}^L_\mathrm{d}]_{00}.
\end{equation}
We thus obtain the large-$q$ OTOC in the DRPM with PBCs
\begin{equation}\label{eq:otoc_drpm_pbc}
    \overline{\mathcal{C}(x,y;t)}=[\hat{S}_\mathrm{d}^{L}]_{00}-[\hat{S}_\mathrm{d}^{L-(x-y)}]_{t0}[\hat{S}^{x-y}_\mathrm{d}]_{0t}.
\end{equation}
Due to dissipation, the OTOC $\overline{\mathcal{C}(x,y;t)}$ now decays exponentially to zero instead of approaching unity at long times. As in the DFF, we are particularly interested in taking the large system size limit $L\to\infty$ first and then sending $\gamma\to0$ (i.e., $\omega\to1$), under which not only the second term in Eq.~\eqref{eq:otoc_drpm_pbc} but also the first term, which is the norm $\overline{\langle \mathcal{O}(y,t)\mathcal{O}(y,t)\rangle}$, decays exponentially.

We now show that the two contributions to the OTOC decay with the \emph{same} rate, which coincides with (twice) the leading RP resonance $2\varepsilon$. This feature of the OTOC was observed before for single-particle systems~\cite{garcia-mata2018}, but our results show that it also extends to the many-body setting, suggesting its universality in quantum dynamics. The transfer-matrix expressions in Eqs.~(\ref{eq:OTOC_OPOP}) and (\ref{eq:OTOC_OO}) are exact but their large-$L$, large-$t$ asymptotics are hard to analyze.
Nevertheless, we have already shown in the previous section that, in the dissipationless limit the decay rate of the four-point contribution $\overline{\langle \mathcal{O}(y,t)P_{\alpha^x}\mathcal{O}(y,t)P_{\alpha^x}\rangle}$ is indeed given by $2\varepsilon$.
\begin{figure}[t]
    \centering
    \includegraphics[width=\columnwidth]{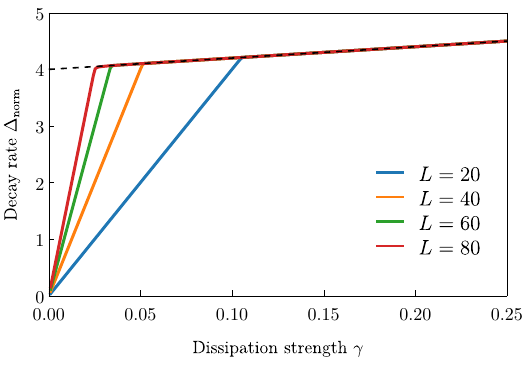}
    \caption{\textbf{Exact transfer-matrix result for the decay of the norm.} The colored lines give the decay rate $\Delta_{\mathrm{norm}}$ of $\overline{\langle\mathcal{O}(y,t)\mathcal{O}(y,t)\rangle}$ for PBCs in units of $\varepsilon=1$, obtained for $120<t<140$ and $20<L<80$, as a function of $\gamma$. The dashed black line is given by the extrapolation of the collapsed $\Delta_{\mathrm{norm}}(\gamma)$ down to $\gamma\to0$, where it intercepts the vertical axis at $\Delta_{\mathrm{norm}}(0)=4$.}
    \label{fig:OO_decay}
\end{figure}

We thus proceed with the operator norm contribution $\overline{\langle \mathcal{O}(y,t)\mathcal{O}(y,t)\rangle}$, Eq.~(\ref{eq:OTOC_OO}). Exact evaluation of $[\hat{S}^{L}_\mathrm{d}]_{00}$ shows that for $L>2(t-1)$ it becomes \emph{independent} of $L$ and we have
\begin{equation}
\label{eq:qdeformation_OO}
    \overline{\langle\mathcal{O}(y,t)\mathcal{O}(y,t)\rangle}
    =(\omega\rho^2)^{(t-1)}\left[\left(1-\rho^{-1};\omega\right)_{t-1}\right]^2,
\end{equation}
where $(a;q)_n=\prod_{k=0}^{n-1}(1-aq^k)$ is the $q$-Pochhammer symbol. Notice that here dissipation acts as a $q$-deformation of the closed-system operator norm. Taking the limit $\omega\to1$ (i.e., $\gamma=0$) reduces it to the result of the previous section, $\overline{\langle\mathcal{O}(y,t)\mathcal{O}(y,t)\rangle}=1$ regardless of whether the thermodynamic limit is taken before or after, since it is independent of $L$. Instead, we are interested in the long-time decay of the correlator and therefore have to take the limit $t\to\infty$ before $L\to\infty$, which is incompatible with the restriction $L>2(t-1)$ in Eq.~(\ref{eq:qdeformation_OO}). In the opposite case, $2(t-1)>L$, we are unable to find a simple closed-form expression. However, we can still evaluate Eq.~(\ref{eq:OTOC_OO}) exactly for arbitrary $t$, $L$, and $\omega$ and extract the decay rate by fitting the late-time decay to an exponential, $\overline{\langle\mathcal{O}(y,t)\mathcal{O}(y,t)\rangle}=A_{\mathrm{norm}} e^{-\Delta_{\mathrm{norm}} t}$, with $A_{\mathrm{norm}}$ and $\Delta_{\mathrm{norm}}$ fitting parameters, as shown in Fig.~\ref{fig:OO_decay}. We find the behavior expected for anomalous relaxation~\cite{Yoshimura2024}: the curves for different $L$ collapse to a universal curve up to a critical value of the dissipation strength below which the system sees the finite spacing of the lattice and, correspondingly, the gap closes. As $L$ increases (but keeping it below the times used for the fit), this critical point decreases until reaching $p=0$ in the thermodynamic limit. Extrapolating the collapsed curve to this limit, we obtain a decay rate of $\Delta_{\mathrm{norm}}\approx4\varepsilon$ for $\gamma=0$, as expected.

Let us next turn to OBCs. Following the same argument as in the dissipationless case, the OTOC reads
\begin{align}
    \label{eq:otoc_drpm_obc}
   & \overline{\mathcal{C}(x,y;t)}\n
   &=\sum_{s,s'}[\hat{S}_\mathrm{d}^{y-1}]_{s0}\omega_{s'}\left([\hat{S}_\mathrm{d}^{L-y}]_{0s'}-[\hat{S}^{x-y}_\mathrm{d}]_{0t}[\hat{S}_\mathrm{d}^{L-x}]_{ts'}\right).
\end{align}
for $x>y$. Similarly to the case of PBCs, we already know that the connected part decays with the rate $2\varepsilon$ for any value of $y$ at late times while $L$ is fixed. This turns out to be not the case for the norm $\overline{\langle\mathcal{O}(y,t)\mathcal{O}(y,t)\rangle}=\sum_{s,s'}[\hat{S}_\mathrm{d}^{y-1}]_{s0}\omega_{s'}[\hat{S}_\mathrm{d}^{L-y}]_{0s'}$, which is sensitive to the initial location of the operator $\mathcal{O}(y,0)$ for OBCs. It can be readily seen that when $y$ is sufficiently close to the boundary $y=1$ (resp.\ $y=L)$ such that $\lim_{L\to\infty} y-1<\infty$ (resp.\ $\lim_{L\to\infty} L-y<\infty$), we have $\lim_{\omega\to1}\lim_{L\to\infty}[\hat{S}_\mathrm{d}^{y-1}]_{s0}=1$ (resp.\ $\lim_{\omega\to1}\lim_{L\to\infty}[\hat{S}_\mathrm{d}^{L-y}]_{0s}\omega_s=1$), yielding
\begin{equation}
    \lim_{\omega\to1}\lim_{L\to\infty}\overline{\langle\mathcal{O}(y,t)\mathcal{O}(y,t)\rangle}=\lim_{\omega\to1}\lim_{L\to\infty}\sum_s\omega_{s}[\hat{S}_\mathrm{d}^{L-y}]_{0s}
\end{equation}
and another similar result for the case where $\mathcal{O}(y,0)$ is localized near $L$. One can again extract the rate with which each sum decays via fitting, and we find that the rate is $2\varepsilon$ for both cases. Instead, if $\mathcal{O}(y,0)$ is supported well within the bulk so that both $y-1$ and $L-y$ go to infinity as $L$ becomes large, both sums that constitute the norm decay with $2\varepsilon$, giving rise to the net decay rate $4\varepsilon$ of the norm.

In summary, the OTOC of the large-$q$ DRPM with OBCs decays with either $2\varepsilon$ or $4\varepsilon$ depending on whether the initial operator $\mathcal{O}(y,0)$ is supported near the boundaries or not (remember that the leading RP resonance is $\varepsilon$ for OBCs). The fact that the behavior of the OTOC is sensitive to whether the operator whose spreading we wish to probe is close to the boundary is reminiscent of what happens for autocorrelation functions of the (D)RPM with OBCs as discussed in Sec.~\ref{sec:corr_open}. This observation raises a natural question: why do only boundary-localized operators tend to overlap with the basis operator $|L_1\rrangle$ so that the leading RP resonance dominates the relaxation? The difference in the number of Haar unitaries that have to be averaged in the correlator makes this phenomenon nontrivial, and a more systematic investigation of it is beyond the scope of the present paper.

\section{Suppression of anomalous relaxation}
\label{sec:suppression}

Having studied in detail the relaxation dynamics of the simplest version of the DRPM, in this section, we inquire what the conditions are for observing anomalous relaxation in quantum many-body systems. To this end, we study minimal extensions that alter some of the properties in the DRPM, namely, extensive (bulk) dissipation, absence of conservation laws, ergodicity, and local interactions. We shall show that in the absence of these properties, anomalous relaxation is suppressed, i.e., certain operators do not relax, and the gap extracted from the DFF is not positive finite when the dissipationless limit is taken after the thermodynamic limit. 

First, in Sec.~\ref{sec:suppression_subextensive}, we consider the DRPM with only partial bulk dissipation (i.e., a circuit in which dissipation acts only on a subset of lattice sites). If the set of dissipated sites is subextensive (i.e., does not grow with $L$), then we cannot suppress the ramp of the SFF and anomalous relaxation is absent. 

We also clarify the role of local interactions in building up the Thouless peak and hence giving rise to anomalous relaxation. In Sec.~\ref{sec:suppression_nonlocal}, we consider a random-matrix Floquet model with fully nonlocal interactions (which can, therefore, be seen as a model of single-body chaos instead of many-body chaos). Here, since the Thouless peak disappears, there is no competition between local chaos and dissipation. When the interactions become nonlocal, operator growth is massively enhanced, and local operators start contributing to the ramp, which cannot be suppressed by adding dissipation. The gap becomes simply proportional to $\gamma L$ for any dissipation strength; in this case, there is no well-defined gap in the thermodynamic limit (and thus no meaningful anomalous relaxation). 

Next, we show that the complex RP resonances merge into the unit disk in the presence of conservation laws. In Sec.~\ref{sec:suppression_u1}, we study the dynamics of both the RPM and the DRPM with a $U(1)$ conservation law. We show that the emergence of long-lived diffusive modes (in all symmetry sectors) leads to the closing of the relaxation gap. 

Finally, in Sec.~\ref{sec:suppression_numerics}, we complement these analytical findings with the numerical computation of the spectral gap for qubit systems with $q=2$. We show that while generic circuits with many-body quantum chaos show finite-size signatures of anomalous relaxation in their spectral gap, these are absent in the presence of conservation laws and many-body localization~\cite{Nandkishore2015,Abanin_MBLreview_2019}. This illustrates the generic character of our results, which are not restricted to toy models such as the RPM at large $q$.

\subsection{Subextensive dissipation}
\label{sec:suppression_subextensive}

Suppose that the RPM is dissipated by quantum channels acting on $\ell\leq L$ sites. Here, we explore the necessary conditions for anomalous relaxation to occur as we vary $\ell$ by focusing on PBCs.

Let us first look at the case $\ell=1$. The boundary-dissipated RPM belongs to this case, but with PBCs the location on which a quantum channel acts has no consequence. Following the same calculation we did for the DRPM, the DFF for this case can be expressed in terms of the transfer matrix $T$, with the following closed expression:
\begin{equation}
\begin{split}
    \overline{F(t)}
    &=\mathrm{Tr}(T^LD)
    =\kappa(t) \overline{K(t)}+(1-\kappa(t))T^L_{00}
    \\
    &=\left(\kappa(t)+\frac{1-\kappa(t)}{t}\right)\left(\lambda_0^L+(t-1)[\lambda(t)]^L\right).
\end{split}
\end{equation}
The late-time domain-wall expansion then gives
\begin{equation}
    \overline{F(t)}\simeq 1+\frac{tL^2}{4}e^{-2\varepsilon t}+te^{-\gamma\varepsilon t}+\frac{L^2t^2}{4}e^{-\varepsilon t(2+\gamma)},
\end{equation}
from which we infer that the gap is given by
\begin{equation}
    \Delta=\varepsilon\min(2,\gamma).
\end{equation} 
Thus, in the limit of small $\gamma$, $\Delta=\varepsilon\gamma\to0$ and anomalous relaxation does not occur in the boundary-dissipated RPM.

Next, let us consider the generic situation $\ell>1$. As in the DRPM, the DFF can be expressed in terms of the transfer matrix,
\begin{equation}
    \overline{F(t)}=\Tr \left(\prod_x TD^{\delta_x}\right),
\end{equation}
where $\delta_x=1$ if dissipation acts on site $x$ and $\delta_x=0$ if it does not.
In this case, the DFF has the following late-time asymptotics,
\begin{equation}
     \overline{F(t)}\simeq 1+te^{-\gamma\ell\varepsilon t}+t\ell e^{-\varepsilon(2+\gamma)t}+\sum_{m=1}^M\frac{\bar{\ell}_m(\bar{\ell}_m+1)}{2}te^{-2\varepsilon t},
\end{equation}
where $m=1,\dots,M$ enumerates the clusters of sites on which quantum channels {\it do not} act, and $\bar{\ell}_m$ is the length of the cluster labeled by $m$. This indicates that the gap is given by 
\begin{equation}
   \Delta=\varepsilon\min(\gamma\ell,2+\gamma),
\end{equation}
and that $\ell$ has to diverge with the system size $L$ for anomalous relaxation to occur.

\subsection{Nonlocal interactions}
\label{sec:suppression_nonlocal}

Next, we consider a chaotic quantum circuit with fully nonlocal interactions. We define the random matrix Floquet (RMF) system as a Floquet system whose time-evolution operator $W$ is a $q^L\times q^L$ matrix drawn from the circular unitary ensemble, and we call the RMF system with local dissipation the dissipative RMF (DRMF) system. The DFF of the DRMF system is then defined in the usual way,
\begin{equation}
\label{eq:DFF_DRMF}
\overline{F(t)}=\sum_{j_1,\dots,j_t}\overline{|\mathrm{Tr}K_{j_t}W\cdots K_{j_1}W|^2},\quad K_{j_\tau}=\bigotimes_{x=1}^L(M_{j_{\tau}})_x,
\end{equation}
where $(M_{j_{\tau}})_x$ are the same local Kraus operators at time $\tau$ and site $x$ as before.
As shown in App.~\ref{sec:nonlocal}, at large $q$, we have
\begin{equation}
\label{eq:subextensive_DFF}
    \overline{F(t)}=1+(t-1)\left(\sum_{i}\eta_i^{2t}\right)^L\simeq 1+(t-1)\eta_M^{2tL},
\end{equation}
where we recall that $\eta_M$ is the largest of $\eta_i$.

For example, for depolarizing channels, this reduces to
\begin{equation}
    \overline{F(t)}=1+(t-1)(1-p)^{2 tL},
\end{equation}
which clearly shows the absence of anomalous relaxation. We thus explicitly demonstrate that local interactions, which in turn induce the Thouless peak, are necessary for anomalous relaxation.

\subsection{Conservation laws}
\label{sec:suppression_u1}

It was recently argued in Ref.~\cite{mori2024} that the absence of conservation laws is required for anomalous relaxation to take place. Indeed, the microscopic origin of anomalous relaxation in the DRPM suggests that the change of the Thouless time, which is usually induced by the presence of a conservation law, would spoil anomalous relaxation. To further substantiate this claim, in the following, we construct a $U(1)$-charge-conserving DRPM and demonstrate explicitly that the conservation law prevents anomalous relaxation in the model.

\subsubsection{RPM with a conserved \texorpdfstring{$U(1)$}{U(1)} charge}

\begin{figure}[t]
\centering
\includegraphics[width=0.7\columnwidth]{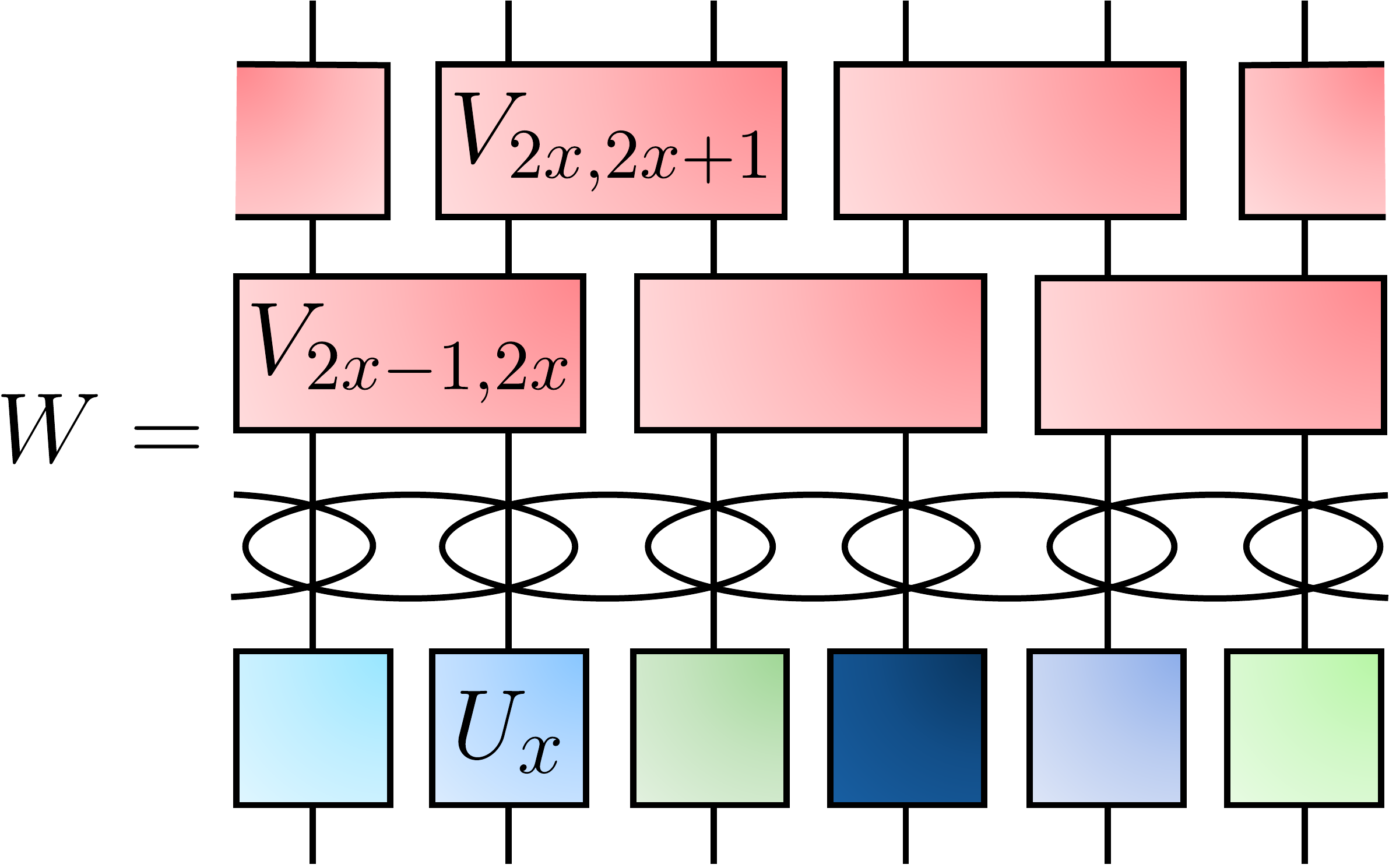}
\caption{\textbf{Diagrammatic representation of Floquet operator of the RPM with a conserved $U(1)$ charge.} The first two layers $W_2W_1$ coincide with the usual RPM operator, see Fig.~\ref{fig:circuits}(c). The third layer $W_3$ has a brickwork architecture with \emph{fixed} gates (red rectangles) given by Eq.~(\ref{def:gate_V}).}
\label{fig:floquet}
\end{figure}

Before introducing the DRPM with a $U(1)$ charge, let us define and study the dissipationless RPM with a $U(1)$-charge-conservation law. We first promote the onsite unitaries $U_x$ that enter in $W_1=\bigotimes_{x=1}^LU_x$ to be a block-diagonal $2q\times2q$ matrix, which acts on the onsite Hilbert space $\mathbb{C}^2\otimes\mathbb{C}^q$ spanned by $|\mathtt{s},a\rangle$ where $\mathtt{s}=\uparrow,\downarrow$ and $a=1,\dots,q$:
\begin{equation}
    U_x=\begin{pmatrix}
        \tilde{U}_{1,x} & 0 \\
        0 & \tilde{U}_{2,x}
    \end{pmatrix},
\end{equation}
where $\tilde{U}_{1,x}$ and $\tilde{U}_{2,x}$ are independently Haar-distribute $q\times q$ random matrices. The basis operator strings now consist of $P_{\mu,\alpha}=\bigotimes_{x=1}^L\sigma_{\mu^x}\otimes P_{\alpha^x}$ where $\sigma_\mu$ are Pauli matrices with $\mu^x=0,\dots,3$ and $\alpha^x=0,\dots,q^2-1$. We choose $\sigma_0=I_2$ and $P_0=I_q$. For brevity, we denote the doublet $(\mu,\alpha)$ by $\underline{\alpha}$. On the other hand, intersite couplings between neighboring sites are induced by $W_2$, which acts on both spin and color degrees of freedom diagonally (hence, no spin transport is induced by the coupling). 
Finally, the third layer $W_3=Z_2Z_1$  introduces spin exchange, where, assuming PBCs with even $L$,
\begin{align}
    &Z_1=V_{1,2}\otimes V_{3,4}\otimes\cdots\otimes V_{L-1,L},
    \\
    &Z_2=V_{2,3}\otimes V_{4,5}\otimes\cdots\otimes V_{L,1},
\end{align}
with 
\begin{equation}
\label{def:gate_V}
    V_{x,x+1}=T_{x,x+1}(I_q)_{x,x+1}, \qquad T_{x,x+1}=e^{\ii\alpha \mathbb{P}_{x,x+1}}.
\end{equation}
Here $ \mathbb{P}_{x,x+1}$ is the \textsc{Swap} operator that acts only on the spin sector and satisfies $e^{\ii\alpha  \mathbb{P}_{x,x+1}}=(\cos\alpha(I_2)_{x,x+1}+\ii\sin\alpha  \mathbb{P}_{x,x+1})$, and $(I_n)_{x,y}=(I_n)_x\otimes(I_n)_y$ where $(I_n)_x$ is the $n\times n$ identity matrix at site $x$. The Floquet operator is therefore composed of these three operators $W=W_3W_2W_1$, see Fig.~\ref{fig:floquet}. 

We will focus on PBCs. Since $[W,S^z]=0$, the Floquet operator is block-diagonal. We are thus interested in computing the SFF at fixed magnetization $N=0,\dots,L$,
\begin{equation}\label{eq:sff}
    \overline{K_N(t)}=\overline{(\Tr_N W^t)(\Tr_N (W^\dagger)^t)}=2^{-L}D_N\sum_{\underline{\alpha}}\overline{C^{[N]}_{\underline{\alpha}\underline{\alpha}}(t)},
\end{equation}
where $C^{[N]}_{\underline{\alpha}\underline{\beta}}(t)=\left(D_Nq^L\right)^{-1}\mathrm{Tr}_N\,((W^\dagger)^t P_{\underline{\alpha}}W^t P_{\underline{\beta}})$ with $D_N:=\begin{pmatrix}
    L\\N
\end{pmatrix}$. Here $\mathrm{Tr}_N\bullet$ is the trace over the sector with the fixed magnetization $N$, and the second equality in Eq.~\eqref{eq:sff} follows from the fact that levels from different $S^z$-sectors do not repel.

Let us compute Eq.~\eqref{eq:sff} at large $q$ using the diagrammatic techniques we developed for the standard RPM. Since the unitary $U_x$ is block diagonal, Haar-averaging induces not only pairings of color degrees of freedom but also those of spin. In particular, the leading pairings are still cyclic: each pairing is characterized by the variable $s=0,\dots,t-1$ that relates the doublet indices $\underline{a}=(s,a)$ labeling $U_x$ and $\underline{b}=(r,b)$ labeling $U^*_x$ as $\underline{a}(t)=\underline{b}(t+s)$. This implies that the domain-wall structure in the RPM remains intact also in the presence of a $U(1)$ charge. Suppose there is no domain wall; i.e., at every site we have the same cyclic pairing. In this case, the pairing for the spin degrees of freedom gives rise to the same spin configurations for both $W$ and $W^*$ due to the $\mathbb{Z}_t$-symmetry of the Fock space orbits $\Tr_N W$ and $\Tr_N W^\dagger$. This suggests that the contribution to the large-$q$ SFF from uniform pairing configurations (i.e., no domain walls) is given by $t\Tr \mathcal{M}^t$, where $\mathcal{M}$ is the time-evolution operator of the discrete symmetric simple exclusion process (d-SSEP)~\cite{Friedman_Spectral_2019}. Namely, $\mathcal{M}$ is made of two layers $\mathcal{M}=\mathcal{M}_2\mathcal{M}_1$ where
\begin{align}
    &\mathcal{M}_1=\Tilde{T}_{1,2}\otimes \Tilde{T}_{3,4}\otimes \cdots\otimes\Tilde{T}_{L-1,L},
    \\
    &\mathcal{M}_2=\Tilde{T}_{2,3}\otimes \Tilde{T}_{4,5}\otimes \cdots\otimes\Tilde{T}_{L-2,L-1}.
\end{align}
Note that the matrix elements of $\Tilde{T}_{x,x+1}$ are given by $(\Tilde{T}_{x,x+1})_{\mathtt{s},\mathtt{s}'}^{\mathtt{r},\mathtt{r}'}:=(T_{x,x+1})_{\mathtt{s},\mathtt{s}'}^{\mathtt{r},\mathtt{r}'}(T^\dagger_{x,x+1})_{\mathtt{s},\mathtt{s}'}^{\mathtt{r},\mathtt{r}'}$, which implies $\Tilde{T}_{x,x+1}=\cos^2\alpha(\mathbb{I}_2)_{x,x+1}+\sin^2\alpha\mathbb{P}_{x,x+1}=(I_2)_{x,x+1}-h_{x,x+1}$ where $h_{x,x+1}=-\frac{\sin^2\alpha}{2}(\Vec{\sigma}_x\cdot\Vec{\sigma}_{x+1}-(I_2)_{x,x+1})$. Therefore, as was already pointed out in Ref.~\cite{Friedman_Spectral_2019}, $\mathcal{M}$ can be thought of as a Trotterization of $e^{-t H_\mathrm{XXX}}$ with $H_\mathrm{XXX}=\sum_xh_{x,x+1}$ the Hamiltonian of the spin-1/2 ferromagnetic Heisenberg spin chain. Since each domain wall comes with the entropic cost $e^{-\varepsilon t}$, the late time behavior of the SFF at large $q$ is simply captured by
\begin{equation}
     \overline{K_N(t)}\simeq t\Tr_N \mathcal{M}^t,
\end{equation}
which exactly coincides with the SFF of the brickwork Floquet circuit with a $U(1)$ charge introduced in Ref.~\cite{Friedman_Spectral_2019} where domain walls are completely suppressed at large $q$. The asymptotics of $\Tr_N \mathcal{M}^t$ was also discussed in Ref.~\cite{Friedman_Spectral_2019} where the authors argued that the asymptotics can be described by the low-lying magnon excitations of the Heisenberg ferromagnet whose dispersion relation is given by $\varepsilon(k)\sim k^2$.  We thus have, at $t\gg L^2$, $\Tr_N \mathcal{M}^t=1+e^{-4\pi^2t/L^2}+\cdots$, yielding the late-time behavior of the SFF of the large-$q$ RPM with a $U(1)$ charge
\begin{equation}
     \overline{K_N(t)}=t(1+e^{-4\pi^2t/L^2}+\cdots),
\end{equation}
which indicates that the Thouless time is 
\begin{equation}
    t_\mathrm{Th}=\frac{L^2}{4\pi^2}.
\end{equation}
This is a diffusive timescale, parametrically larger than $\tth=\log L/\varepsilon$ for the standard RPM. Given the relation between the Thouless physics and the RP resonances, we can thus expect a parametrically smaller leading RP resonance (and, correspondingly, a parametrically slower relaxation). We explicitly confirm these expectations below.

Next, we consider the corrections to the leading asymptotic behavior that stems from domain walls. A domain wall essentially serves as an ``impenetrable'' wall, across which spin configurations need not be compatible (i.e., magnetization is conserved separately in each region). For example, suppose we have two domain walls that separate the system into a region of size $\ell$ and $L-\ell$.
The two-domain-wall contribution to the SFF is then given by
\begin{equation}
    \begin{pmatrix}
        t\\
        2
    \end{pmatrix}e^{-2\varepsilon t}\sum_{\ell=1}^{L-1}\sum_{N'=0}^N(L-\ell+1)\Tr_{N'} \mathcal{M}_\ell^t\Tr_{N-N'}\mathcal{M}_{L-\ell}^t,
\end{equation}
where $\mathcal{M}_\ell$ is the d-SSEP generator acting on $\ell$-contiguous sites. Note that the precise form of $\mathcal{M}_\ell$ varies depending on whether the two ends of the region on which it acts correspond to even/odd sites, but for our purpose below it need not be specified. 
Since clearly both $\Tr_{N'} \mathcal{M}_\ell^t$ and $\Tr_{N-N'} \mathcal{M}_{L-\ell}^t$ are suppressed by some constant $c$ at late times, over the Thouless time the two-domain-wall contribution is upper-bounded by, at large $L$,
\begin{equation}
\begin{split}
    \begin{pmatrix}
        t\\
        2
    \end{pmatrix}e^{-2\varepsilon t}\sum_{\ell=1}^{L-1}\sum_{N'=0}^N&(L-\ell+1)\Tr_{N'} \mathcal{M}_\ell^t
    \\
    &\times\Tr_{N-N'}\mathcal{M}_{L-\ell}^t\lesssim c^2N\frac{L^2t^2}{4}e^{-2\varepsilon t}.
\end{split}
\end{equation}
This value becomes larger than the leading correction from the no-domain-wall contribution $te^{-4\pi^2t/L^2}$ only when
\begin{equation}
    t\lesssim \frac{\log (c^2NL^2)}{2\varepsilon-4\pi^2/L^2}\ll t_\mathrm{Th},
\end{equation}
thus we confirm that the subleading corrections from domain-wall configurations remain negligible at late times.

\subsubsection{DRPM with a conserved \texorpdfstring{$U(1)$}{U(1)} charge}

Now we turn to the RPM with a $U(1)$ charge, which is subject to bulk dissipation. We choose Kraus operators such that the whole system including quantum channels still retains a $U(1)$ conservation law (i.e., the circuit has a strong U(1) symmetry~\cite{buca2012}: the spin component of the Kraus operators $M_i$ is either $\sigma_0=I_2$ or $\sigma_3=\sigma_z$ with normalization $(2q)^{-1}\Tr(M_iM^\dagger_j)=\eta_i\delta_{ij}$. For instance, if the quantum channel is a depolarizing one, we have
\begin{align}
\label{eq:U(1)jumpops}
    &M_0=\sqrt{1-\frac{p(q^2-1)}{q^2}}I_q\otimes I_2\simeq\sqrt{1-p}I_q\otimes I_2,
    \nonumber\\
    &M_{\mu,i}=\sqrt{\frac{p}{2q^2}}\sigma_\mu\otimes P_i.
\end{align}
with $\mu=0,3$ and $i=0,\dots,q^2-1$.

With this in mind, it is readily seen that the cyclic pairings are again the leading pairings upon Haar averaging, and, in particular, the $s\neq0$ pairings are accompanied by the factor $\kappa(t)=\sum_i\eta_i^t$ whereas $s=0$ comes with weight one. We thus arrive at a simple conclusion that the late-time asymptotics of the DFF $\overline{F(t)}$ is given by the no-domain-wall configurations, which leads to
\begin{align}
\label{eq:DFF_U1}
    \overline{F(t)}&= (1+(t-1)\kappa^L)\Tr_N\mathcal{M}^t+\cdots \n
    &\simeq(1+(t-1)\kappa^L)(1+e^{-4\pi^2t/L^2})+\cdots \n
    &\simeq 1+(t-1)e^{-\gamma L \varepsilon t}+e^{-4\pi^2t/L^2}+\cdots,
\end{align}
whence it follows that
\begin{equation}
    \Delta=\min\{\varepsilon\gamma L,\frac{4\pi^2}{L^2}\}.
\end{equation}
We observe that the third term in Eq.~(\ref{eq:DFF_U1}) is greater than the second one when 
\begin{equation}
    \gamma\gtrsim\frac{4\pi^2}{\varepsilon L^3},
\end{equation}
which is always true in the thermodynamic limit. In the thermodynamic limit, the third term is always dominant at late times and the gap becomes $\Delta=t_\mathrm{Th}^{-1}=4\pi^2/L^2$. Since the domain-wall corrections do not affect this asymptotic behavior from the discussion above, we conclude that $\lim_{\gamma\to0}\lim_{L\to\infty}\Delta=\lim_{L\to\infty}\lim_{\gamma\to0}\Delta=0$ and anomalous relaxation is absent in the DRPM with a $U(1)$ conservation law.

So far we have demonstrated that the presence of a $U(1)$ charge in the DRPM induces a change of the Thouless time. Since anomalous relaxation in the standard DRPM takes place due to the overlap of the Thouless time and the timescale associated with the dissipative peaks, the change of the scaling of the Thouless time in the $U(1)$-charge-conserving DRPM spoils anomalous relaxation. We believe that this microscopic mechanism, i.e., the shift of the Thouless time due to the presence of conserved charges, is also at the root of the absence of anomalous relaxation in generic quantum many-body systems with conservation laws. 

That being said, the behavior of the gap $\Delta$, which should be regarded as the leading RP resonance here, behaves somewhat differently from that evaluated for the kicked Ising model, which is a Hamiltonian system. It was numerically confirmed in Ref.~\cite{mori2024} that its leading RP resonance, which was identified as the {\it projected} Liouvillian gap, decays as $\sim L^{-1}$ even if the energy transport is purely diffusive. We currently have no explanation for this dichotomy, which might stem from a hitherto overlooked difference between the energy and $U(1)$ conservation laws.

We also expect that systems with Poissonian spectral statistics, such as integrable systems and systems in a many-body localized (MBL) phase, do not show anomalous relaxation. To verify this, in the next section, we numerically simulate a system in an MBL phase and observe the absence of anomalous relaxation.

\begin{figure*}[t]
    \centering
    \includegraphics[width=0.8\textwidth]{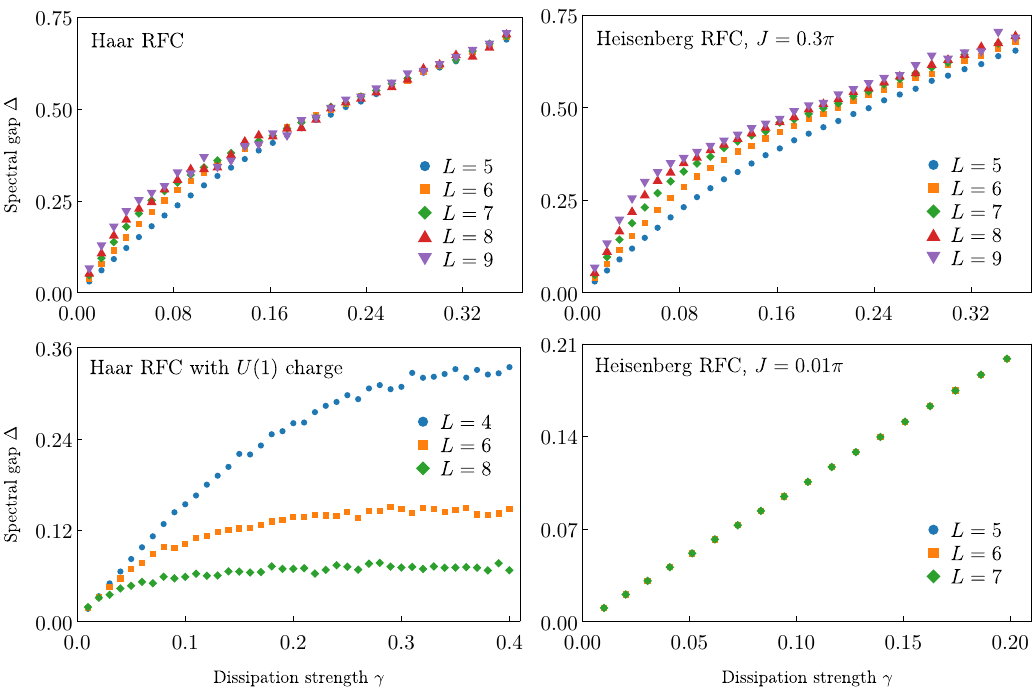}
    \caption{\textbf{Numerical spectral gap of qubit brickwork RFCs as a function of dissipation strength $\gamma$ for different system sizes.} The Haar-random RFC (top left) and Heisenberg RFC at strong coupling (top right) are quantum chaotic and show the finite-size signatures of anomalous relaxation. The RFC with U(1) conservation (bottom left) and the Heisenberg RFC in the localized phase (bottom right) do not show anomalous relaxation, as the gap closes in the thermodynamic limit at small $\gamma$. The spectral gap was computed by a power iteration method~\cite{Yoshimura2024} for the Haar and Heisenberg RFCs and by ED for the U(1) RFC. We have checked that for system sizes amenable to ED, the gaps obtained by power iteration and ED coincide for the Haar RFC until $\gamma\approx 0.25$ (above which there is a small systematic deviation of about $3\%$) and for the Heisenberg RFC with $J=0.01\pi$ at all $\gamma$. For the U(1) circuit, there was a large systematic deviation, hence we resorted only to the ED-computed gap.}
    \label{fig:gap_numerics}
\end{figure*}

\subsection{Numerical results for \texorpdfstring{$q=2$}{q2}}
\label{sec:suppression_numerics}

In this section, we numerically compute the spectral gap for four different types of qubit brickwork RFCs ($q=2$). Besides the generic Haar RFC described in Sec.~\ref{sec:models_brickwork}, we can also endow the dynamics with some additional structure by restricting the gates. 

First, we consider a Haar RFC with a $U(1)$ conservation law. 
To that end, we restrict the circuit to conserve particle number, $Q=\sum_{j=1}^L \sigma^z_j$. We thus replace the $4\times4$ two-site Haar-random unitary by the block-diagonal gate $U_{\uparrow\uparrow}\oplus U_{\uparrow\downarrow,\downarrow\uparrow}\oplus U_{\downarrow\downarrow}$, where $U_{\uparrow\uparrow}$, $U_{\downarrow\downarrow}$, and $U_{\uparrow\downarrow,\downarrow\uparrow}$ are independent Haar-random unitaries of dimension $1$, $1$, and $2$, respectively. The-particle-number-conserving jump operators are chosen as $M_0=\sqrt{1-p}I_2$ and $M_z=\sqrt{p}\sigma^z$.
The unitary evolution matrix and the Kraus operators are block diagonal, with the block of dimension $\binom{L}{m}$ labeled by the particle number $Q=m$, with $m=0,\dots,L$. For the numerical results below, we considered the block at half-filling, $m=L/2$.

Second, to study the fate of RP resonances when passing from an ergodic to a localized phase, we consider the following restricted two-site gate~\cite{Garratt_ManyBody_2021}
\begin{equation}
\label{eq:gate_heisenberg}
    U(J)=(u_1\otimes u_2)\left[
    \cos(J)I_q+i \sin(J)\mathbb{P}_q
    \right](u_3\otimes u_4),
\end{equation}
where $u_{1,2,3,4}$ are independent $q\times q$ Haar-random unitaries, $J$ is the nearest-neighbor interaction strength, and $\mathbb{P}_q$ is the two-site \textsc{Swap} gate. At $q=2$, Eq.~(\ref{eq:gate_heisenberg}) is equivalent to a Trotterized Heisenberg interaction in random fields, and with an abuse of nomenclature, we refer to it as a Heisenberg gate for arbitrary $q$. For small enough $J$, neighboring sites are weakly coupled and the system localizes, while for larger $J$ the interactions dominate and we enter an ergodic phase (for $q=2$, the transition occurs at $J_c\approx0.07\pi$~\cite{Garratt_ManyBody_2021}). We considered the Heisenberg RFC in the ergodic phase for $J=0.3\pi$ and in the localized phase for $J=0.01\pi$.

To compute the spectral gap at finite dissipation strength, we employed exact diagonalization (ED) and a power iteration method described in Ref.~\cite{Yoshimura2024}. In Fig.~\ref{fig:gap_numerics} we show the numerical results for the spectral gap computed for the four different qubit brickwork RFCs and different system sizes as a function of dissipation strength. For the ergodic circuits (the Haar RFC and the Heisenberg RFC at large $J$), we see the finite-size signatures of anomalous relaxation (i.e., a nonvanishing leading RP resonance): although the gap must close at any finite system size in the $\gamma\to0$ limit, the slope at small $\gamma$ is increasing with $L$. At larger $\gamma$, curves for different $L$ collapse to a single one, which persists until $\gamma$ is small enough to probe the finite spacings between individual eigenvalues. Accordingly, as $L$ increases, the single collapsed curve extends until smaller values of $\gamma$, and, in the thermodynamic limit, it extends to $\gamma=0$, consistent with a diverging slope at the origin in this limit. 

On the other hand, in the presence of conservation laws---either the explicit $U(1)$ conservation in the block-diagonal Haar RFC or the emergent conserved quantities in the localized Heisenberg RFC---the gap closes even in the thermodynamic limit (i.e., the leading RP resonance merges with the unit circle). 
For the $U(1)$-conserving circuit, $\Delta\to0$ for all $\gamma$ as $L\to\infty$. A finite-size analysis of the gap shows that the expected $1/L^2$ scaling might be reached for larger values of $L$, but the small sizes available render the results inconclusive in this regard. Nevertheless, the absence of anomalous relaxation as $L\to\infty$ is clear from Fig.~\ref{fig:gap_numerics}. 
For the localized Heisenberg circuit, the gap is $L$-independent and linear in $\gamma$, also closing as $\gamma\to0$. A similar behavior (which is qualitatively distinct from both chaotic circuits and circuits with an explicit conservation law) was observed in the integrable dissipative SYK model~\cite{garcia2023PRD2}.

\section{Discussion}
\label{sec:conclusion}

In this paper, we laid down a general theory that describes how the quantum many-body analogs of RP resonances govern the emergence of irreversible dynamics in isolated quantum many-body systems.
Building upon a recent work by the same authors of the present paper~\cite{Yoshimura2024}, we first explained why it is possible to extract the quantum RP resonances from the long-time behavior of the DFF by identifying it with the sum of autocorrelation functions whose asymptotic decay in time is controlled by the RP resonances on the $O(L)$ timescale, after which a linear ramp emerges~\cite{yoshimura2023operator}. In particular, we clarified the importance of taking the thermodynamic limit before the dissipationless limit in obtaining the leading RP resonance from the DFF; this, in turn, gives rise to a nonzero Liouvillian gap of the system in the limit of vanishing dissipation~\cite{mori2024}, a phenomenon dubbed {\it anomalous relaxation}~\cite{garcia2023PRD2}. At the level of the sum of autocorrelation functions, dissipation suppresses the ramp of the SFF, which starts at shorter timescales for highly nonlocal operators. This observation also allowed us to demonstrate that operator truncation, which is another approach to defining quantum RP resonances~\cite{prosen2002,prosen2004,prosen2007,znidaric2024,znidaric2024b}, is essentially equivalent to the above prescription using weak dissipation and therefore yields the same set of quantum RP resonances. To establish these claims in a concrete setting, we fully worked out the random phase model~\cite{Chan_Spectral_2018} and the dissipative random phase model~\cite{Yoshimura2024} at large $q$.

To further substantiate our claim that the RP resonances dictate the relaxation of quantum many-body dynamics, we looked into the OTOC for the RPM at large $q$ and showed that with PBCs it undergoes two-stage relaxation. In the second stage, the relaxation is fully controlled by the leading RP resonance, as expected. Furthermore, we evaluated the OTOC in the large-$q$ DRPM exactly and demonstrated that it again decays with the leading RP resonance in the dissipationless limit after taking the thermodynamic limit, similarly to the DFF. Finally, we examined several scenarios in which quantum RP resonances are constrained to be on the unit circle, including the case where the system has a $U(1)$ charge, both in solvable models and numerically.

The notion of quantum RP resonances has also been discussed recently for systems whose time-evolution operator is random in time, e.g., random unitary circuits (RUCs). However, the important difference from Floquet systems is that since the dynamics lacks a fixed time-evolution operator, there is no obvious operator for which one can study the spectrum. This problem was avoided in Refs.~\cite{zhang2024,jacoby2024} by focusing on the time-evolution of the ensemble-averaged density $\overline{n(x)}$ of the operators with the right endpoint at position $x$. The density is known to satisfy a recurrence equation that defines the transition matrix in RUCs, and its subleading eigenvalues constitute the quantum RP resonances for this particular observable. This is, therefore, in stark contrast to Floquet systems, where the RP resonances are directly related to the dynamics and thus observable-independent. For this reason, we emphasize that the RP resonances defined in this way for Floquet circuits, which was conjectured to be possible in Refs.~\cite{zhang2024,jacoby2024}, in principle need not agree with those defined for the original time-evolution operator in systems with time translation symmetry.  

Our work offers a unifying viewpoint on quantum many-body RP resonances and thus naturally leads to several new directions. First, it would be very interesting to adopt our formalism to classical many-body systems, in particular classical circuits (e.g., random permutation circuits~\cite{bertini2024quantumclassicaldynamicsrandom}), and evaluate {\it classical} RP resonances. In classical systems, the RP resonances are usually computed by diagonalizing the Fokker-Planck operator and taking the noiseless limit subsequently, much like what we did for quantum systems using weak dissipation. We expect that the operator-based perspective we introduced in this paper provides a useful complementary interpretation of classical RP resonances.

Another important direction that warrants further study is to look into the system with conserved charges, in particular Hamiltonian systems, in more detail. While quantum circuits with a $U(1)$-charge have been routinely used as toy models that mimic Hamiltonian systems in certain aspects, it remains unclear whether the quantum RP resonances behave similarly in both systems. Indeed, in Ref.~\cite{mori2024}, it was discovered that the RP resonance in the kicked Ising model, which conserves the energy, is proportional to $L^{-1}$ despite the naive expectation that it decays diffusively, i.e., $L^{-2}$. A systematic comparison between systems that conserve a $U(1)$ charge and energy is, therefore, highly desired.

\section*{Acknowledgements}
We gratefully acknowledge insightful discussions with Peter Abbamonte, Antonio Garc\'ia-Garc\'ia, Takashi Mori, Toma\v{z} Prosen, Jac Verbaarschot, and Marko \v{Z}nidari\v{c}. T.Y. also thanks J.\ T. Chalker for useful discussions on domain walls in Floquet circuits with a $U(1)$ charge. We further thank Marko \v{Z}nidari\v{c} for comments on the manuscript.
L.S.\ was supported by a Research Fellowship from the Royal Commission for the Exhibition of 1851.

\appendix

\section{Partial dissipative form factor}
\label{sec:PSFF}

In this appendix, we consider a quantity that is closely related to the SFF and has been studied in the context of quantum many-body chaos. The partial spectral form factor (PSFF) is defined as
\begin{equation}
    K_A(t)=q^{-(L-L_A)}\Tr_{\bar{A}}[|\Tr_A W^t|^2],
\end{equation}
where the system is partitioned into the region $A$ and its complement $\bar{A}$. The natural generalization of this in the presence of quantum channels is to promote the unitary time-evolution operator to the global Kraus operators
\begin{equation}
    F_A(t)=q^{-(L-L_A)}\sum_{j_1,\dots,j_t}\Tr_{\bar{A}}[|\Tr_A K_{j_t}\cdots K_{j_1}|^2].
\end{equation}
Now, let us compute its ensemble average in the DRPM. Following the same argument as in the DFF, it can be readily shown that the averaged PDFF $\overline{F_A(t)}$ can be neatly expressed as
\begin{equation}\label{eq:pdff}
    \overline{F_A(t)}=\langle 0|\hat{T}^{L_A+1}|0\rangle=\hat{T}^{L_A+1}_{00}.
\end{equation}
While a closed expression of Eq.~\eqref{eq:pdff} is still not feasible, its late-time behavior can be inferred by the domain-wall expansion as before, yielding
\begin{equation}
    \overline{F_A(t)}\simeq 1+L_A\kappa(t)e^{-2\varepsilon t}.
\end{equation}
The consequence of dissipation is therefore not as stark as that in the DFF, and it merely induces a prefactor $\kappa(t)$. 

\section{Systems with nonlocal interactions}
\label{sec:nonlocal}

In Sec.~\ref{sec:suppression_nonlocal}, we consider the RMF model (a simple Floquet model whose time-evolution operator acting on the whole system itself is Haar-distributed) as an example of systems without local structure. The DFF in the DRMF system is given by Eq.~(\ref{eq:DFF_DRMF}). To evaluate it, we expand the following building blocks over Fock space:
\begin{widetext}
\begin{equation}
\mathrm{Tr}\prod_{\tau=1,\dots,t}K_{j_\tau}W=\sum_{\substack{a_0,\dots,a_{t-1}\\b_0,\dots,b_{t-1}}}[K_{j_t}]_{a_{t-1}b_{t-1}}W_{b_{t-1}a_{t-1}}[K_{j_{t-1}}]_{a_{t-1}b_{t-2}}W_{b_{t-2}a_{t-2}}\cdots [K_{j_1}]_{a_1b_0}W_{b_0a_0},
\end{equation}
\begin{equation}
\mathrm{Tr}\prod_{\tau=1,\dots,t}K_{j_\tau}^*W^*=\sum_{\substack{a^*_0,\dots,a^*_{t-1}\\b^*_0,\dots,b^*_{t-1}}}[K_{j_t}^*]_{a^*_0b^*_{t-1}}W^*_{b^*_{t-1}a^*_{t-1}}[K_{j_{t-1}}^*]_{a^*_0b^*_{t-2}}W^*_{b^*_{t-2}a^*_{t-2}}\cdots [K_{j_1}^*]_{a^*_1b^*_0}W^*_{b^*_0a^*_0}.
\end{equation}
As in the DRPM, upon Haar averaging, only cyclic pairings of indices contribute at large $q$, which we parametrize as $(a(\tau),b(\tau))=(a^*(\tau+s),b^*(\tau+s))$ for $s=0,\dots,t-1$. As a result, for example at $t=2$, we obtain the large-$q$ DFF
\begin{align}
\overline{F(2)}&=q^{-2L}\sum_{j_1,j_2}\sum_{a_0,a_1,b_0,b_1}\left(
[K_{j_2}]_{a_0b_1}[K_{j_2}^*]_{a_0b_1}[K_{j_1}]_{a_1b_0}[K_{j_1}^*]_{a_1b_0}+[K_{j_2}]_{a_0b_1}[K_{j_2}^*]_{a_1b_0}[K_{j_1}]_{a_1b_0}[K_{j_1}^*]_{a_0b_1}\right) \n
&=q^{-2L}\sum_{j_1,j_2}\left(\mathrm{Tr}(K_{j_2}K_{j_2}^\dagger)\mathrm{Tr}(K_{j_1}K_{j_1}^\dagger)+\mathrm{Tr}(K_{j_2}K_{j_1}^\dagger)\mathrm{Tr}(K_{j_1}K_{j_2}^\dagger)\right).
\end{align}
Noting that $q^{-L}\sum_{j_\tau}\mathrm{Tr}(K_{j_\tau}K_{j_\tau}^\dagger)=1$ and $q^{-2L}\sum_{j_1,j_2}|\mathrm{Tr}(K_{j_1}K_{j_2}^\dagger)|^2=(\sum_{i}\eta_i^2)^L$, the DFF $\overline{F(2)}$ thus has a simple expression $\overline{F(2)}=1+(\sum_{i}\eta_i^2)^L$. For generic $t$ it can be straightforwardly generalized and, for large $q$, we recover Eq.~(\ref{eq:subextensive_DFF}).
\end{widetext}

\bibliography{main_v7.bbl}

\begin{thebibliography}{87}%
\makeatletter
\providecommand \@ifxundefined [1]{%
 \@ifx{#1\undefined}
}%
\providecommand \@ifnum [1]{%
 \ifnum #1\expandafter \@firstoftwo
 \else \expandafter \@secondoftwo
 \fi
}%
\providecommand \@ifx [1]{%
 \ifx #1\expandafter \@firstoftwo
 \else \expandafter \@secondoftwo
 \fi
}%
\providecommand \natexlab [1]{#1}%
\providecommand \enquote  [1]{``#1''}%
\providecommand \bibnamefont  [1]{#1}%
\providecommand \bibfnamefont [1]{#1}%
\providecommand \citenamefont [1]{#1}%
\providecommand \href@noop [0]{\@secondoftwo}%
\providecommand \href [0]{\begingroup \@sanitize@url \@href}%
\providecommand \@href[1]{\@@startlink{#1}\@@href}%
\providecommand \@@href[1]{\endgroup#1\@@endlink}%
\providecommand \@sanitize@url [0]{\catcode `\\12\catcode `\$12\catcode `\&12\catcode `\#12\catcode `\^12\catcode `\_12\catcode `\%12\relax}%
\providecommand \@@startlink[1]{}%
\providecommand \@@endlink[0]{}%
\providecommand \url  [0]{\begingroup\@sanitize@url \@url }%
\providecommand \@url [1]{\endgroup\@href {#1}{\urlprefix }}%
\providecommand \urlprefix  [0]{URL }%
\providecommand \Eprint [0]{\href }%
\providecommand \doibase [0]{https://doi.org/}%
\providecommand \selectlanguage [0]{\@gobble}%
\providecommand \bibinfo  [0]{\@secondoftwo}%
\providecommand \bibfield  [0]{\@secondoftwo}%
\providecommand \translation [1]{[#1]}%
\providecommand \BibitemOpen [0]{}%
\providecommand \bibitemStop [0]{}%
\providecommand \bibitemNoStop [0]{.\EOS\space}%
\providecommand \EOS [0]{\spacefactor3000\relax}%
\providecommand \BibitemShut  [1]{\csname bibitem#1\endcsname}%
\let\auto@bib@innerbib\@empty
\bibitem [{\citenamefont {Brush}(2003)}]{brush2003}%
  \BibitemOpen
  \bibfield  {author} {\bibinfo {author} {\bibfnamefont {S.~G.}\ \bibnamefont {Brush}},\ }\href@noop {} {\emph {\bibinfo {title} {{The Kinetic Theory of Gases: An Anthology of Classic Papers with Historical Commentary}}}}\ (\bibinfo  {publisher} {Imperial College Press},\ \bibinfo {address} {London},\ \bibinfo {year} {2003})\BibitemShut {NoStop}%
\bibitem [{\citenamefont {Gaspard}(1998)}]{Gaspard_1998}%
  \BibitemOpen
  \bibfield  {author} {\bibinfo {author} {\bibfnamefont {P.}~\bibnamefont {Gaspard}},\ }\href@noop {} {\emph {\bibinfo {title} {Chaos, Scattering and Statistical Mechanics}}}\ (\bibinfo  {publisher} {Cambridge University Press},\ \bibinfo {address} {Cambridge},\ \bibinfo {year} {1998})\BibitemShut {NoStop}%
\bibitem [{\citenamefont {Gaspard}(2005)}]{gaspardReview}%
  \BibitemOpen
  \bibfield  {author} {\bibinfo {author} {\bibfnamefont {P.}~\bibnamefont {Gaspard}},\ }\bibfield  {title} {\bibinfo {title} {Dynamical systems theory of irreversibility},\ }in\ \href@noop {} {\emph {\bibinfo {booktitle} {Chaotic Dynamics and Transport in Classical and Quantum Systems}}},\ \bibinfo {editor} {edited by\ \bibinfo {editor} {\bibfnamefont {P.}~\bibnamefont {Collet}}, \bibinfo {editor} {\bibfnamefont {M.}~\bibnamefont {Courbage}}, \bibinfo {editor} {\bibfnamefont {S.}~\bibnamefont {M{\'e}tens}}, \bibinfo {editor} {\bibfnamefont {A.}~\bibnamefont {Neishtadt}},\ and\ \bibinfo {editor} {\bibfnamefont {G.}~\bibnamefont {Zaslavsky}}}\ (\bibinfo  {publisher} {Springer Netherlands},\ \bibinfo {address} {Dordrecht},\ \bibinfo {year} {2005})\ pp.\ \bibinfo {pages} {107--157}\BibitemShut {NoStop}%
\bibitem [{\citenamefont {Ruelle}(1986{\natexlab{a}})}]{ruelle1985}%
  \BibitemOpen
  \bibfield  {author} {\bibinfo {author} {\bibfnamefont {D.}~\bibnamefont {Ruelle}},\ }\bibfield  {title} {\bibinfo {title} {{Resonances of chaotic dynamical systems}},\ }\href {https://doi.org/10.1103/PhysRevLett.56.405} {\bibfield  {journal} {\bibinfo  {journal} {Phys. Rev. Lett.}\ }\textbf {\bibinfo {volume} {56}},\ \bibinfo {pages} {405} (\bibinfo {year} {1986}{\natexlab{a}})}\BibitemShut {NoStop}%
\bibitem [{\citenamefont {Ruelle}(1986{\natexlab{b}})}]{ruelle1986}%
  \BibitemOpen
  \bibfield  {author} {\bibinfo {author} {\bibfnamefont {D.}~\bibnamefont {Ruelle}},\ }\bibfield  {title} {\bibinfo {title} {{Locating resonances for Axiom A dynamical systems}},\ }\href {https://doi.org/https://doi.org/10.1007/BF01011300} {\bibfield  {journal} {\bibinfo  {journal} {J. Stat. Phys.}\ }\textbf {\bibinfo {volume} {44}},\ \bibinfo {pages} {281} (\bibinfo {year} {1986}{\natexlab{b}})}\BibitemShut {NoStop}%
\bibitem [{\citenamefont {Pollicott}(1985)}]{pollicott1985}%
  \BibitemOpen
  \bibfield  {author} {\bibinfo {author} {\bibfnamefont {M.}~\bibnamefont {Pollicott}},\ }\bibfield  {title} {\bibinfo {title} {{On the rate of mixing of Axiom A flows}},\ }\href {https://doi.org/https://doi.org/10.1007/BF01388579} {\bibfield  {journal} {\bibinfo  {journal} {Invent. Math.}\ }\textbf {\bibinfo {volume} {81}},\ \bibinfo {pages} {413} (\bibinfo {year} {1985})}\BibitemShut {NoStop}%
\bibitem [{\citenamefont {Pollicott}(1986)}]{pollicott1986}%
  \BibitemOpen
  \bibfield  {author} {\bibinfo {author} {\bibfnamefont {M.}~\bibnamefont {Pollicott}},\ }\bibfield  {title} {\bibinfo {title} {{Meromorphic extensions of generalised zeta functions}},\ }\href {https://doi.org/https://doi.org/10.1007/BF01388795} {\bibfield  {journal} {\bibinfo  {journal} {Invent. Math.}\ }\textbf {\bibinfo {volume} {85}},\ \bibinfo {pages} {147} (\bibinfo {year} {1986})}\BibitemShut {NoStop}%
\bibitem [{\citenamefont {Manderfeld}\ \emph {et~al.}(2001)\citenamefont {Manderfeld}, \citenamefont {Weber},\ and\ \citenamefont {Haake}}]{manderfeld2001JPA}%
  \BibitemOpen
  \bibfield  {author} {\bibinfo {author} {\bibfnamefont {C.}~\bibnamefont {Manderfeld}}, \bibinfo {author} {\bibfnamefont {J.}~\bibnamefont {Weber}},\ and\ \bibinfo {author} {\bibfnamefont {F.}~\bibnamefont {Haake}},\ }\bibfield  {title} {\bibinfo {title} {{Classical versus quantum time evolution of (quasi-) probability densities at limited phase-space resolution}},\ }\href {https://doi.org/10.1088/0305-4470/34/46/312} {\bibfield  {journal} {\bibinfo  {journal} {J. Phys. A}\ }\textbf {\bibinfo {volume} {34}},\ \bibinfo {pages} {9893} (\bibinfo {year} {2001})}\BibitemShut {NoStop}%
\bibitem [{\citenamefont {Nonnenmacher}(2003)}]{nonnenmacher2003}%
  \BibitemOpen
  \bibfield  {author} {\bibinfo {author} {\bibfnamefont {S.}~\bibnamefont {Nonnenmacher}},\ }\bibfield  {title} {\bibinfo {title} {{Spectral properties of noisy classical and quantum propagators}},\ }\href {https://doi.org/10.1088/0951-7715/16/5/309} {\bibfield  {journal} {\bibinfo  {journal} {Nonlinearity}\ }\textbf {\bibinfo {volume} {16}},\ \bibinfo {pages} {1685} (\bibinfo {year} {2003})}\BibitemShut {NoStop}%
\bibitem [{\citenamefont {Garc\'{\i}a-Mata}\ \emph {et~al.}(2003)\citenamefont {Garc\'{\i}a-Mata}, \citenamefont {Saraceno},\ and\ \citenamefont {Spina}}]{garcia-mata2003PRL}%
  \BibitemOpen
  \bibfield  {author} {\bibinfo {author} {\bibfnamefont {I.}~\bibnamefont {Garc\'{\i}a-Mata}}, \bibinfo {author} {\bibfnamefont {M.}~\bibnamefont {Saraceno}},\ and\ \bibinfo {author} {\bibfnamefont {M.~E.}\ \bibnamefont {Spina}},\ }\bibfield  {title} {\bibinfo {title} {{Classical Decays in Decoherent Quantum Maps}},\ }\href {https://doi.org/10.1103/PhysRevLett.91.064101} {\bibfield  {journal} {\bibinfo  {journal} {Phys. Rev. Lett.}\ }\textbf {\bibinfo {volume} {91}},\ \bibinfo {pages} {064101} (\bibinfo {year} {2003})}\BibitemShut {NoStop}%
\bibitem [{\citenamefont {Garc\'{\i}a-Mata}\ and\ \citenamefont {Saraceno}(2004)}]{garcia-mata2004PRE}%
  \BibitemOpen
  \bibfield  {author} {\bibinfo {author} {\bibfnamefont {I.}~\bibnamefont {Garc\'{\i}a-Mata}}\ and\ \bibinfo {author} {\bibfnamefont {M.}~\bibnamefont {Saraceno}},\ }\bibfield  {title} {\bibinfo {title} {{Spectral properties and classical decays in quantum open systems}},\ }\href {https://doi.org/10.1103/PhysRevE.69.056211} {\bibfield  {journal} {\bibinfo  {journal} {Phys. Rev. E}\ }\textbf {\bibinfo {volume} {69}},\ \bibinfo {pages} {056211} (\bibinfo {year} {2004})}\BibitemShut {NoStop}%
\bibitem [{\citenamefont {Garc\'{\i}a-Mata}\ \emph {et~al.}(2018)\citenamefont {Garc\'{\i}a-Mata}, \citenamefont {Saraceno}, \citenamefont {Jalabert}, \citenamefont {Roncaglia},\ and\ \citenamefont {Wisniacki}}]{garcia-mata2018}%
  \BibitemOpen
  \bibfield  {author} {\bibinfo {author} {\bibfnamefont {I.}~\bibnamefont {Garc\'{\i}a-Mata}}, \bibinfo {author} {\bibfnamefont {M.}~\bibnamefont {Saraceno}}, \bibinfo {author} {\bibfnamefont {R.~A.}\ \bibnamefont {Jalabert}}, \bibinfo {author} {\bibfnamefont {A.~J.}\ \bibnamefont {Roncaglia}},\ and\ \bibinfo {author} {\bibfnamefont {D.~A.}\ \bibnamefont {Wisniacki}},\ }\bibfield  {title} {\bibinfo {title} {{Chaos Signatures in the Short and Long Time Behavior of the Out-of-Time Ordered Correlator}},\ }\href {https://doi.org/10.1103/PhysRevLett.121.210601} {\bibfield  {journal} {\bibinfo  {journal} {Phys. Rev. Lett.}\ }\textbf {\bibinfo {volume} {121}},\ \bibinfo {pages} {210601} (\bibinfo {year} {2018})}\BibitemShut {NoStop}%
\bibitem [{\citenamefont {Prosen}(2002)}]{prosen2002}%
  \BibitemOpen
  \bibfield  {author} {\bibinfo {author} {\bibfnamefont {T.}~\bibnamefont {Prosen}},\ }\bibfield  {title} {\bibinfo {title} {{Ruelle resonances in quantum many-body dynamics}},\ }\href {https://doi.org/10.1088/0305-4470/35/48/102} {\bibfield  {journal} {\bibinfo  {journal} {J. Phys. A}\ }\textbf {\bibinfo {volume} {35}},\ \bibinfo {pages} {L737} (\bibinfo {year} {2002})}\BibitemShut {NoStop}%
\bibitem [{\citenamefont {Prosen}(2004)}]{prosen2004}%
  \BibitemOpen
  \bibfield  {author} {\bibinfo {author} {\bibfnamefont {T.}~\bibnamefont {Prosen}},\ }\bibfield  {title} {\bibinfo {title} {{Ruelle resonances in kicked quantum spin chain}},\ }\href {https://doi.org/https://doi.org/10.1016/j.physd.2003.09.017} {\bibfield  {journal} {\bibinfo  {journal} {Physica D}\ }\textbf {\bibinfo {volume} {187}},\ \bibinfo {pages} {244} (\bibinfo {year} {2004})}\BibitemShut {NoStop}%
\bibitem [{\citenamefont {Prosen}(2007)}]{prosen2007}%
  \BibitemOpen
  \bibfield  {author} {\bibinfo {author} {\bibfnamefont {T.}~\bibnamefont {Prosen}},\ }\bibfield  {title} {\bibinfo {title} {{Chaos and complexity of quantum motion}},\ }\href {https://doi.org/10.1088/1751-8113/40/28/S02} {\bibfield  {journal} {\bibinfo  {journal} {J. Phys. A}\ }\textbf {\bibinfo {volume} {40}},\ \bibinfo {pages} {7881} (\bibinfo {year} {2007})}\BibitemShut {NoStop}%
\bibitem [{\citenamefont {Mori}(2024)}]{mori2024}%
  \BibitemOpen
  \bibfield  {author} {\bibinfo {author} {\bibfnamefont {T.}~\bibnamefont {Mori}},\ }\bibfield  {title} {\bibinfo {title} {{Liouvillian-gap analysis of open quantum many-body systems in the weak dissipation limit}},\ }\href {https://doi.org/10.1103/PhysRevB.109.064311} {\bibfield  {journal} {\bibinfo  {journal} {Phys. Rev. B}\ }\textbf {\bibinfo {volume} {109}},\ \bibinfo {pages} {064311} (\bibinfo {year} {2024})}\BibitemShut {NoStop}%
\bibitem [{\citenamefont {Yoshimura}\ and\ \citenamefont {Sá}(2024)}]{Yoshimura2024}%
  \BibitemOpen
  \bibfield  {author} {\bibinfo {author} {\bibfnamefont {T.}~\bibnamefont {Yoshimura}}\ and\ \bibinfo {author} {\bibfnamefont {L.}~\bibnamefont {Sá}},\ }\bibfield  {title} {\bibinfo {title} {{Robustness of quantum chaos and anomalous relaxation in open quantum circuits}},\ }\href {http://dx.doi.org/10.1038/s41467-024-54164-7} {\bibfield  {journal} {\bibinfo  {journal} {Nat. Commun.}\ }\textbf {\bibinfo {volume} {15}},\ \bibinfo {pages} {9808} (\bibinfo {year} {2024})}\BibitemShut {NoStop}%
\bibitem [{\citenamefont {\v{Z}nidari\v{c}}(2024)}]{znidaric2024}%
  \BibitemOpen
  \bibfield  {author} {\bibinfo {author} {\bibfnamefont {M.}~\bibnamefont {\v{Z}nidari\v{c}}},\ }\bibfield  {title} {\bibinfo {title} {{Momentum-dependent quantum Ruelle-Pollicott resonances in translationally invariant many-body systems}},\ }\href {https://doi.org/10.1103/PhysRevE.110.054204} {\bibfield  {journal} {\bibinfo  {journal} {Phys. Rev. E}\ }\textbf {\bibinfo {volume} {110}},\ \bibinfo {pages} {054204} (\bibinfo {year} {2024})}\BibitemShut {NoStop}%
\bibitem [{\citenamefont {\v{Z}nidari\v{c}}\ \emph {et~al.}(2024)\citenamefont {\v{Z}nidari\v{c}}, \citenamefont {Duh},\ and\ \citenamefont {Zadnik}}]{znidaric2024b}%
  \BibitemOpen
  \bibfield  {author} {\bibinfo {author} {\bibfnamefont {M.}~\bibnamefont {\v{Z}nidari\v{c}}}, \bibinfo {author} {\bibfnamefont {U.}~\bibnamefont {Duh}},\ and\ \bibinfo {author} {\bibfnamefont {L.}~\bibnamefont {Zadnik}},\ }\bibfield  {title} {\bibinfo {title} {{Integrability is generic in homogeneous U(1)-invariant nearest-neighbor qubit circuits}},\ }\href {https://arxiv.org/abs/2410.06760} {\bibfield  {journal} {\bibinfo  {journal} {arXiv:2410.06760}\ } (\bibinfo {year} {2024})}\BibitemShut {NoStop}%
\bibitem [{\citenamefont {Jacoby}\ \emph {et~al.}(2025)\citenamefont {Jacoby}, \citenamefont {Huse},\ and\ \citenamefont {Gopalakrishnan}}]{jacoby2024}%
  \BibitemOpen
  \bibfield  {author} {\bibinfo {author} {\bibfnamefont {J.~A.}\ \bibnamefont {Jacoby}}, \bibinfo {author} {\bibfnamefont {D.~A.}\ \bibnamefont {Huse}},\ and\ \bibinfo {author} {\bibfnamefont {S.}~\bibnamefont {Gopalakrishnan}},\ }\bibfield  {title} {\bibinfo {title} {{Spectral gaps of local quantum channels in the weak-dissipation limit}},\ }\href {https://doi.org/10.1103/PhysRevB.111.104303} {\bibfield  {journal} {\bibinfo  {journal} {Phys. Rev. B}\ }\textbf {\bibinfo {volume} {111}},\ \bibinfo {pages} {104303} (\bibinfo {year} {2025})}\BibitemShut {NoStop}%
\bibitem [{\citenamefont {Zhang}\ \emph {et~al.}(2024)\citenamefont {Zhang}, \citenamefont {Nie},\ and\ \citenamefont {von Keyserlingk}}]{zhang2024}%
  \BibitemOpen
  \bibfield  {author} {\bibinfo {author} {\bibfnamefont {C.}~\bibnamefont {Zhang}}, \bibinfo {author} {\bibfnamefont {L.}~\bibnamefont {Nie}},\ and\ \bibinfo {author} {\bibfnamefont {C.}~\bibnamefont {von Keyserlingk}},\ }\bibfield  {title} {\bibinfo {title} {{Thermalization rates and quantum Ruelle-Pollicott resonances: insights from operator hydrodynamics}},\ }\href {https://arxiv.org/abs/2409.17251} {\bibfield  {journal} {\bibinfo  {journal} {arXiv:2409.17251}\ } (\bibinfo {year} {2024})}\BibitemShut {NoStop}%
\bibitem [{\citenamefont {Bohigas}\ \emph {et~al.}(1984)\citenamefont {Bohigas}, \citenamefont {Giannoni},\ and\ \citenamefont {Schmit}}]{Bohigas_Characterization_1984}%
  \BibitemOpen
  \bibfield  {author} {\bibinfo {author} {\bibfnamefont {O.}~\bibnamefont {Bohigas}}, \bibinfo {author} {\bibfnamefont {M.~J.}\ \bibnamefont {Giannoni}},\ and\ \bibinfo {author} {\bibfnamefont {C.}~\bibnamefont {Schmit}},\ }\bibfield  {title} {\bibinfo {title} {{Characterization of Chaotic Quantum Spectra and Universality of Level Fluctuation Laws}},\ }\href {https://doi.org/10.1103/PhysRevLett.52.1} {\bibfield  {journal} {\bibinfo  {journal} {Phys. Rev. Lett.}\ }\textbf {\bibinfo {volume} {52}},\ \bibinfo {pages} {1} (\bibinfo {year} {1984})}\BibitemShut {NoStop}%
\bibitem [{\citenamefont {Haake}(2013)}]{haake2013}%
  \BibitemOpen
  \bibfield  {author} {\bibinfo {author} {\bibfnamefont {F.}~\bibnamefont {Haake}},\ }\href@noop {} {\emph {\bibinfo {title} {{Quantum Signatures of Chaos}}}}\ (\bibinfo  {publisher} {Springer},\ \bibinfo {address} {Cham},\ \bibinfo {year} {2013})\BibitemShut {NoStop}%
\bibitem [{\citenamefont {Rakovszky}\ \emph {et~al.}(2022)\citenamefont {Rakovszky}, \citenamefont {von Keyserlingk},\ and\ \citenamefont {Pollmann}}]{Rakovszky_Dissipation_2022}%
  \BibitemOpen
  \bibfield  {author} {\bibinfo {author} {\bibfnamefont {T.}~\bibnamefont {Rakovszky}}, \bibinfo {author} {\bibfnamefont {C.~W.}\ \bibnamefont {von Keyserlingk}},\ and\ \bibinfo {author} {\bibfnamefont {F.}~\bibnamefont {Pollmann}},\ }\bibfield  {title} {\bibinfo {title} {{Dissipation-assisted operator evolution method for capturing hydrodynamic transport}},\ }\href {https://doi.org/10.1103/PhysRevB.105.075131} {\bibfield  {journal} {\bibinfo  {journal} {Phys. Rev. B}\ }\textbf {\bibinfo {volume} {105}},\ \bibinfo {pages} {075131} (\bibinfo {year} {2022})}\BibitemShut {NoStop}%
\bibitem [{\citenamefont {von Keyserlingk}\ \emph {et~al.}(2022)\citenamefont {von Keyserlingk}, \citenamefont {Pollmann},\ and\ \citenamefont {Rakovszky}}]{Keyserlingk_Operator_2022}%
  \BibitemOpen
  \bibfield  {author} {\bibinfo {author} {\bibfnamefont {C.}~\bibnamefont {von Keyserlingk}}, \bibinfo {author} {\bibfnamefont {F.}~\bibnamefont {Pollmann}},\ and\ \bibinfo {author} {\bibfnamefont {T.}~\bibnamefont {Rakovszky}},\ }\bibfield  {title} {\bibinfo {title} {{Operator backflow and the classical simulation of quantum transport}},\ }\href {https://doi.org/10.1103/PhysRevB.105.245101} {\bibfield  {journal} {\bibinfo  {journal} {Phys. Rev. B}\ }\textbf {\bibinfo {volume} {105}},\ \bibinfo {pages} {245101} (\bibinfo {year} {2022})}\BibitemShut {NoStop}%
\bibitem [{\citenamefont {S\'a}\ \emph {et~al.}(2022)\citenamefont {S\'a}, \citenamefont {Ribeiro},\ and\ \citenamefont {Prosen}}]{sa2022PRR}%
  \BibitemOpen
  \bibfield  {author} {\bibinfo {author} {\bibfnamefont {L.}~\bibnamefont {S\'a}}, \bibinfo {author} {\bibfnamefont {P.}~\bibnamefont {Ribeiro}},\ and\ \bibinfo {author} {\bibfnamefont {T.}~\bibnamefont {Prosen}},\ }\bibfield  {title} {\bibinfo {title} {{Lindbladian dissipation of strongly-correlated quantum matter}},\ }\href {https://doi.org/10.1103/PhysRevResearch.4.L022068} {\bibfield  {journal} {\bibinfo  {journal} {Phys. Rev. Res.}\ }\textbf {\bibinfo {volume} {4}},\ \bibinfo {pages} {L022068} (\bibinfo {year} {2022})}\BibitemShut {NoStop}%
\bibitem [{\citenamefont {Garc\'{\i}a-Garc\'{\i}a}\ \emph {et~al.}(2023{\natexlab{a}})\citenamefont {Garc\'{\i}a-Garc\'{\i}a}, \citenamefont {S\'a},\ and\ \citenamefont {Verbaarschot}}]{garcia2023PRD}%
  \BibitemOpen
  \bibfield  {author} {\bibinfo {author} {\bibfnamefont {A.~M.}\ \bibnamefont {Garc\'{\i}a-Garc\'{\i}a}}, \bibinfo {author} {\bibfnamefont {L.}~\bibnamefont {S\'a}},\ and\ \bibinfo {author} {\bibfnamefont {J.~J.~M.}\ \bibnamefont {Verbaarschot}},\ }\bibfield  {title} {\bibinfo {title} {{Universality and its limits in non-Hermitian many-body quantum chaos using the Sachdev-Ye-Kitaev model}},\ }\href {https://doi.org/10.1103/PhysRevD.107.066007} {\bibfield  {journal} {\bibinfo  {journal} {Phys. Rev. D}\ }\textbf {\bibinfo {volume} {107}},\ \bibinfo {pages} {066007} (\bibinfo {year} {2023}{\natexlab{a}})}\BibitemShut {NoStop}%
\bibitem [{\citenamefont {Shackleton}\ and\ \citenamefont {Scheurer}(2024)}]{scheurer2023ARXIV}%
  \BibitemOpen
  \bibfield  {author} {\bibinfo {author} {\bibfnamefont {H.}~\bibnamefont {Shackleton}}\ and\ \bibinfo {author} {\bibfnamefont {M.~S.}\ \bibnamefont {Scheurer}},\ }\bibfield  {title} {\bibinfo {title} {{Exactly solvable dissipative spin liquid}},\ }\href {https://doi.org/10.1103/PhysRevB.109.085115} {\bibfield  {journal} {\bibinfo  {journal} {Phys. Rev. B}\ }\textbf {\bibinfo {volume} {109}},\ \bibinfo {pages} {085115} (\bibinfo {year} {2024})}\BibitemShut {NoStop}%
\bibitem [{\citenamefont {Garc\'{\i}a-Garc\'{\i}a}\ \emph {et~al.}(2023{\natexlab{b}})\citenamefont {Garc\'{\i}a-Garc\'{\i}a}, \citenamefont {S\'a}, \citenamefont {Verbaarschot},\ and\ \citenamefont {Zheng}}]{garcia2023PRD2}%
  \BibitemOpen
  \bibfield  {author} {\bibinfo {author} {\bibfnamefont {A.~M.}\ \bibnamefont {Garc\'{\i}a-Garc\'{\i}a}}, \bibinfo {author} {\bibfnamefont {L.}~\bibnamefont {S\'a}}, \bibinfo {author} {\bibfnamefont {J.~J.~M.}\ \bibnamefont {Verbaarschot}},\ and\ \bibinfo {author} {\bibfnamefont {J.~P.}\ \bibnamefont {Zheng}},\ }\bibfield  {title} {\bibinfo {title} {{Keldysh wormholes and anomalous relaxation in the dissipative Sachdev-Ye-Kitaev model}},\ }\href {https://doi.org/10.1103/PhysRevD.107.106006} {\bibfield  {journal} {\bibinfo  {journal} {Phys. Rev. D}\ }\textbf {\bibinfo {volume} {107}},\ \bibinfo {pages} {106006} (\bibinfo {year} {2023}{\natexlab{b}})}\BibitemShut {NoStop}%
\bibitem [{\citenamefont {Dodelson}(2024)}]{dodelson2024}%
  \BibitemOpen
  \bibfield  {author} {\bibinfo {author} {\bibfnamefont {M.}~\bibnamefont {Dodelson}},\ }\bibfield  {title} {\bibinfo {title} {{Ringdown in the SYK model}},\ }\href {https://arxiv.org/abs/2408.05790} {\bibfield  {journal} {\bibinfo  {journal} {arXiv:2408.05790}\ } (\bibinfo {year} {2024})}\BibitemShut {NoStop}%
\bibitem [{\citenamefont {Jak{\v{s}}i{\'c}}\ and\ \citenamefont {Pillet}(2002)}]{jaksic2002}%
  \BibitemOpen
  \bibfield  {author} {\bibinfo {author} {\bibfnamefont {V.}~\bibnamefont {Jak{\v{s}}i{\'c}}}\ and\ \bibinfo {author} {\bibfnamefont {C.-A.}\ \bibnamefont {Pillet}},\ }\bibfield  {title} {\bibinfo {title} {{Mathematical theory of non-equilibrium quantum statistical mechanics}},\ }\href {https://doi.org/https://doi.org/10.1023/A:1019818909696} {\bibfield  {journal} {\bibinfo  {journal} {J. Stat. Phys.}\ }\textbf {\bibinfo {volume} {108}},\ \bibinfo {pages} {787} (\bibinfo {year} {2002})}\BibitemShut {NoStop}%
\bibitem [{\citenamefont {Chan}\ \emph {et~al.}(2018)\citenamefont {Chan}, \citenamefont {De~Luca},\ and\ \citenamefont {Chalker}}]{Chan_Spectral_2018}%
  \BibitemOpen
  \bibfield  {author} {\bibinfo {author} {\bibfnamefont {A.}~\bibnamefont {Chan}}, \bibinfo {author} {\bibfnamefont {A.}~\bibnamefont {De~Luca}},\ and\ \bibinfo {author} {\bibfnamefont {J.~T.}\ \bibnamefont {Chalker}},\ }\bibfield  {title} {\bibinfo {title} {{Spectral Statistics in Spatially Extended Chaotic Quantum Many-Body Systems}},\ }\href {https://doi.org/10.1103/PhysRevLett.121.060601} {\bibfield  {journal} {\bibinfo  {journal} {Phys. Rev. Lett.}\ }\textbf {\bibinfo {volume} {121}},\ \bibinfo {pages} {060601} (\bibinfo {year} {2018})}\BibitemShut {NoStop}%
\bibitem [{\citenamefont {Larkin}\ and\ \citenamefont {Ovchinnikov}(1969)}]{larkin1969}%
  \BibitemOpen
  \bibfield  {author} {\bibinfo {author} {\bibfnamefont {A.~I.}\ \bibnamefont {Larkin}}\ and\ \bibinfo {author} {\bibfnamefont {Y.~N.}\ \bibnamefont {Ovchinnikov}},\ }\bibfield  {title} {\bibinfo {title} {{Quasiclassical method in the theory of superconductivity}},\ }\href {http://jetp.ras.ru/cgi-bin/e/index/e/28/6/p1200?a=list} {\bibfield  {journal} {\bibinfo  {journal} {Sov. Phys. JETP}\ }\textbf {\bibinfo {volume} {28}},\ \bibinfo {pages} {1200} (\bibinfo {year} {1969})}\BibitemShut {NoStop}%
\bibitem [{\citenamefont {Berman}\ and\ \citenamefont {Zaslavsky}(1978)}]{berman1978}%
  \BibitemOpen
  \bibfield  {author} {\bibinfo {author} {\bibfnamefont {G.~P.}\ \bibnamefont {Berman}}\ and\ \bibinfo {author} {\bibfnamefont {G.~M.}\ \bibnamefont {Zaslavsky}},\ }\bibfield  {title} {\bibinfo {title} {{Condition of stochasticity in quantum nonlinear systems}},\ }\href {https://doi.org/http://dx.doi.org/10.1016/0378-4371(78)90190-5} {\bibfield  {journal} {\bibinfo  {journal} {Physica A}\ }\textbf {\bibinfo {volume} {91}},\ \bibinfo {pages} {450 } (\bibinfo {year} {1978})}\BibitemShut {NoStop}%
\bibitem [{\citenamefont {Jalabert}\ \emph {et~al.}(2018)\citenamefont {Jalabert}, \citenamefont {Garc\'{\i}a-Mata},\ and\ \citenamefont {Wisniacki}}]{jalabert2018}%
  \BibitemOpen
  \bibfield  {author} {\bibinfo {author} {\bibfnamefont {R.~A.}\ \bibnamefont {Jalabert}}, \bibinfo {author} {\bibfnamefont {I.}~\bibnamefont {Garc\'{\i}a-Mata}},\ and\ \bibinfo {author} {\bibfnamefont {D.~A.}\ \bibnamefont {Wisniacki}},\ }\bibfield  {title} {\bibinfo {title} {{Semiclassical theory of out-of-time-order correlators for low-dimensional classically chaotic systems}},\ }\href {https://doi.org/10.1103/PhysRevE.98.062218} {\bibfield  {journal} {\bibinfo  {journal} {Phys. Rev. E}\ }\textbf {\bibinfo {volume} {98}},\ \bibinfo {pages} {062218} (\bibinfo {year} {2018})}\BibitemShut {NoStop}%
\bibitem [{\citenamefont {Kitaev}(2015)}]{kitaev2015}%
  \BibitemOpen
  \bibfield  {author} {\bibinfo {author} {\bibfnamefont {A.}~\bibnamefont {Kitaev}},\ }\href {http://online.kitp.ucsb.edu/online/entangled15/} {\bibinfo {title} {A simple model of quantum holography}} (\bibinfo {year} {2015}),\ \bibinfo {note} {string seminar at KITP and Entanglement 2015 program, 12 February, 7 April and 27 May 2015, http://online.kitp.ucsb.edu/online/entangled15/}\BibitemShut {NoStop}%
\bibitem [{\citenamefont {Maldacena}\ \emph {et~al.}(2016)\citenamefont {Maldacena}, \citenamefont {Shenker},\ and\ \citenamefont {Stanford}}]{maldacena2016JHEP}%
  \BibitemOpen
  \bibfield  {author} {\bibinfo {author} {\bibfnamefont {J.}~\bibnamefont {Maldacena}}, \bibinfo {author} {\bibfnamefont {S.~H.}\ \bibnamefont {Shenker}},\ and\ \bibinfo {author} {\bibfnamefont {D.}~\bibnamefont {Stanford}},\ }\bibfield  {title} {\bibinfo {title} {A bound on chaos},\ }\href {https://doi.org/10.1007/JHEP08(2016)106} {\bibfield  {journal} {\bibinfo  {journal} {J. High Energy Phys{.}}\ }\textbf {\bibinfo {volume} {2016}},\ \bibinfo {pages} {106} (\bibinfo {year} {2016})}\BibitemShut {NoStop}%
\bibitem [{\citenamefont {Polchinski}(2015)}]{polchinski2015}%
  \BibitemOpen
  \bibfield  {author} {\bibinfo {author} {\bibfnamefont {J.}~\bibnamefont {Polchinski}},\ }\bibfield  {title} {\bibinfo {title} {{Chaos in the black hole S-matrix}},\ }\href {https://arxiv.org/abs/1505.08108} {\bibfield  {journal} {\bibinfo  {journal} {arXiv:1505.08108}\ } (\bibinfo {year} {2015})}\BibitemShut {NoStop}%
\bibitem [{\citenamefont {Hasegawa}\ and\ \citenamefont {Saphir}(1992)}]{Hasegawa_1992_unitarity}%
  \BibitemOpen
  \bibfield  {author} {\bibinfo {author} {\bibfnamefont {H.~H.}\ \bibnamefont {Hasegawa}}\ and\ \bibinfo {author} {\bibfnamefont {W.~C.}\ \bibnamefont {Saphir}},\ }\bibfield  {title} {\bibinfo {title} {{Unitarity and irreversibility in chaotic systems}},\ }\href {https://doi.org/10.1103/PhysRevA.46.7401} {\bibfield  {journal} {\bibinfo  {journal} {Phys. Rev. A}\ }\textbf {\bibinfo {volume} {46}},\ \bibinfo {pages} {7401} (\bibinfo {year} {1992})}\BibitemShut {NoStop}%
\bibitem [{\citenamefont {Malinin}\ and\ \citenamefont {Chernyak}(2008)}]{Malinin_nonlinear_2008}%
  \BibitemOpen
  \bibfield  {author} {\bibinfo {author} {\bibfnamefont {S.~V.}\ \bibnamefont {Malinin}}\ and\ \bibinfo {author} {\bibfnamefont {V.~Y.}\ \bibnamefont {Chernyak}},\ }\bibfield  {title} {\bibinfo {title} {{Classical nonlinear response of a chaotic system. II. Langevin dynamics and spectral decomposition}},\ }\href {https://doi.org/10.1103/PhysRevE.77.056202} {\bibfield  {journal} {\bibinfo  {journal} {Phys. Rev. E}\ }\textbf {\bibinfo {volume} {77}},\ \bibinfo {pages} {056202} (\bibinfo {year} {2008})}\BibitemShut {NoStop}%
\bibitem [{\citenamefont {Gaspard}\ \emph {et~al.}(1995)\citenamefont {Gaspard}, \citenamefont {Nicolis}, \citenamefont {Provata},\ and\ \citenamefont {Tasaki}}]{Gaspard_bifurcation_1995}%
  \BibitemOpen
  \bibfield  {author} {\bibinfo {author} {\bibfnamefont {P.}~\bibnamefont {Gaspard}}, \bibinfo {author} {\bibfnamefont {G.}~\bibnamefont {Nicolis}}, \bibinfo {author} {\bibfnamefont {A.}~\bibnamefont {Provata}},\ and\ \bibinfo {author} {\bibfnamefont {S.}~\bibnamefont {Tasaki}},\ }\bibfield  {title} {\bibinfo {title} {{Spectral signature of the pitchfork bifurcation: Liouville equation approach}},\ }\href {https://doi.org/10.1103/PhysRevE.51.74} {\bibfield  {journal} {\bibinfo  {journal} {Phys. Rev. E}\ }\textbf {\bibinfo {volume} {51}},\ \bibinfo {pages} {74} (\bibinfo {year} {1995})}\BibitemShut {NoStop}%
\bibitem [{\citenamefont {Pance}\ \emph {et~al.}(2000)\citenamefont {Pance}, \citenamefont {Lu},\ and\ \citenamefont {Sridhar}}]{pance2000PRL}%
  \BibitemOpen
  \bibfield  {author} {\bibinfo {author} {\bibfnamefont {K.}~\bibnamefont {Pance}}, \bibinfo {author} {\bibfnamefont {W.}~\bibnamefont {Lu}},\ and\ \bibinfo {author} {\bibfnamefont {S.}~\bibnamefont {Sridhar}},\ }\bibfield  {title} {\bibinfo {title} {{Quantum Fingerprints of Classical Ruelle-Pollicott Resonances}},\ }\href {https://doi.org/10.1103/PhysRevLett.85.2737} {\bibfield  {journal} {\bibinfo  {journal} {Phys. Rev. Lett.}\ }\textbf {\bibinfo {volume} {85}},\ \bibinfo {pages} {2737} (\bibinfo {year} {2000})}\BibitemShut {NoStop}%
\bibitem [{\citenamefont {Sridhar}\ and\ \citenamefont {Lu}(2002)}]{sridhar2002JSP}%
  \BibitemOpen
  \bibfield  {author} {\bibinfo {author} {\bibfnamefont {S.}~\bibnamefont {Sridhar}}\ and\ \bibinfo {author} {\bibfnamefont {W.}~\bibnamefont {Lu}},\ }\bibfield  {title} {\bibinfo {title} {{Sinai Billiards, Ruelle Zeta-functions and Ruelle Resonances: Microwave Experiments}},\ }\href {https://doi.org/10.1023/A:1019714808787} {\bibfield  {journal} {\bibinfo  {journal} {J. Stat. Phys.}\ }\textbf {\bibinfo {volume} {108}},\ \bibinfo {pages} {755} (\bibinfo {year} {2002})}\BibitemShut {NoStop}%
\bibitem [{\citenamefont {Gottesman}(1999)}]{gottesman1998}%
  \BibitemOpen
  \bibfield  {author} {\bibinfo {author} {\bibfnamefont {D.}~\bibnamefont {Gottesman}},\ }\bibfield  {title} {\bibinfo {title} {{Fault-Tolerant Quantum Computation with Higher-Dimensional Systems}},\ }in\ \href {https://doi.org/https://doi.org/10.1007/3-540-49208-9_27} {\emph {\bibinfo {booktitle} {Quantum Computing and Quantum Communications}}},\ \bibinfo {editor} {edited by\ \bibinfo {editor} {\bibfnamefont {C.~P.}\ \bibnamefont {Williams}}}\ (\bibinfo  {publisher} {Springer Berlin},\ \bibinfo {address} {Berlin},\ \bibinfo {year} {1999})\ p.\ \bibinfo {pages} {302–313}\BibitemShut {NoStop}%
\bibitem [{\citenamefont {Berry}(1985)}]{berry1985}%
  \BibitemOpen
  \bibfield  {author} {\bibinfo {author} {\bibfnamefont {M.~V.}\ \bibnamefont {Berry}},\ }\bibfield  {title} {\bibinfo {title} {{Semiclassical theory of spectral rigidity}},\ }\href {https://doi.org/10.1098/rspa.1985.0078} {\bibfield  {journal} {\bibinfo  {journal} {Proc. R. Soc. London A}\ }\textbf {\bibinfo {volume} {400}},\ \bibinfo {pages} {229} (\bibinfo {year} {1985})}\BibitemShut {NoStop}%
\bibitem [{\citenamefont {Sieber}\ and\ \citenamefont {Richter}(2001)}]{sieber2001}%
  \BibitemOpen
  \bibfield  {author} {\bibinfo {author} {\bibfnamefont {M.}~\bibnamefont {Sieber}}\ and\ \bibinfo {author} {\bibfnamefont {K.}~\bibnamefont {Richter}},\ }\bibfield  {title} {\bibinfo {title} {{Correlations between periodic orbits and their r\^{o}le in spectral statistics}},\ }\href {https://doi.org/10.1238/Physica.Topical.090a00128} {\bibfield  {journal} {\bibinfo  {journal} {Phys. Scr.}\ }\textbf {\bibinfo {volume} {2001}},\ \bibinfo {pages} {128} (\bibinfo {year} {2001})}\BibitemShut {NoStop}%
\bibitem [{\citenamefont {Sieber}(2002)}]{sieber2002}%
  \BibitemOpen
  \bibfield  {author} {\bibinfo {author} {\bibfnamefont {M.}~\bibnamefont {Sieber}},\ }\bibfield  {title} {\bibinfo {title} {{Leading off-diagonal approximation for the spectral form factor for uniformly hyperbolic systems}},\ }\href {https://doi.org/10.1088/0305-4470/35/42/104} {\bibfield  {journal} {\bibinfo  {journal} {J. Phys. A}\ }\textbf {\bibinfo {volume} {35}},\ \bibinfo {pages} {L613} (\bibinfo {year} {2002})}\BibitemShut {NoStop}%
\bibitem [{\citenamefont {Heusler}\ \emph {et~al.}(2004)\citenamefont {Heusler}, \citenamefont {M\"{u}ller}, \citenamefont {Braun},\ and\ \citenamefont {Haake}}]{heusler2004}%
  \BibitemOpen
  \bibfield  {author} {\bibinfo {author} {\bibfnamefont {S.}~\bibnamefont {Heusler}}, \bibinfo {author} {\bibfnamefont {S.}~\bibnamefont {M\"{u}ller}}, \bibinfo {author} {\bibfnamefont {P.}~\bibnamefont {Braun}},\ and\ \bibinfo {author} {\bibfnamefont {F.}~\bibnamefont {Haake}},\ }\bibfield  {title} {\bibinfo {title} {{Universal spectral form factor for chaotic dynamics}},\ }\href {https://doi.org/10.1088/0305-4470/37/3/L02} {\bibfield  {journal} {\bibinfo  {journal} {J. Phys. A}\ }\textbf {\bibinfo {volume} {37}},\ \bibinfo {pages} {L31} (\bibinfo {year} {2004})}\BibitemShut {NoStop}%
\bibitem [{\citenamefont {M\"uller}\ \emph {et~al.}(2004)\citenamefont {M\"uller}, \citenamefont {Heusler}, \citenamefont {Braun}, \citenamefont {Haake},\ and\ \citenamefont {Altland}}]{muller2004}%
  \BibitemOpen
  \bibfield  {author} {\bibinfo {author} {\bibfnamefont {S.}~\bibnamefont {M\"uller}}, \bibinfo {author} {\bibfnamefont {S.}~\bibnamefont {Heusler}}, \bibinfo {author} {\bibfnamefont {P.}~\bibnamefont {Braun}}, \bibinfo {author} {\bibfnamefont {F.}~\bibnamefont {Haake}},\ and\ \bibinfo {author} {\bibfnamefont {A.}~\bibnamefont {Altland}},\ }\bibfield  {title} {\bibinfo {title} {{Semiclassical Foundation of Universality in Quantum Chaos}},\ }\href {https://doi.org/10.1103/PhysRevLett.93.014103} {\bibfield  {journal} {\bibinfo  {journal} {Phys. Rev. Lett.}\ }\textbf {\bibinfo {volume} {93}},\ \bibinfo {pages} {014103} (\bibinfo {year} {2004})}\BibitemShut {NoStop}%
\bibitem [{\citenamefont {M\"uller}\ \emph {et~al.}(2005)\citenamefont {M\"uller}, \citenamefont {Heusler}, \citenamefont {Braun}, \citenamefont {Haake},\ and\ \citenamefont {Altland}}]{muller2005}%
  \BibitemOpen
  \bibfield  {author} {\bibinfo {author} {\bibfnamefont {S.}~\bibnamefont {M\"uller}}, \bibinfo {author} {\bibfnamefont {S.}~\bibnamefont {Heusler}}, \bibinfo {author} {\bibfnamefont {P.}~\bibnamefont {Braun}}, \bibinfo {author} {\bibfnamefont {F.}~\bibnamefont {Haake}},\ and\ \bibinfo {author} {\bibfnamefont {A.}~\bibnamefont {Altland}},\ }\bibfield  {title} {\bibinfo {title} {{Periodic-orbit theory of universality in quantum chaos}},\ }\href {https://doi.org/10.1103/PhysRevE.72.046207} {\bibfield  {journal} {\bibinfo  {journal} {Phys. Rev. E}\ }\textbf {\bibinfo {volume} {72}},\ \bibinfo {pages} {046207} (\bibinfo {year} {2005})}\BibitemShut {NoStop}%
\bibitem [{\citenamefont {Kos}\ \emph {et~al.}(2018)\citenamefont {Kos}, \citenamefont {Ljubotina},\ and\ \citenamefont {Prosen}}]{Kos_ManyBody_2018}%
  \BibitemOpen
  \bibfield  {author} {\bibinfo {author} {\bibfnamefont {P.}~\bibnamefont {Kos}}, \bibinfo {author} {\bibfnamefont {M.}~\bibnamefont {Ljubotina}},\ and\ \bibinfo {author} {\bibfnamefont {T.}~\bibnamefont {Prosen}},\ }\bibfield  {title} {\bibinfo {title} {{Many-Body Quantum Chaos: Analytic Connection to Random Matrix Theory}},\ }\href {https://doi.org/10.1103/PhysRevX.8.021062} {\bibfield  {journal} {\bibinfo  {journal} {Phys. Rev. X}\ }\textbf {\bibinfo {volume} {8}},\ \bibinfo {pages} {021062} (\bibinfo {year} {2018})}\BibitemShut {NoStop}%
\bibitem [{\citenamefont {Garratt}\ and\ \citenamefont {Chalker}(2021{\natexlab{a}})}]{Garratt_Local_2021}%
  \BibitemOpen
  \bibfield  {author} {\bibinfo {author} {\bibfnamefont {S.~J.}\ \bibnamefont {Garratt}}\ and\ \bibinfo {author} {\bibfnamefont {J.~T.}\ \bibnamefont {Chalker}},\ }\bibfield  {title} {\bibinfo {title} {{Local Pairing of Feynman Histories in Many-Body Floquet Models}},\ }\href {https://doi.org/10.1103/PhysRevX.11.021051} {\bibfield  {journal} {\bibinfo  {journal} {Phys. Rev. X}\ }\textbf {\bibinfo {volume} {11}},\ \bibinfo {pages} {021051} (\bibinfo {year} {2021}{\natexlab{a}})}\BibitemShut {NoStop}%
\bibitem [{\citenamefont {Can}(2019)}]{can2019JPhysA}%
  \BibitemOpen
  \bibfield  {author} {\bibinfo {author} {\bibfnamefont {T.}~\bibnamefont {Can}},\ }\bibfield  {title} {\bibinfo {title} {{Random Lindblad dynamics}},\ }\href {https://doi.org/10.1088/1751-8121/ab4d26} {\bibfield  {journal} {\bibinfo  {journal} {J. Phys. A}\ }\textbf {\bibinfo {volume} {52}},\ \bibinfo {pages} {485302} (\bibinfo {year} {2019})}\BibitemShut {NoStop}%
\bibitem [{\citenamefont {Kawabata}\ \emph {et~al.}(2023)\citenamefont {Kawabata}, \citenamefont {Kulkarni}, \citenamefont {Li}, \citenamefont {Numasawa},\ and\ \citenamefont {Ryu}}]{kawabata2023PRB}%
  \BibitemOpen
  \bibfield  {author} {\bibinfo {author} {\bibfnamefont {K.}~\bibnamefont {Kawabata}}, \bibinfo {author} {\bibfnamefont {A.}~\bibnamefont {Kulkarni}}, \bibinfo {author} {\bibfnamefont {J.}~\bibnamefont {Li}}, \bibinfo {author} {\bibfnamefont {T.}~\bibnamefont {Numasawa}},\ and\ \bibinfo {author} {\bibfnamefont {S.}~\bibnamefont {Ryu}},\ }\bibfield  {title} {\bibinfo {title} {{Dynamical quantum phase transitions in Sachdev-Ye-Kitaev Lindbladians}},\ }\href {https://doi.org/10.1103/PhysRevB.108.075110} {\bibfield  {journal} {\bibinfo  {journal} {Phys. Rev. B}\ }\textbf {\bibinfo {volume} {108}},\ \bibinfo {pages} {075110} (\bibinfo {year} {2023})}\BibitemShut {NoStop}%
\bibitem [{\citenamefont {Braun}(2001)}]{braun2001}%
  \BibitemOpen
  \bibfield  {author} {\bibinfo {author} {\bibfnamefont {D.}~\bibnamefont {Braun}},\ }\href@noop {} {\emph {\bibinfo {title} {{Dissipative Quantum Chaos and Decoherence}}}}\ (\bibinfo  {publisher} {Springer},\ \bibinfo {address} {Heidelberg},\ \bibinfo {year} {2001})\BibitemShut {NoStop}%
\bibitem [{\citenamefont {Fyodorov}\ \emph {et~al.}(1997)\citenamefont {Fyodorov}, \citenamefont {Khoruzhenko},\ and\ \citenamefont {Sommers}}]{fyodorov1997}%
  \BibitemOpen
  \bibfield  {author} {\bibinfo {author} {\bibfnamefont {Y.~V.}\ \bibnamefont {Fyodorov}}, \bibinfo {author} {\bibfnamefont {B.~A.}\ \bibnamefont {Khoruzhenko}},\ and\ \bibinfo {author} {\bibfnamefont {H.-J.}\ \bibnamefont {Sommers}},\ }\bibfield  {title} {\bibinfo {title} {{Almost Hermitian Random Matrices: Crossover from Wigner-Dyson to Ginibre Eigenvalue Statistics}},\ }\href {https://doi.org/10.1103/PhysRevLett.79.557} {\bibfield  {journal} {\bibinfo  {journal} {Phys. Rev. Lett.}\ }\textbf {\bibinfo {volume} {79}},\ \bibinfo {pages} {557} (\bibinfo {year} {1997})}\BibitemShut {NoStop}%
\bibitem [{\citenamefont {Li}\ \emph {et~al.}(2021)\citenamefont {Li}, \citenamefont {Prosen},\ and\ \citenamefont {Chan}}]{li2021PRL}%
  \BibitemOpen
  \bibfield  {author} {\bibinfo {author} {\bibfnamefont {J.}~\bibnamefont {Li}}, \bibinfo {author} {\bibfnamefont {T.}~\bibnamefont {Prosen}},\ and\ \bibinfo {author} {\bibfnamefont {A.}~\bibnamefont {Chan}},\ }\bibfield  {title} {\bibinfo {title} {{Spectral Statistics of Non-Hermitian Matrices and Dissipative Quantum Chaos}},\ }\href {https://doi.org/10.1103/PhysRevLett.127.170602} {\bibfield  {journal} {\bibinfo  {journal} {Phys. Rev. Lett.}\ }\textbf {\bibinfo {volume} {127}},\ \bibinfo {pages} {170602} (\bibinfo {year} {2021})}\BibitemShut {NoStop}%
\bibitem [{\citenamefont {Shivam}\ \emph {et~al.}(2023)\citenamefont {Shivam}, \citenamefont {De~Luca}, \citenamefont {Huse},\ and\ \citenamefont {Chan}}]{shivam2023PRL}%
  \BibitemOpen
  \bibfield  {author} {\bibinfo {author} {\bibfnamefont {S.}~\bibnamefont {Shivam}}, \bibinfo {author} {\bibfnamefont {A.}~\bibnamefont {De~Luca}}, \bibinfo {author} {\bibfnamefont {D.~A.}\ \bibnamefont {Huse}},\ and\ \bibinfo {author} {\bibfnamefont {A.}~\bibnamefont {Chan}},\ }\bibfield  {title} {\bibinfo {title} {{Many-Body Quantum Chaos and Emergence of Ginibre Ensemble}},\ }\href {https://doi.org/10.1103/PhysRevLett.130.140403} {\bibfield  {journal} {\bibinfo  {journal} {Phys. Rev. Lett.}\ }\textbf {\bibinfo {volume} {130}},\ \bibinfo {pages} {140403} (\bibinfo {year} {2023})}\BibitemShut {NoStop}%
\bibitem [{\citenamefont {Ghosh}\ \emph {et~al.}(2022)\citenamefont {Ghosh}, \citenamefont {Gupta},\ and\ \citenamefont {Kulkarni}}]{gosh2022PRB}%
  \BibitemOpen
  \bibfield  {author} {\bibinfo {author} {\bibfnamefont {S.}~\bibnamefont {Ghosh}}, \bibinfo {author} {\bibfnamefont {S.}~\bibnamefont {Gupta}},\ and\ \bibinfo {author} {\bibfnamefont {M.}~\bibnamefont {Kulkarni}},\ }\bibfield  {title} {\bibinfo {title} {{Spectral properties of disordered interacting non-Hermitian systems}},\ }\href {https://doi.org/10.1103/PhysRevB.106.134202} {\bibfield  {journal} {\bibinfo  {journal} {Phys. Rev. B}\ }\textbf {\bibinfo {volume} {106}},\ \bibinfo {pages} {134202} (\bibinfo {year} {2022})}\BibitemShut {NoStop}%
\bibitem [{\citenamefont {Chan}\ \emph {et~al.}(2022)\citenamefont {Chan}, \citenamefont {Shivam}, \citenamefont {Huse},\ and\ \citenamefont {De~Luca}}]{chan2022NatComm}%
  \BibitemOpen
  \bibfield  {author} {\bibinfo {author} {\bibfnamefont {A.}~\bibnamefont {Chan}}, \bibinfo {author} {\bibfnamefont {S.}~\bibnamefont {Shivam}}, \bibinfo {author} {\bibfnamefont {D.~A.}\ \bibnamefont {Huse}},\ and\ \bibinfo {author} {\bibfnamefont {A.}~\bibnamefont {De~Luca}},\ }\bibfield  {title} {\bibinfo {title} {{Many-body quantum chaos and space-time translational invariance}},\ }\href {https://doi.org/10.1038/s41467-022-34318-1} {\bibfield  {journal} {\bibinfo  {journal} {Nat. Commun.}\ }\textbf {\bibinfo {volume} {13}},\ \bibinfo {pages} {7484} (\bibinfo {year} {2022})}\BibitemShut {NoStop}%
\bibitem [{\citenamefont {Xu}\ \emph {et~al.}(2019)\citenamefont {Xu}, \citenamefont {Garc\'{\i}a-Pintos}, \citenamefont {Chenu},\ and\ \citenamefont {del Campo}}]{xu2019PRL}%
  \BibitemOpen
  \bibfield  {author} {\bibinfo {author} {\bibfnamefont {Z.}~\bibnamefont {Xu}}, \bibinfo {author} {\bibfnamefont {L.~P.}\ \bibnamefont {Garc\'{\i}a-Pintos}}, \bibinfo {author} {\bibfnamefont {A.}~\bibnamefont {Chenu}},\ and\ \bibinfo {author} {\bibfnamefont {A.}~\bibnamefont {del Campo}},\ }\bibfield  {title} {\bibinfo {title} {{Extreme Decoherence and Quantum Chaos}},\ }\href {https://doi.org/10.1103/PhysRevLett.122.014103} {\bibfield  {journal} {\bibinfo  {journal} {Phys. Rev. Lett.}\ }\textbf {\bibinfo {volume} {122}},\ \bibinfo {pages} {014103} (\bibinfo {year} {2019})}\BibitemShut {NoStop}%
\bibitem [{\citenamefont {Xu}\ \emph {et~al.}(2021)\citenamefont {Xu}, \citenamefont {Chenu}, \citenamefont {Prosen},\ and\ \citenamefont {del Campo}}]{xu2021PRB}%
  \BibitemOpen
  \bibfield  {author} {\bibinfo {author} {\bibfnamefont {Z.}~\bibnamefont {Xu}}, \bibinfo {author} {\bibfnamefont {A.}~\bibnamefont {Chenu}}, \bibinfo {author} {\bibfnamefont {T.}~\bibnamefont {Prosen}},\ and\ \bibinfo {author} {\bibfnamefont {A.}~\bibnamefont {del Campo}},\ }\bibfield  {title} {\bibinfo {title} {{Thermofield dynamics: Quantum chaos versus decoherence}},\ }\href {https://doi.org/10.1103/PhysRevB.103.064309} {\bibfield  {journal} {\bibinfo  {journal} {Phys. Rev. B}\ }\textbf {\bibinfo {volume} {103}},\ \bibinfo {pages} {064309} (\bibinfo {year} {2021})}\BibitemShut {NoStop}%
\bibitem [{\citenamefont {Cornelius}\ \emph {et~al.}(2022)\citenamefont {Cornelius}, \citenamefont {Xu}, \citenamefont {Saxena}, \citenamefont {Chenu},\ and\ \citenamefont {del Campo}}]{cornelius2022PRL}%
  \BibitemOpen
  \bibfield  {author} {\bibinfo {author} {\bibfnamefont {J.}~\bibnamefont {Cornelius}}, \bibinfo {author} {\bibfnamefont {Z.}~\bibnamefont {Xu}}, \bibinfo {author} {\bibfnamefont {A.}~\bibnamefont {Saxena}}, \bibinfo {author} {\bibfnamefont {A.}~\bibnamefont {Chenu}},\ and\ \bibinfo {author} {\bibfnamefont {A.}~\bibnamefont {del Campo}},\ }\bibfield  {title} {\bibinfo {title} {{Spectral Filtering Induced by Non-Hermitian Evolution with Balanced Gain and Loss: Enhancing Quantum Chaos}},\ }\href {https://doi.org/10.1103/PhysRevLett.128.190402} {\bibfield  {journal} {\bibinfo  {journal} {Phys. Rev. Lett.}\ }\textbf {\bibinfo {volume} {128}},\ \bibinfo {pages} {190402} (\bibinfo {year} {2022})}\BibitemShut {NoStop}%
\bibitem [{\citenamefont {Matsoukas-Roubeas}\ \emph {et~al.}(2024)\citenamefont {Matsoukas-Roubeas}, \citenamefont {Prosen},\ and\ \citenamefont {del Campo}}]{matsoukas-roubeas2023ARXIV}%
  \BibitemOpen
  \bibfield  {author} {\bibinfo {author} {\bibfnamefont {A.~S.}\ \bibnamefont {Matsoukas-Roubeas}}, \bibinfo {author} {\bibfnamefont {T.}~\bibnamefont {Prosen}},\ and\ \bibinfo {author} {\bibfnamefont {A.}~\bibnamefont {del Campo}},\ }\bibfield  {title} {\bibinfo {title} {Quantum {C}haos and {C}oherence: {R}andom {P}arametric {Q}uantum {C}hannels},\ }\href {https://doi.org/10.22331/q-2024-08-27-1446} {\bibfield  {journal} {\bibinfo  {journal} {{Quantum}}\ }\textbf {\bibinfo {volume} {8}},\ \bibinfo {pages} {1446} (\bibinfo {year} {2024})}\BibitemShut {NoStop}%
\bibitem [{\citenamefont {Kos}\ \emph {et~al.}(2021{\natexlab{a}})\citenamefont {Kos}, \citenamefont {Bertini},\ and\ \citenamefont {Prosen}}]{kos2021PRL}%
  \BibitemOpen
  \bibfield  {author} {\bibinfo {author} {\bibfnamefont {P.}~\bibnamefont {Kos}}, \bibinfo {author} {\bibfnamefont {B.}~\bibnamefont {Bertini}},\ and\ \bibinfo {author} {\bibfnamefont {T.}~\bibnamefont {Prosen}},\ }\bibfield  {title} {\bibinfo {title} {Chaos and ergodicity in extended quantum systems with noisy driving},\ }\href {https://doi.org/10.1103/PhysRevLett.126.190601} {\bibfield  {journal} {\bibinfo  {journal} {Phys. Rev. Lett.}\ }\textbf {\bibinfo {volume} {126}},\ \bibinfo {pages} {190601} (\bibinfo {year} {2021}{\natexlab{a}})}\BibitemShut {NoStop}%
\bibitem [{\citenamefont {Kos}\ \emph {et~al.}(2021{\natexlab{b}})\citenamefont {Kos}, \citenamefont {Prosen},\ and\ \citenamefont {Bertini}}]{kos2021PRB}%
  \BibitemOpen
  \bibfield  {author} {\bibinfo {author} {\bibfnamefont {P.}~\bibnamefont {Kos}}, \bibinfo {author} {\bibfnamefont {T.}~\bibnamefont {Prosen}},\ and\ \bibinfo {author} {\bibfnamefont {B.}~\bibnamefont {Bertini}},\ }\bibfield  {title} {\bibinfo {title} {{Thermalization dynamics and spectral statistics of extended systems with thermalizing boundaries}},\ }\href {https://doi.org/10.1103/PhysRevB.104.214303} {\bibfield  {journal} {\bibinfo  {journal} {Phys. Rev. B}\ }\textbf {\bibinfo {volume} {104}},\ \bibinfo {pages} {214303} (\bibinfo {year} {2021}{\natexlab{b}})}\BibitemShut {NoStop}%
\bibitem [{\citenamefont {Vikram}\ and\ \citenamefont {Galitski}(2024)}]{vikram2022ARXIV}%
  \BibitemOpen
  \bibfield  {author} {\bibinfo {author} {\bibfnamefont {A.}~\bibnamefont {Vikram}}\ and\ \bibinfo {author} {\bibfnamefont {V.}~\bibnamefont {Galitski}},\ }\bibfield  {title} {\bibinfo {title} {{Exact Universal Bounds on Quantum Dynamics and Fast Scrambling}},\ }\href {https://doi.org/10.1103/PhysRevLett.132.040402} {\bibfield  {journal} {\bibinfo  {journal} {Phys. Rev. Lett.}\ }\textbf {\bibinfo {volume} {132}},\ \bibinfo {pages} {040402} (\bibinfo {year} {2024})}\BibitemShut {NoStop}%
\bibitem [{\citenamefont {Roberts}\ and\ \citenamefont {Yoshida}(2017)}]{roberts2017JHEP}%
  \BibitemOpen
  \bibfield  {author} {\bibinfo {author} {\bibfnamefont {D.~A.}\ \bibnamefont {Roberts}}\ and\ \bibinfo {author} {\bibfnamefont {B.}~\bibnamefont {Yoshida}},\ }\bibfield  {title} {\bibinfo {title} {{Chaos and complexity by design}},\ }\href {https://doi.org/10.1007/JHEP04(2017)121} {\bibfield  {journal} {\bibinfo  {journal} {J. High Energy Phys{.}}\ }\textbf {\bibinfo {volume} {2017}},\ \bibinfo {pages} {121} (\bibinfo {year} {2017})}\BibitemShut {NoStop}%
\bibitem [{\citenamefont {Cotler}\ \emph {et~al.}(2017)\citenamefont {Cotler}, \citenamefont {Hunter-Jones}, \citenamefont {Liu},\ and\ \citenamefont {Yoshida}}]{cotler2017JHEP}%
  \BibitemOpen
  \bibfield  {author} {\bibinfo {author} {\bibfnamefont {J.}~\bibnamefont {Cotler}}, \bibinfo {author} {\bibfnamefont {N.}~\bibnamefont {Hunter-Jones}}, \bibinfo {author} {\bibfnamefont {J.}~\bibnamefont {Liu}},\ and\ \bibinfo {author} {\bibfnamefont {B.}~\bibnamefont {Yoshida}},\ }\bibfield  {title} {\bibinfo {title} {{Chaos, complexity, and random matrices}},\ }\href {https://doi.org/10.1007/JHEP11(2017)048} {\bibfield  {journal} {\bibinfo  {journal} {J. High Energy Phys{.}}\ }\textbf {\bibinfo {volume} {2017}},\ \bibinfo {pages} {48} (\bibinfo {year} {2017})}\BibitemShut {NoStop}%
\bibitem [{\citenamefont {Gharibyan}\ \emph {et~al.}(2018)\citenamefont {Gharibyan}, \citenamefont {Hanada}, \citenamefont {Shenker},\ and\ \citenamefont {Tezuka}}]{Gharibyan_Onset_2018}%
  \BibitemOpen
  \bibfield  {author} {\bibinfo {author} {\bibfnamefont {H.}~\bibnamefont {Gharibyan}}, \bibinfo {author} {\bibfnamefont {M.}~\bibnamefont {Hanada}}, \bibinfo {author} {\bibfnamefont {S.~H.}\ \bibnamefont {Shenker}},\ and\ \bibinfo {author} {\bibfnamefont {M.}~\bibnamefont {Tezuka}},\ }\bibfield  {title} {\bibinfo {title} {{Onset of random matrix behavior in scrambling systems}},\ }\href {https://doi.org/10.1007/JHEP07(2018)124} {\bibfield  {journal} {\bibinfo  {journal} {J. High Energy Phys{.}}\ }\textbf {\bibinfo {volume} {2018}},\ \bibinfo {pages} {124} (\bibinfo {year} {2018})}\BibitemShut {NoStop}%
\bibitem [{\citenamefont {Yoshimura}\ \emph {et~al.}(2025)\citenamefont {Yoshimura}, \citenamefont {Garratt},\ and\ \citenamefont {Chalker}}]{yoshimura2023operator}%
  \BibitemOpen
  \bibfield  {author} {\bibinfo {author} {\bibfnamefont {T.}~\bibnamefont {Yoshimura}}, \bibinfo {author} {\bibfnamefont {S.~J.}\ \bibnamefont {Garratt}},\ and\ \bibinfo {author} {\bibfnamefont {J.~T.}\ \bibnamefont {Chalker}},\ }\bibfield  {title} {\bibinfo {title} {{Operator dynamics in Floquet many-body systems}},\ }\href {https://doi.org/10.1103/PhysRevB.111.094316} {\bibfield  {journal} {\bibinfo  {journal} {Phys. Rev. B}\ }\textbf {\bibinfo {volume} {111}},\ \bibinfo {pages} {094316} (\bibinfo {year} {2025})}\BibitemShut {NoStop}%
\bibitem [{\citenamefont {Schuster}\ and\ \citenamefont {Yao}(2023)}]{schuster2023PRL}%
  \BibitemOpen
  \bibfield  {author} {\bibinfo {author} {\bibfnamefont {T.}~\bibnamefont {Schuster}}\ and\ \bibinfo {author} {\bibfnamefont {N.~Y.}\ \bibnamefont {Yao}},\ }\bibfield  {title} {\bibinfo {title} {{Operator Growth in Open Quantum Systems}},\ }\href {https://doi.org/10.1103/PhysRevLett.131.160402} {\bibfield  {journal} {\bibinfo  {journal} {Phys. Rev. Lett.}\ }\textbf {\bibinfo {volume} {131}},\ \bibinfo {pages} {160402} (\bibinfo {year} {2023})}\BibitemShut {NoStop}%
\bibitem [{\citenamefont {Bensa}\ and\ \citenamefont {\v{Z}nidari\v{c}}(2021)}]{Bensa_phantom_2021}%
  \BibitemOpen
  \bibfield  {author} {\bibinfo {author} {\bibfnamefont {J.}~\bibnamefont {Bensa}}\ and\ \bibinfo {author} {\bibfnamefont {M.}~\bibnamefont {\v{Z}nidari\v{c}}},\ }\bibfield  {title} {\bibinfo {title} {{Fastest Local Entanglement Scrambler, Multistage Thermalization, and a Non-Hermitian Phantom}},\ }\href {https://doi.org/10.1103/PhysRevX.11.031019} {\bibfield  {journal} {\bibinfo  {journal} {Phys. Rev. X}\ }\textbf {\bibinfo {volume} {11}},\ \bibinfo {pages} {031019} (\bibinfo {year} {2021})}\BibitemShut {NoStop}%
\bibitem [{\citenamefont {Bensa}\ and\ \citenamefont {\v{Z}nidari\v{c}}(2022)}]{Bensa_relaxation_2022}%
  \BibitemOpen
  \bibfield  {author} {\bibinfo {author} {\bibfnamefont {J.}~\bibnamefont {Bensa}}\ and\ \bibinfo {author} {\bibfnamefont {M.}~\bibnamefont {\v{Z}nidari\v{c}}},\ }\bibfield  {title} {\bibinfo {title} {{Two-step phantom relaxation of out-of-time-ordered correlations in random circuits}},\ }\href {https://doi.org/10.1103/PhysRevResearch.4.013228} {\bibfield  {journal} {\bibinfo  {journal} {Phys. Rev. Res.}\ }\textbf {\bibinfo {volume} {4}},\ \bibinfo {pages} {013228} (\bibinfo {year} {2022})}\BibitemShut {NoStop}%
\bibitem [{\citenamefont {\v{Z}nidari\v{c}}(2022)}]{znidaric2022}%
  \BibitemOpen
  \bibfield  {author} {\bibinfo {author} {\bibfnamefont {M.}~\bibnamefont {\v{Z}nidari\v{c}}},\ }\bibfield  {title} {\bibinfo {title} {{Solvable non-Hermitian skin effect in many-body unitary dynamics}},\ }\href {https://doi.org/10.1103/PhysRevResearch.4.033041} {\bibfield  {journal} {\bibinfo  {journal} {Phys. Rev. Res.}\ }\textbf {\bibinfo {volume} {4}},\ \bibinfo {pages} {033041} (\bibinfo {year} {2022})}\BibitemShut {NoStop}%
\bibitem [{\citenamefont {Žnidarič}(2023)}]{Znidaric_relaxation_2023}%
  \BibitemOpen
  \bibfield  {author} {\bibinfo {author} {\bibfnamefont {M.}~\bibnamefont {Žnidarič}},\ }\bibfield  {title} {\bibinfo {title} {{Two-step relaxation in local many-body Floquet systems}},\ }\href {https://doi.org/10.1088/1751-8121/acfc05} {\bibfield  {journal} {\bibinfo  {journal} {J. Phys. A}\ }\textbf {\bibinfo {volume} {56}},\ \bibinfo {pages} {434001} (\bibinfo {year} {2023})}\BibitemShut {NoStop}%
\bibitem [{\citenamefont {\v{Z}nidari\\v{c}}(2023)}]{znidaric2023_PRR}%
  \BibitemOpen
  \bibfield  {author} {\bibinfo {author} {\bibfnamefont {M.}~\bibnamefont {\v{Z}nidari\\v{c}}},\ }\bibfield  {title} {\bibinfo {title} {{Phantom relaxation rate of the average purity evolution in random circuits due to Jordan non-Hermitian skin effect and magic sums}},\ }\href {https://doi.org/10.1103/PhysRevResearch.5.033145} {\bibfield  {journal} {\bibinfo  {journal} {Phys. Rev. Res.}\ }\textbf {\bibinfo {volume} {5}},\ \bibinfo {pages} {033145} (\bibinfo {year} {2023})}\BibitemShut {NoStop}%
\bibitem [{\citenamefont {Bensa}(2024)}]{bensa2024}%
  \BibitemOpen
  \bibfield  {author} {\bibinfo {author} {\bibfnamefont {J.}~\bibnamefont {Bensa}},\ }\bibfield  {title} {\bibinfo {title} {{Arbitrary relaxation rate under non-Hermitian matrix iterations}},\ }\href {https://doi.org/10.1103/PhysRevResearch.6.023331} {\bibfield  {journal} {\bibinfo  {journal} {Phys. Rev. Res.}\ }\textbf {\bibinfo {volume} {6}},\ \bibinfo {pages} {023331} (\bibinfo {year} {2024})}\BibitemShut {NoStop}%
\bibitem [{\citenamefont {Bensa}\ and\ \citenamefont {\v{Z}nidari\v{c}}(2023)}]{bensa2024_PRA}%
  \BibitemOpen
  \bibfield  {author} {\bibinfo {author} {\bibfnamefont {J.}~\bibnamefont {Bensa}}\ and\ \bibinfo {author} {\bibfnamefont {M.}~\bibnamefont {\v{Z}nidari\v{c}}},\ }\bibfield  {title} {\bibinfo {title} {{Purity decay rate in random circuits with different configurations of gates}},\ }\href {https://doi.org/10.1103/PhysRevA.107.022604} {\bibfield  {journal} {\bibinfo  {journal} {Phys. Rev. A}\ }\textbf {\bibinfo {volume} {107}},\ \bibinfo {pages} {022604} (\bibinfo {year} {2023})}\BibitemShut {NoStop}%
\bibitem [{\citenamefont {Jonay}\ and\ \citenamefont {Zhou}(2024)}]{Jonay_thermalization_2024}%
  \BibitemOpen
  \bibfield  {author} {\bibinfo {author} {\bibfnamefont {C.}~\bibnamefont {Jonay}}\ and\ \bibinfo {author} {\bibfnamefont {T.}~\bibnamefont {Zhou}},\ }\bibfield  {title} {\bibinfo {title} {Physical theory of two-stage thermalization},\ }\href {https://doi.org/10.1103/PhysRevB.110.L020306} {\bibfield  {journal} {\bibinfo  {journal} {Phys. Rev. B}\ }\textbf {\bibinfo {volume} {110}},\ \bibinfo {pages} {L020306} (\bibinfo {year} {2024})}\BibitemShut {NoStop}%
\bibitem [{\citenamefont {Jonay}\ \emph {et~al.}(2024)\citenamefont {Jonay}, \citenamefont {Li},\ and\ \citenamefont {Zhou}}]{jonay2024twostagerelaxationoperatorsdomain}%
  \BibitemOpen
  \bibfield  {author} {\bibinfo {author} {\bibfnamefont {C.}~\bibnamefont {Jonay}}, \bibinfo {author} {\bibfnamefont {C.}~\bibnamefont {Li}},\ and\ \bibinfo {author} {\bibfnamefont {T.}~\bibnamefont {Zhou}},\ }\bibfield  {title} {\bibinfo {title} {{Two-stage relaxation of operators through domain wall and magnon dynamics}},\ }\href {https://arxiv.org/abs/2411.07298} {\bibfield  {journal} {\bibinfo  {journal} {arXiv:2411.07298}\ } (\bibinfo {year} {2024})}\BibitemShut {NoStop}%
\bibitem [{\citenamefont {Nandkishore}\ and\ \citenamefont {Huse}(2015)}]{Nandkishore2015}%
  \BibitemOpen
  \bibfield  {author} {\bibinfo {author} {\bibfnamefont {R.}~\bibnamefont {Nandkishore}}\ and\ \bibinfo {author} {\bibfnamefont {D.~A.}\ \bibnamefont {Huse}},\ }\bibfield  {title} {\bibinfo {title} {Many-body localization and thermalization in quantum statistical mechanics},\ }\href {https://doi.org/10.1146/annurev-conmatphys-031214-014726} {\bibfield  {journal} {\bibinfo  {journal} {Annual Review of Condensed Matter Physics}\ }\textbf {\bibinfo {volume} {6}},\ \bibinfo {pages} {15–38} (\bibinfo {year} {2015})}\BibitemShut {NoStop}%
\bibitem [{\citenamefont {Abanin}\ \emph {et~al.}(2019)\citenamefont {Abanin}, \citenamefont {Altman}, \citenamefont {Bloch},\ and\ \citenamefont {Serbyn}}]{Abanin_MBLreview_2019}%
  \BibitemOpen
  \bibfield  {author} {\bibinfo {author} {\bibfnamefont {D.~A.}\ \bibnamefont {Abanin}}, \bibinfo {author} {\bibfnamefont {E.}~\bibnamefont {Altman}}, \bibinfo {author} {\bibfnamefont {I.}~\bibnamefont {Bloch}},\ and\ \bibinfo {author} {\bibfnamefont {M.}~\bibnamefont {Serbyn}},\ }\bibfield  {title} {\bibinfo {title} {Colloquium: Many-body localization, thermalization, and entanglement},\ }\href {https://doi.org/10.1103/RevModPhys.91.021001} {\bibfield  {journal} {\bibinfo  {journal} {Rev. Mod. Phys.}\ }\textbf {\bibinfo {volume} {91}},\ \bibinfo {pages} {021001} (\bibinfo {year} {2019})}\BibitemShut {NoStop}%
\bibitem [{\citenamefont {Friedman}\ \emph {et~al.}(2019)\citenamefont {Friedman}, \citenamefont {Chan}, \citenamefont {De~Luca},\ and\ \citenamefont {Chalker}}]{Friedman_Spectral_2019}%
  \BibitemOpen
  \bibfield  {author} {\bibinfo {author} {\bibfnamefont {A.~J.}\ \bibnamefont {Friedman}}, \bibinfo {author} {\bibfnamefont {A.}~\bibnamefont {Chan}}, \bibinfo {author} {\bibfnamefont {A.}~\bibnamefont {De~Luca}},\ and\ \bibinfo {author} {\bibfnamefont {J.~T.}\ \bibnamefont {Chalker}},\ }\bibfield  {title} {\bibinfo {title} {{Spectral Statistics and Many-Body Quantum Chaos with Conserved Charge}},\ }\href {https://doi.org/10.1103/PhysRevLett.123.210603} {\bibfield  {journal} {\bibinfo  {journal} {Phys. Rev. Lett.}\ }\textbf {\bibinfo {volume} {123}},\ \bibinfo {pages} {210603} (\bibinfo {year} {2019})}\BibitemShut {NoStop}%
\bibitem [{\citenamefont {Buča}\ and\ \citenamefont {Prosen}(2012)}]{buca2012}%
  \BibitemOpen
  \bibfield  {author} {\bibinfo {author} {\bibfnamefont {B.}~\bibnamefont {Buča}}\ and\ \bibinfo {author} {\bibfnamefont {T.}~\bibnamefont {Prosen}},\ }\bibfield  {title} {\bibinfo {title} {{A note on symmetry reductions of the Lindblad equation: transport in constrained open spin chains}},\ }\href {https://doi.org/10.1088/1367-2630/14/7/073007} {\bibfield  {journal} {\bibinfo  {journal} {New J. Phys.}\ }\textbf {\bibinfo {volume} {14}},\ \bibinfo {pages} {073007} (\bibinfo {year} {2012})}\BibitemShut {NoStop}%
\bibitem [{\citenamefont {Garratt}\ and\ \citenamefont {Chalker}(2021{\natexlab{b}})}]{Garratt_ManyBody_2021}%
  \BibitemOpen
  \bibfield  {author} {\bibinfo {author} {\bibfnamefont {S.~J.}\ \bibnamefont {Garratt}}\ and\ \bibinfo {author} {\bibfnamefont {J.~T.}\ \bibnamefont {Chalker}},\ }\bibfield  {title} {\bibinfo {title} {Many-body delocalization as symmetry breaking},\ }\href {https://doi.org/10.1103/PhysRevLett.127.026802} {\bibfield  {journal} {\bibinfo  {journal} {Phys. Rev. Lett.}\ }\textbf {\bibinfo {volume} {127}},\ \bibinfo {pages} {026802} (\bibinfo {year} {2021}{\natexlab{b}})}\BibitemShut {NoStop}%
\bibitem [{\citenamefont {Bertini}\ \emph {et~al.}(2025)\citenamefont {Bertini}, \citenamefont {Klobas}, \citenamefont {Kos},\ and\ \citenamefont {Malz}}]{bertini2024quantumclassicaldynamicsrandom}%
  \BibitemOpen
  \bibfield  {author} {\bibinfo {author} {\bibfnamefont {B.}~\bibnamefont {Bertini}}, \bibinfo {author} {\bibfnamefont {K.}~\bibnamefont {Klobas}}, \bibinfo {author} {\bibfnamefont {P.}~\bibnamefont {Kos}},\ and\ \bibinfo {author} {\bibfnamefont {D.}~\bibnamefont {Malz}},\ }\bibfield  {title} {\bibinfo {title} {{Quantum and Classical Dynamics with Random Permutation Circuits}},\ }\href {https://doi.org/10.1103/PhysRevX.15.011015} {\bibfield  {journal} {\bibinfo  {journal} {Phys. Rev. X}\ }\textbf {\bibinfo {volume} {15}},\ \bibinfo {pages} {011015} (\bibinfo {year} {2025})}\BibitemShut {NoStop}%
\end{thebibliography}%

\end{document}